\documentclass[12pt]{article}
\usepackage{graphicx}
\usepackage{amssymb}
\usepackage[round]{natbib}
\def\includefigs{\let\ifincfigs=\iftrue}
\def\noincludefigs{\let\ifincfigs=\iffalse}
\includefigs

\newbox\epsfvertlab
\newbox\epsfhorlab
\newbox\epsffiglab

\newdimen\epsfvlabsize
\newdimen\scott

\def\setvlabel#1{\setbox\epsfvertlab=\vbox{\hbox{#1}}}%
\def\sethlabel#1{\setbox\epsfhorlab=\vbox{\hbox{#1}}}%
\def\figlab#1 #2 #3{\setbox\epsffiglab=\vbox to 0pt{%
\ifvoid\epsffiglab\else\box\epsffiglab\fi\vss\hbox to 0pt{\raise #2 \hbox{\hskip #1 #3}\hss}}}

\newdimen\fighor
\newdimen\figver
\newbox\rotbox
\long\def\lrlap#1{\hbox to 0pt{#1\hss}}
\long\def\verttex#1#2#3{{\fighor = #1\figver = #2\vbox to \figver{\vss%
\hbox to \fighor{\hfill\hsize=\fighor%
\lrlap{\rotstart{-90 rotate}\vbox to \fighor{#3\vfil}\rotfinish}}}}}

\def\dvipsvspec#1{\special{ps:#1}}
\def\dvipsrotstart#1{\dvipsvspec{gsave currentpoint currentpoint translate
   #1 neg exch neg exch translate}}
\def\dvipsrotfinish{\dvipsvspec{currentpoint grestore moveto}}

\def\rotstart#1{\dvipsrotstart{#1}}
\def\rotfinish{\dvipsrotfinish}

\def\epsfsetlab{%
\ifvoid\epsfvertlab%
\else%
\verttex{\epsfvlabsize}{\epsfysize}%
{\hbox to \epsfysize{\hss\box\epsfvertlab\hss}}%
\fi%
\ifvoid\epsfhorlab%
\else%
\scott=\epsfxsize%
\advance\scott by \epsfvlabsize%
\rlap{\vtop{\hrule height0pt\hbox to \scott{\hss\box\epsfhorlab\hss}}}%
\fi%
}

\def\epsfsetover{\ifvoid\epsffiglab\else\box\epsffiglab\fi}

\newread\epsffilein    
\newif\ifepsffileok    
\newif\ifepsfbbfound   
\newif\ifepsfverbose   
\newdimen\epsfxsize    
\newdimen\epsfysize    
\newdimen\epsftsize    
\newdimen\epsfrsize    
\newdimen\epsftmp      
\newdimen\pspoints     
\def\epsfbox#1{
   \ifvoid\epsfvertlab%
   \else\epsfvlabsize=\ht\epsfvertlab \advance\epsfvlabsize by \dp\epsfvertlab\fi%
   \leavevmode\global\def\epsfllx{72}\global\def\epsflly{72}%
   \global\def\epsfurx{540}\global\def\epsfury{720}%
   \def\lbracket{[}\def\testit{#1}\ifx\testit\lbracket
   \let\next=\epsfgetlitbb\else\let\next=\epsfnormal\fi\next{#1}}%
\def\epsfgetlitbb#1#2 #3 #4 #5]#6{\epsfgrab #2 #3 #4 #5 .\\%
   \epsfsetgraph{#6}}%
\def\epsfnormal#1{\epsfgetbb{#1}\epsfsetgraph{#1}}%
\def\epsfgetbb#1{%
\openin\epsffilein=#1
\ifeof\epsffilein\errmessage{I couldn't open #1, will ignore it}\else
   {\epsffileoktrue \chardef\other=12
    \def\do##1{\catcode`##1=\other}\dospecials \catcode`\ =10
    \loop
       \read\epsffilein to \epsffileline
       \ifeof\epsffilein\epsffileokfalse\else
          \expandafter\epsfaux\epsffileline:. \\%
       \fi
   \ifepsffileok\repeat
   \ifepsfbbfound\else
    \ifepsfverbose\message{No bounding box comment in #1; using defaults}\fi\fi
   }\closein\epsffilein\fi}%
\def\epsfsetgraph#1{%
   \epsfrsize=\epsfury\pspoints
   \advance\epsfrsize by-\epsflly\pspoints
   \epsftsize=\epsfurx\pspoints
   \advance\epsftsize by-\epsfllx\pspoints
   \epsfxsize\epsfsize\epsftsize\epsfrsize
   \ifnum\epsfxsize=0 \ifnum\epsfysize=0
      \epsfxsize=\epsftsize \epsfysize=\epsfrsize
     \else\epsftmp=\epsftsize \divide\epsftmp\epsfrsize
       \epsfxsize=\epsfysize \multiply\epsfxsize\epsftmp
       \multiply\epsftmp\epsfrsize \advance\epsftsize-\epsftmp
       \epsftmp=\epsfysize
       \loop \advance\epsftsize\epsftsize \divide\epsftmp 2
       \ifnum\epsftmp>0
          \ifnum\epsftsize<\epsfrsize\else
             \advance\epsftsize-\epsfrsize \advance\epsfxsize\epsftmp \fi
       \repeat
     \fi
   \else\epsftmp=\epsfrsize \divide\epsftmp\epsftsize
     \epsfysize=\epsfxsize \multiply\epsfysize\epsftmp   
     \multiply\epsftmp\epsftsize \advance\epsfrsize-\epsftmp
     \epsftmp=\epsfxsize
     \loop \advance\epsfrsize\epsfrsize \divide\epsftmp 2
     \ifnum\epsftmp>0
        \ifnum\epsfrsize<\epsftsize\else
           \advance\epsfrsize-\epsftsize \advance\epsfysize\epsftmp \fi
     \repeat     
   \fi
   \ifepsfverbose\message{#1: width=\the\epsfxsize, height=\the\epsfysize}\fi
   \epsftmp=10\epsfxsize \divide\epsftmp\pspoints
   \epsfsetlab%
   \ifincfigs%
     \vbox to\epsfysize{\vfil\hbox to\epsfxsize{%
        \includegraphics{#1}%
        \epsfsetover\hfil}}%
   \else%
     \epsfsetover%
     \vbox to\epsfysize{\hrule\vss\hbox to\epsfxsize{\vrule height
                        \epsfysize\hfil\vrule}\vss\hrule}%
   \fi%
\epsfxsize=0pt\epsfysize=0pt}%

{\catcode`\%=12 \global\let\epsfpercent=
\long\def\epsfaux#1#2:#3\\{\ifx#1\epsfpercent
   \def\testit{#2}\ifx\testit\epsfbblit
      \epsfgrab #3 . . . \\%
      \epsffileokfalse
      \global\epsfbbfoundtrue
   \fi\else\ifx#1\par\else\epsffileokfalse\fi\fi}%
\def\epsfgrab #1 #2 #3 #4 #5\\{%
   \global\def\epsfllx{#1}\ifx\epsfllx\empty
      \epsfgrab #2 #3 #4 #5 .\\\else
   \global\def\epsflly{#2}%
   \global\def\epsfurx{#3}\global\def\epsfury{#4}\fi}%
\def\epsfsize#1#2{\epsfxsize}
\def\ifspace{\ifcat\issp.\else~\fi}
\def\tspace{\futurelet\issp\ifspace}
\def\a{({\it a\kern 1pt})\tspace}
\def\b{({\it b\kern 1pt})\tspace}
\def\c{({\it c\kern 1pt})\tspace}
\def\d{({\it d\kern 1pt})\tspace}
\def\e{({\it e\kern 1pt})\tspace}
\def\f{({\it f\kern 1pt})\tspace}
\def\g{({\it g\kern 1pt})\tspace}
\def\h{({\it h\kern 1pt})\tspace}
\def\i{({\it i\kern 1pt})\tspace}
\def\j{({\it j\kern 1pt})\tspace}
\def\abc#1{({\it #1\kern 1pt})\tspace}

\newcount\ndots
\def\drawline#1#2{\raise 2.5pt\vbox{\hrule width #1pt height #2pt}}

\def\trian{\raise 1.25pt\hbox{$\scriptscriptstyle\triangle$}\nobreak\ }
\def\solidtrian{\raise 1.25pt
\hbox to 3bp{
\hfill}\nobreak\ }
\def\dsolidtrian{\raise 1.25pt
\hbox to 3bp{
\hfill}\nobreak\ }
\def\soliddiamond{\raise 1.25pt
\hbox to 4bp{
\hfill}\nobreak\ }

\def\square{${\vcenter{\hrule height .4pt 
              \hbox{\vrule width .4pt height 3pt \kern 3pt \vrule width .4pt}
          \hrule height .4pt}}$\nobreak\ }

\def\plus{\raise 1.25pt \hbox{$\scriptscriptstyle +$}\nobreak\ }
\def\x{\raise 1.25pt \hbox{$\scriptscriptstyle \times$}\nobreak\ }
\def\legendtable#1{\vbox{\baselineskip=10pt\tabskip=0pt\let\\=\cr\halign{\hfil##\hskip 3pt&##\hfil\cr#1\crcr}}}
\def\lllegend#1 #2 #3{\figlab {#1} {#2} {\legendtable{#3}}}
\def\lrlegend#1 #2 #3{\figlab {#1} {#2} {\llap{\legendtable{#3}}}}
\def\ullegend#1 #2 #3{\figlab {#1} {#2} {\vtop{\hrule height 0pt\legendtable{#3}}}}
\def\urlegend#1 #2 #3{\figlab {#1} {#2} {\llap{\vtop{\hrule height 0pt\legendtable{#3}}}}}

\newdimen\xorigon
\newdimen\yorigon
\newdimen\scaleval
\newdimen\scaleorigon

\def\setxscale#1 #2 #3 #4 #5 {%
    \xorigon=#1\yorigon=#3%
    \scaleval=#2\advance\scaleval by -\xorigon%
    \tempdimen=#5 pt\advance\tempdimen by -#4pt%
    \divide\tempdimen by 1000%
    \divide\scaleval by \tempdimen%
    \scaleorigon=-#4pt\divide\scaleorigon by 1000%
    \multiply\scaleorigon by \scaleval}
\def\xtickup#1 #2{\tempdimen=#1pt\divide\tempdimen by 1000%
    \multiply\tempdimen by \scaleval\advance\tempdimen by \scaleorigon%
    \advance\tempdimen by \xorigon%
    \figlab {\tempdimen} {\yorigon} {\vbox {\hbox to 0pt{\hss #2\hss}%
        \baselineskip=8pt\lineskiplimit=-5pt%
        \hbox to 0pt{\hss \vrule height 3pt\hss}}}}
\def\xtickdown#1 #2{\tempdimen=#1pt\divide\tempdimen by 1000%
    \multiply\tempdimen by \scaleval\advance\tempdimen by \scaleorigon%
    \advance\tempdimen by \xorigon%
    \figlab {\tempdimen} {\yorigon} {\vbox to 0pt {\hbox to 0pt{\hss \vrule height 3pt\hss}%
        \nointerlineskip\vskip 3pt%
        \hbox to 0pt{\hss #2\hss}\vss}}}
\def\nofig#1#2{\leavevmode{\vbox {\hrule \hbox to #1{\vrule height #2 \hfill \vrule} \hrule}} }

\def\ts{t_s}
\def\te{t_e}
\def\rhog{\rho_g}
\def\rhop{\rho_p}
\def\rhos{\rho_s}

\def\be{\begin{equation}}
\def\ee{\end{equation}}
\def\ba{\begin{eqnarray}}
\def\ea{\end{eqnarray}}

\begin{document}
\begin{titlepage}
\begin{center}
\begin{large}
       {\bf Towards Initial Mass Functions for Asteroids and Kuiper Belt Objects} \\
Jeffrey N. Cuzzi, Robert C. Hogan, and William F. Bottke  \\
\end{large}
\today
\end{center}
\vspace{0.1 in}

\begin{abstract}

Our goal is to understand primary accretion of the first planetesimals. Some examples are seen today in the asteroid belt, providing the parent bodies for the primitive meteorites. The primitive meteorite record suggests that sizeable planetesimals formed over a period longer than a million years, each of which being composed entirely of an unusual, but homogeneous, mixture of mm-size particles. We sketch a scenario that might help explain how this occurred, in which primary accretion of 10-100km size planetesimals proceeds directly, if sporadically, from aerodynamically-sorted mm-size particles (generically ``chondrules"). These planetesimal sizes are in general agreement with the currently observed asteroid mass peak near 100km diameter, which has been identified as a ``fossil" property of the pre-erosion, pre-depletion population. We extend our primary accretion theory to make predictions for outer solar system planetesimals, which may also have a preferred size in the 100km diameter range. We estimate formation {\it rates} of planetesimals and explore parameter space to assess the conditions needed to match estimates of both asteroid and Kuiper Belt Object (KBO) formation rates. For parameters that satisfy observed mass accretion rates of Myr-old protoplanetary nebulae, the scenario is roughly consistent with not only the ``fossil" sizes of the asteroids, and their estimated production rates, but also with the observed spread in formation ages of chondrules in a given chondrite, and with a tolerably small radial diffusive mixing during this time between formation and accretion. As previously noted, the model naturally helps explain the peculiar size distribution of chondrules within such objects. The optimum range of parameters, however, represents a higher gas density and fractional abundance of solids, and a smaller difference between keplerian and pressure-supported orbital velocities, than ``canonical" models of the solar nebula. We discuss several potential explanations for these differences. The scenario also produces 10-100km diameter primary KBOs, and also requires an enhanced abundance of solids to match the mass production rate estimates for KBOs (and presumably the planetesimal precursors of the ice giants themselves). We discuss the advantages and plausibility of the scenario, outstanding issues, and future directions of research. 

\end{abstract} 

\end{titlepage}
\section{Introduction}

Primary accretion is the stage of growth in which tiny protoplanetary nebula dust grains grow into objects of 10-100 km size, such as most asteroids, Kuiper Belt Objects (KBOs), and comets.  The most well known, traditional approaches to modeling primary accretion are {\it incremental growth} by simple sticking to ever larger sizes (Weidenschilling 1997, 2000; Dullemond and Dominik 2004, 2005; reviewed by Dominik et al 2007; most recently Brauer et al 2008) and {\it midplane instabilities} of various types, going back to Goldreich and Ward (1973) ({\it cf.} reviews by Cuzzi and Weidenschilling 2006; henceforth CW06, and more recently Chiang and Youdin 2009). 

Several important clues as to the nature of primary accretion, which can help us assess these different hypotheses, are to be found in primitive meteorites and asteroids (discussed in more detail in section 2.1). The most primitive chondritic meteorites display a characteristic texture: predominance of mm-sized, once-molten silicate chondrules, metal grains, and refractory oxide particles, each surrounded by fine-grained dust rims and all embedded in a granular matrix. The size distribution of the chondrules in all classes of chondrite is quite narrow and nearly universal in shape, but with a mean size distinctive of each class.  At least two entire chondrite classes are each thought to derive from only one or two planetesimals, roughly 100 km in size and originally composed largely of chondrules with very similar properties. This ubiquitous and unusual texture is surely telling us something important about primary accretion, but there is no explanation for it at present. The Myr duration of meteorite parent body formation as revealed in isotopic age-dating, and the prevalence of unmelted asteroids, suggest that primary accretion went on for a long time (section 2.1).

The observations suggest that primary accretion was inefficient, and took a long time to complete (CW06; Cuzzi et al 2008, henceforth CHS08; see section 2.1). If the nebula were nonturbulent, as required for traditional midplane instabilities to play a role, particles settle into a dense midplane layer and growth by incremental accretion comes to completion too quickly - numerous 100km planetesimals and even lunar-size objects form in $10^5$ years (Weidenschilling 2000), all of which would melt due to short-lived radionuclides such as $^{26}$Al. Nebula gas turbulence can frustrate primary accretion if simple ``incremental accretion" stalls at roughly dm to m-size in turbulence, depending on gas density (the so-called ``m-size barrier"; see Cuzzi and Weidenschilling 2006, Dominik et al 2007, Brauer et al 2008). Moreover, recent work has identified a new ``km-size barrier" for incremental accretion in turbulence (Ida et al 2009). A challenge for primary accretion in turbulence is to leapfrog not only the meter-size barrier, but perhaps also the km-size barrier - and create 10-100km asteroids entirely from ``chondrules" with similar properties. It is in this sense that the first planetesimals might indeed have been 10-100km in diameter. If this happens in a temporally extended fashion, nebula chemical and physical properties can change slowly, perhaps helping explain the variable chemical and isotopic properties of chondrites ({\it eg} Cuzzi et al 2005). 

In previous work we have emphasized intriguing connections between these properties of primitive meteorites and asteroids, and the general scenario we present here. We have shown how well-sorted, chondrule-sized mineral particles are concentrated, by orders of magnitude, into dense zones in weak nebula turbulence (sections 2.2-2.3). This {\it turbulent concentration} can explain the characteristic size and size distribution of chondrules in a natural way. We developed a {\it cascade model} of the statistics of dense zones and their correlation with gas vorticity, which incorporates the effects of particle mass loading on the gas and predicts the fractional volume of particle-rich zones which can evolve directly into objects with some physical cohesiveness. Here we derive threshold conditions (combinations of the density and lengthscale of particle clumps, and the density, pressure gradient, and local vorticity of the gas) which allow dense clumps to proceed to become actual planetesimals. Combination of these thresholds with our cascade models leads to a prediction of the relative abundance of primary planetesimals as a function of mass - their initial mass functions - and even (with uncertainties) their production rate (sections 3.1-3.4). 

In this paper we explore, in a preliminary way, primary accretion initial mass functions (IMFs) at two disparate locations in the early solar system.  These predictions may be compared with both the known asteroid size distribution (which is thought to be a ``fossil" representing the actual size of primary planetesimals), and also with (limited) knowledge for Kuiper Belt Objects in the 30AU region.  Under different assumptions regarding nebula properties, we estimate not only the characteristic planetesimal size or mass which results, but also the planetesimal formation {\it rate}, which can itself be compared with crude estimates in the asteroid and KBO regions (section 3.4).  Our IMFs are consistent with previous suggestions that ``asteroids were born big" (Bottke et al 2005; Ida et al 2008; Morbidelli et al 2009a; Weidenschilling 2009). The scenario we envision for primary accretion, based on turbulent concentration, might occur continuously - and inefficiently - over an extended time, but when it does occur it is highly selective as to constituents and bypasses the problematic meter-size range (and even the km-size range) entirely, leading directly to 10-100km size objects composed of aerodynamically sorted particles. An independent study along these lines has also been done by Chambers (2010). A different scenario has been advanced to explain direct growth to 100km or larger diameter bodies, starting with meter-size bodies (Johansen et al 2007). This alternate pathway occurs in environments similar to that described here, and could proceed simultaneously (see sections 2.2 and 4 for more discussion). It will become apparent that current uncertainties in both the observations and the theory render our scenario more of a suggestive roadmap for, rather than an exhaustive explanation of, primary accretion. 

\section{Background}
\subsection{Clues from meteorites, asteroids, and KBOs:} 

{\it Meteorites:} Several different isotope systems (Al-Mg and Pb-Pb primarily) testify that the bulk of chondrites (their dominant iron-magnesium-silicate chondrules and matrix) was last processed in the nebula 1-3 Myr after the formation of the oldest, highest-temperature minerals (the refractory Calcium-Aluminum-rich Inclusions or CAIs) found in the same meteorites (Russell et al 2006; Kita et al 2000, 2005). The rare, even later-forming CH and CB chondrites probably resulted from an entirely different process, in an entirely different environment (Wasson and Kallemeyn 1990, Krot et al 2005). Yet, {\it some} parent bodies apparently accreted and melted nearly contemporaneously with CAIs, forming achondrites and metal cores (Kleine et al 2005, Markowski et al 2007). Primary accretion thus lasted several million years, suggesting that it was {\it inefficient}. Moreover, isotopic age-dating has recently progressed in accuracy and quantity to the point where several different groups find, for several different chondrite classes (carbonaceous and ordinary), that the formation ages of chondrules {\it within a given chondrite} range over almost 1 Myr (Kita et al 2000, 2005; Mostefaoui et al 2002, Sugiura and Krot 2007, Kurahashi et al 2008, Villenueve et al 2009). The nominal two-sigma error bars on individual chondrite ages in the best of these data are roughly 0.3-0.4 Myr, so a range of perhaps a half-million years can't be ruled out, but it appears from taking these results at face value that a range as short as $10^3-10^4$ years is unlikely in spite of qualitative thinking in the past that chondrules had to be accreted into chondrites ``rapidly" after their formation. Cautionary notes have been raised that some or all of these apparent age spreads might be the result of mineral-specific parent body alteration processes ({\it eg.} Alexander 2005, section 7.2); it is of vital importance to continue to make measurements of this type while addressing questions of alteration because, as we will show, they provide powerful constraints on models of primary accretion. 

The texture of primitive chondrites is unusual, and suggests a role for aerodynamical effects in most cases (for reviews see Scott and Krot 2005 or Brearley and Jones 1998; Cuzzi 2004 and CW06 present more discussion of the evidence for aerodynamical effects). The sizes of silicate and metal particles in the young CH and CB chondrites are counter-indicative of aerodynamic sorting, showing the prevalent evidence from normal chondrites to be non-trivial. The most primitive chondrites - especially those containing unbrecciated ``primary texture" (Metzler et al 1992, Brearley 1993) - look like collections of dust-rimmed chondrules and other mm-size particles, directly accumulated and merely compressed and compacted. Individual constituents of chondrites (chondrules in particular) have a size distribution that, while centered at different sizes from class to class, has a not-quite-lognormal {\it shape} that appears universal (CHPD01, Teitler et al 2009; see section 2.2). The H-type ordinary chondrite class is believed to derive from a single 80-100 km radius parent body, initially composed of a homogeneous collection of similarly well-defined chondrules which experienced post-accretional heating, metamorphism, and cooling to different degrees at different depths (Trieloff et al 2003, Grimm et al 2005). A similar (but less clear) story can be told for the L and LL-type ordinary chondrites (Marti and Graf 1992). It's reasonable to suspect that chondrite parent bodies may all be large ($\sim$100 km) objects, each initially composed primarily and homogeneously of chondrules (and other associated mineral particles) with {\it average} chemical, physical, and isotopic properties which are well-defined in any parent body, but differ dramatically from one parent body to another. Thus, primary accretion may be inefficient, but when it operates, it is highly selective. We return to an assessment of the situation in our concluding remarks.

{\it Asteroids:} Most of the S-type asteroids are probably related to ordinary (unmelted) chondrites (Binzel et al 2002, Clark et al 2002). This is not to say their interiors were never heated, or even partially melted (Elkins-Tanton and Weiss 2009) but there are only a few asteroid surfaces manifesting widespread and complete melting, as on Vesta. For instance, Sunshine et al (2004) show that in addition to Vesta and the unrelated, but similarly differentiated basaltic object 1489Magnaya, three other S-type family parents (17Thetis, 847Agnia, and 808Merxia) have igneous surfaces. Others of this type might yet be found. However, {\it all} objects larger than 50 km radius would melt extensively if they accreted earlier than 1.5-2.5 Myr after CAIs, because of radiogenic heating by live $^{26}$Al (LaTourrette and Wasserburg 1998, Woolum and Cassen 1999, McSween et al 2002, Hevey and Sanders 2006). The combination of few thoroughly melted asteroids and many unmelted ones, like the spread in meteorite age dates, points to a temporally extended primary accretion process. 
	
The observed asteroid population shows a distinct mode in the distribution of mass as a function of size {\bf Figure 1}. Bottke et al (2005) locate the peak of the observed asteroid mass distribution at 100km diameter, using cumulative distributions. They make a case that this mass peak is not explainable by erosive processes, and instead testifies to an initial mass function deficient in smaller objects. {\bf Figure 1} shows a differential presentation of the asteroid data that suggests the mass peak may lie at 140km diameter. For diameters larger than 350km, there are two or fewer asteroids per bin, so the details of the distribution are highly uncertain; for comparison there are about 50 asteroids in the mass bin at 140 km diameter. Nevertheless, it is a fact that the asteroid belt mass is dominated by the few largest asteroids. This is generally taken as evidence for runaway accretion into even larger objects, of which more than 99\% have been subsequently removed by size-independent dynamical depletion processes (Chambers 2004). The peristence of the 140km bump testifies to the vast number of asteroids of these sizes in the pre-depletion population; the diameter at the peak is thought to be an unbiased estimate of the primitive asteroid mass distribution at the time dynamical stirring and removal began (presumably at the time the nebula gas was removed and/or Jupiter formed; see Bottke et al 2005 or Morbidelli et al 2009a for a discussion). Whether the actual {\it primary} bodies needed to be just the same size as the {\it current} fossil population (Morbidelli et al 2009a), or a factor of 3-10 smaller in diameter, incurring some subsequent growth before the start of the erosive regime (Weidenschilling 2009) remains a subject of debate. Either way, our predictions of the IMF and other physical properties of primary bodies provide initial conditions for, and are testable by, models such as these. 

\begin{figure*}[t]                                 
\centering                                                                   
\includegraphics[angle=0,width=3.5in,height=2.7in]{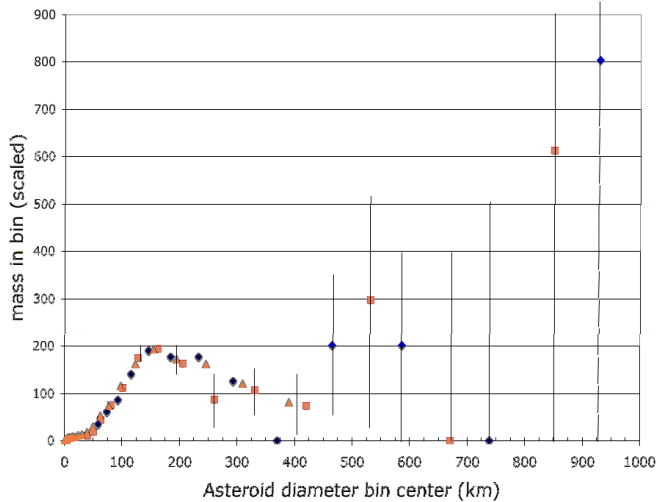}
\caption{\footnotesize{A histogram indicating where the bulk of mass lies in the current asteroids. Diameter bin centers are spaced by a factor of two in mass ({\it ie}, in $D^3$), starting at the largest asteroid (Ceres) and ranging downwards. Diameter boundaries between bins are taken midway between diameters at bin centers. Results are shown from several asteroid databases: the IRAS albedo-diameter data posted on the PDS Small Bodies node (IRAS-A-FPA-3-RDR-IMPS-V6-0; red squares), a tabulation by Farinella and Davis (1992, ascribed to Tedesco 1989; blue diamonds), and a tabulation by Jedicke et al (2003; green triangles). Poisson statistics error bars are indicated. In this representation, most of the mass seems to lie at around 140km diameter. It is the conclusion of Bottke et al (2005) that the depletion shortward of 140km diameter cannot be accomplished by erosion during post-accretional collisional evolution, but must be a primordial signature.}}
\end{figure*}

{\it Kuiper Belt Objects:} The KBO size distribution, and indeed the entire KBO formation scenario, is less well constrained. It is generally agreed that there is a KBO ``mass bump" as in the asteroid case (figure 1), but the modal peak may lie anywhere between 20 and 100km diameter based on the same (magnitude) data, given uncertainties in the observations and assumed albedos. Values close to the low end of this range might be ascribed to collisional erosion, for weak objects (Kenyon et al 2008), but values at the high end of this range would probably represent ``fossil" signatures of the primary accretion process, as in the asteroid case (Bottke et al 2005). KBOs come in several dynamical classes, which have different size distributions at sizes larger than the mass bump (Bernstein et al 2004, Morbidelli et al 2009b); of these the most abundant are the ``hot" and ``cold" classical objects, named for their relatively high and low eccentricities and inclinations, respectively.  

It is not even known for certain where the currently observed KBOs were formed. Traditional incremental accretion models form them in place (Stern and Colwell 1997, Kenyon and Luu 1998). This scenario requires a massive local source population of solids, of which more than 99\% must be subsequently removed by erosive collisions or by dynamics. Cleanup by dynamics alone is slow, unless augmented by local embryo-size objects which are, as in the asteroid belt, subsequently lost themselves (Chiang et al 2007). Ford and Chiang (2007) explored the excitation of KBOs by local icy embryos which were subsequently lost, with mixed results, but detailed studies of dynamical clearing {\it per se} in this scenario have not yet been done. Kenyon et al (2008) believe that cleanup by erosion (with planetesimals stirred only by Neptune) can remove more than 90\% of the bodies by grinding and drag loss of small particles (assuming a size distribution with plentiful 1-10km bodies); however, removal by erosion would be much less efficient if planetesimals are typically ``born big" as implied by the larger end of the diameter interpretations of the observations (Bernstein et al 2004). 

The outward dynamical evolution of giant planets by interactions with planetesimals (Malhotra 1995, Fernandez and Ip 1996), more recently refined into the so-called  ``Nice model", has several implications. One is that a massive indigenous population of planetesimals outside of 30AU would lead to greater migration of Neptune than observed, and its existence has been questioned on these grounds  (Gomes et al 2004). Another suggestion is that the current crop of KBOs (of all dynamical types) may have been formed at smaller distances - specifically between 16-30AU, and emplaced into their current locations by dynamical processes (Gomes 2003, Levison et al 2008). Supportive connections have been made between this emplacement and other related primitive body populations (D-type and Trojan asteroids) emplaced at the same time and in the same general way (Levison et al 2009, Morbidelli et al 2009).  In this scenario, there is no ``cleanup" problem - the mass emplaced into the current Kuiper belt is only about 0.1\% of the mass in its 16-30 AU source region (and most nebula models contain more than enough total mass in this region). However, the observed ``cold classical" KBO population is less eccentric than the model results predict; also, relative to the hot population, it is four times richer in binaries which might be easily disrupted during extended dynamical evolution from closer to the sun (Stephens and Noll 2006). 

Other issues regard timescales. Some {\it in situ}, incremental growth, massive source scenarios take 10-30Myr to build 10-100km radius KBOs (Kenyon 2002, Kenyon et al 2008), which probably precludes any thermal alteration by short-lived radioisotopes such as $^{26}$Al. On the other hand, Weidenschilling (1997; his figure 12) includes gas-drag augmentation of growth, and grows 10-100km objects at 30 AU in $<$1 Myr; this short accretion timescale would seem to predict widespread melting. 

McKinnon et al (2008) note that the emerging diversity of KBO albedos and densities (including the apparent differentiation of Pluto and Haumea, formerly 2003EL61, and now probably Quaoar as well (Fraser and Brown, 2009) might be hard to explain without short-lived isotopes; yet the need to preserve amorphous ice and supervolatiles like CO in other objects seems to preclude them.  This paradox is reminiscent of the meteoritical arguments for drawn-out accretion over a period spanning a little longer than the lifetime of the short-lived isotopes. 

Finally, none of the current KBO incremental growth scenarios involve nebula turbulence, which slows growth beyond a meter or frustrates it entirely (stalling probably occurs at even smaller sizes at these distances, as noted below), and yet it is generally agreed that, if any part of the nebula were robustly turbulent, the $>$20 AU region would be (see next subsection). Even if the meter-(or smaller) size barrier could be overcome in a turbulent environment at 30 AU, Ida et al (2008) have shown that expected levels of turbulence may excite random velocities that render the entire region erosive for 1-10km sized objects. No detailed models have been run for growth of planetesimals in the 16-30 AU source region, with or without turbulence, but at least without turbulence, timescales would probably be faster than in the traditional {\it in situ} models (Weidenschilling 1997, Kenyon 2002, Kenyon et al 2008) because of higher solids densities and shorter timescales. 

As in the asteroid belt region, it is plausible that a primary accretion scenario in which large planetesimals are created sporadically, over this period of time, and perhaps with a different efficiency than incremental growth models, might help resolve some of these KBO puzzles. In this paper we will make representative calculations at 30 AU, assuming the primordial KBO mass needed between 16-30 AU was about 40$M_{\oplus}$. Future refinements of this preliminary study are discussed in section 3.4.2. 

\begin{table}
{\footnotesize 
\begin{center}
\begin{tabular}{ c c c c}
{\bf Symbol} & {\bf Definition} & {\bf Equation or section} \\
\hline \\*[0.02in]
$l, v(l), t_e(l), \omega(l)$ &  eddy scale, velocity, lifetime, and frequency &  sec. 2.2 \\
$L, V_L, \Omega_L$ &  largest eddy scale, velocity, and frequency &  sec. 2.2 \\
$\eta, t_{\eta}$ &  Kolmogorov (smallest) scale and lifetime&  sec. 2.2 \\
$a, \Omega, V_K=a\Omega$ & distance from sun, orbital frequency, Kepler velocity &  sec. 2.2 \\
$H, c, \rho_g$ & gas vertical scale height, sound speed, and density &  sec. 2.2 \\
$Re$ &  Reynolds number &  sec. 2.2 \\
$\alpha, \nu_T$ &  kinematic viscosity $\nu_t=\alpha c H$ &  sec. 2.2 \\
$t_s$ & particle stopping time &  eqn. 1\\
$\rho_p$ & local mass density in particles &  sec. 2.3 \\
$\Phi$ & local mass loading factor $=\rho_p/\rho_g$  &  sec. 2.3 \\
$S$ & normalized gas enstrophy $\omega^2(l)/\left< \omega^2(l) \right>$ &  sec. 2.3 \\
$N$ & cascade level corresponding to lengthscale $l$ &  eqn. 2 \\
$m, p(m)$ & cascade multiplier and its PDF &  sec. 2.3 \\
$P(\Phi,S)$ & joint PDF of mass loading and enstrophy &  sec. 2.3 \\
$P^*$ & $P(\Phi,S)$ at the peak of an IMF &  sec. 3.3 \\
$P_{goal}$ & value of $P^*$ needed to create $\dot{M}_{pa}$ & eqns. 10-15 \\
$t_{pa}$ & conversion timescale of mass into planetesimals &  sec. 3.3.1\\
$\Phi^*, N^*$ & values of $\Phi,N$ at $P^*$  &  sec. 3.3 \\
$\dot{M} $ & mass accretion rate of gas  &  sec. 3.3.1 \\
$\dot{M}_{pa}$ & mass accretion rate of planetesimals &  eqn. 9 \\
$A, A_o$ & actual and canonical solids abundance relative to gas &  sec. 3.3 \\
$t_G$ & dynamical collapse time of a dense clump &  eqn. 3 \\
$t_{sed}$ & sedimentation time of a dense clump &  eqn. 4 \\
$We_G, We_G^*$ & Gravitational Weber number and its critical value &  sec. 3.1 \\
$\beta$ & pressure gradient parameter &  sec. 3.2 \\
$\sigma(a),\rho_g(a),H(a),\beta(a)$ & radially dependent nebula properties &  eqns. 5\\
$a_o$ & reference distance from sun (2.5 AU) &  sec. 3.2  \\
$\rho_R$ & Roche density &  sec. 3.2.1 \\
$\Phi_1, \Phi_2, S_{min}$ & thresholds for primary accretion &  eqns. 6-8 \\
$F_V,F_p,F_t(>T)$ & volume, particle, and time fractions exceeding threshold $T$ &  sec. 3.5.1 \\
$t_{enc}$ & particle encounter time with planetesimal-forming clump &  eqn. 17 \\
$\Delta a$ & radial diffusion (mixing) extent in $t_{enc}$ &  sec. 3.5.1 \\
\hline
\end{tabular}
\end{center} 
}
\caption{Symbols, parameters, and functions used in this paper}
\end{table}

\subsection{Turbulence and particle-gas interactions} While the ultimate cause and intensity of nebula turbulence remain subjects of debate on theoretical grounds (Fleming and Stone 2003; Johnson and Gammie 2005; Turner et al 2007), observational arguments suggest it was indeed present at interesting levels throughout the primary accretion stage (Dullemond and Dominik 2004, 2005, Dominik et al 2007).  The most generally accepted (although perhaps not the only) way to drive nebula turbulence is the magnetorotational instability (MRI), in which the turbulent intensity is considerably higher in the dilute gas of the outer (and upper) nebula  than in the terrestrial planet region (Turner and Sano 2008). In contrast to the original idea of a ``dead zone" near the nebula midplane (Gammie 1996), Turner and Sano (2008) dub the midplane region the ``undead zone" because it can be excited in as-yet poorly understood ways by strong turbulence in the rarified layers at high altitudes. Moreover, even without considering MHD turbulence, other possibilities remain open ({\it cf} CW06). Here we assume weak, but widespread turbulence throughout the asteroid formation region. 

Turbulence is an essentially lossless cascade of energy from large, slowly rotating eddies with lengthscale $L$ and velocity $V_L$, which are forced by (currently unknown) nebula-scale processes, through smaller and smaller scales of size $l$, having correspondingly shorter eddy timescales $\te(l)$, to some minimum lengthscale $\eta$, called the Kolmogorov scale, where molecular
viscosity $\nu_m$ can dissipate the macroscopic gas motions and turbulence ceases.  We characterize the intensity of turbulence by the parameter $\alpha$ which sets the disk turbulent viscosity $\nu_T=L V_L \equiv \alpha c H$, where $c$ is the gas sound speed, $H$ is the nebula vertical scale height, $L = H \alpha^{1/2}$, and $V_L = c
\alpha^{1/2}$. Then the turbulent Reynolds number $Re = (L/\eta)^{4/3} = \alpha c H /\nu_m$. A typical T Tauri-like nebula with mass accretion rate $\dot{M} \sim$ a few$\times 10^{-8}M_{\odot}$/yr, channeling 2-3\% of its accretional energy into turbulence, would have $\alpha \sim 10^{-4} - 10^{-3}$ or $Re = 10^7-10^8$ at 3 AU (Cuzzi et al 2001; henceforth CHPD01; also CW06). One may distinguish between turbulent {\it viscosity} and turbulent {\it diffusivity} (Prinn 1990): the former is problematic in, for instance, convective turbulence (Ryu and Goodman 1992) but the latter is robust in turbulence of all kinds, and it is the latter that drives our primary accretion scenario. A significant nebula turbulent diffusivity can also help explain the persistence of ancient, refractory inclusions in chondrites (Cuzzi et al 2003, 2005) and the abundance of crystalline, moderate volatility silicates in the STARDUST sample (Ciesla 2009; {\it cf.} also Bokelee-Morvan et al 2002). That is, nebula turbulence can mix material radially by significant distances over time. 

In most cases of realistic, high-$Re$ turbulence, the Kolmogorov energy spectrum is a good approximation, where for a wide range of lengthscales $\eta < l <L$, the turbulent kinetic energy density $E(l)$ is given by the {\it inertial range} expression $E(l) = (V_L^2/2L)(l/L)^{-1/3}$. The
eddy frequencies then scale as $\omega(l) = 1/\te(l) = v(l)/l = (2 l E(l))^{1/2}/l = \Omega_L (l/L)^{-2/3}$, where $v(l)$ is the velocity of an eddy of size $l$, and  the large eddy frequency $\Omega_L$ is usually identified as the local orbit frequency $\Omega$ (CHPD01, Johansen et al 2007). These properties tend to be independent of the forcing mechanism and even of the Reynolds number of the turbulence. Even if the initial forcing is anisotropic (as perhaps for MRI turbulence), smaller eddies become more isotropic as the 3D nonlinear cascade proceeds (Kato and Yoshizawa 1997). High $Re$, inertial range turbulence is sufficiently scale-free (Falkovich and Sreenivasan 2006) that using statistical and spectral properties from limited inertial ranges to characterize more extensive ones (those at higher $Re$) is an appealing approach. We make extensive practical use of this ``cascade" property, as described in section 2.3. We note here for future use in section 3.3.1 that the Kolmogorov eddy timescale $t_{\eta} = 1/(\Omega_L (L/\eta)^{2/3}) = 1/(\Omega Re^{1/2})$. 

{\it Particle-gas interactions:} Particle interactions with the gas are characterized by the particle stopping time $\ts$ which, for particles of interest here, is defined by the Epstein drag law:
\begin{equation}
t_s=r \rhos / c \rhog,
\end{equation}
where $r$ and $\rhos$ are particle radius and internal density, and $c$ and $\rhog$ are the gas sound speed and density (see CW06 for more discussion). 

Particles interact with the gas, turbulent or not, within their stopping time and acquire inertial space (absolute) and random (relative) velocities accordingly. The relative velocities between particles determine the outcome of their collisions (sticking, erosion, or breakup), and the inertial space velocities determine the degree to which they diffuse radially and vertically, thus controlling their settling to the midplane (see eg Dubrulle et al 1995, Weidenschilling 1997, Ormel et al 2008, Brauer et al 2008). In the dense midplane layers of cm-m size particles which can form in {\it nonturbulent} nebulae, the local gas is driven to near-Keplerian speeds and relative velocities between particles remain low enough for continued growth to planetesimal size to occur very rapidly, with or without the help of various midplane instabilities, on timescales of $10^3 - 10^5$ years (Cuzzi et al 1993, Weidenschilling 1997, 2000; Youdin and Goodman 2005). This is actually problematic in view of the extended formation epoch of primitive bodies discussed above - the process may go to completion too rapidly (section 2.1; CW06). 

Even weak turbulence, however, frustrates growth at some limiting size which depends on the local gas density (Dominik et al 2007). As particles grow they become more vulnerable to mutual destruction, because their increasing stopping time couples them to eddies of increasing size and velocity. In turbulence this coupling is captured by the Stokes number $St = t_s \omega$ where $\omega$ can represent the eddy frequency on any scale - commonly either the large eddy scale $L$ ($\Omega_L\sim \Omega$) or the Kolmogorov scale $\eta$. Particles with $t_s$ comparable to the lifetime of the largest eddies ($t_s \Omega \sim 1$) achieve the highest velocities $V_L = \alpha^{1/2}c$. For $\alpha \sim 10^{-4}$, such particles collide at relative velocities $V_{rel} \sim V_L \sim 10^3$ cm/sec - which are probably disruptive (Stewart and Leinhardt 2009). For a range of nebula properties, particles in the dm-m radius range have this property (see CW06, figure 1, or Ormel et al 2008) and this problem is commonly referred to as part of the ``meter-size barrier". However Brauer et al (2008), who assume a relatively low gas density, see growth frustrated at an even smaller size because the lower gas density leads to longer $t_s$.  Recent lab work is challenging some of the sticking assumptions of prior years at even lower velocities (G{\"u}ttler et al 2010); models using these new results even suggest that a ``bouncing barrier" might preclude growth beyond objects having masses not much larger than those of chondrule precursors (Zsom et al 2010) - again, depending on nebula properties. Obtaining and retaining objects with $t_s \Omega \sim 1$ is perhaps the major issue in the primary accretion scenario of Johansen et al (2007), which relies on an abundance of such particles in moderate-intensity turbulence because they drift rapidly into high-pressure ridges to become concentrated. Progress in this area will be interesting to follow. Meanwhile, we focus on a different accretion pathway, that also relies on turbulence but acts on particles much smaller than a meter, which are excited to small relative (collision) velocities well below the disruption threshold (Hogan and Cuzzi 2003, Ormel and Cuzzi 2007), and have sizes directly relevant to meteorites. 
\subsection{Turbulent concentration and the cascade model} Small particles diffuse in turbulence, but the trajectories of particles of a certain well-defined aerodynamic stopping time avoid fluid zones of high vorticity and converge in zones of low vorticity. Here, concentration factors $C\equiv\rhop/\overline{\rhop}$ may be $ \gg 1$,
where $\rhop$ and $\overline{\rhop}$ are the local and nebula-averaged particle
mass density, respectively.  We define the local mass loading $\Phi \equiv
\rhop/\rhog$, where $\Phi$ can thus also be $\gg 1$. The maximally
concentrated particles have a stopping time $t_s$ equal to the overturn time $t_{\eta}$ of the smallest eddies (which have size $\eta$, the Kolmogorov scale). Two ``fingerprints" of this {\it turbulent concentration} (TC) seem evident in the meteorite record. The typical chondrule size (crudely, mm-diameter) is naturally explained by TC merely by requiring $t_s=t_{\eta}$ (CHPD01). An equally compelling fingerprint is the very characteristic chondrule size {\it distribution}, which is very similar across meteorite groups when scaled to the mean size, and is an excellent fit to the distribution predicted by TC (CHPD01). It has recently been shown that the TC size distribution is even statistically preferable to a lognormal distribution (Teitler et al 2009).  It should be noted that, in the outer nebula where gas densities are lower and turbulent intensities plausibly larger, much smaller solid grains, or, more likely, porous aggregates of grains, such as seen in cometary IDPs, become the preferred candidates for TC rather than chondrules, for which $t_s$ would be too large (CHPD01; see their section 3 and figure 1). Of course, it would be nearly impossible to extract ``fingerprints" of the process in, for instance, returned KBO samples, after porous aggregates had become compacted in a parent body. 

{\it Cascade model:} The spatial distribution of $\Phi=\rho_p/\rho_g$ is determined only statistically, and must be studied with Probability Distribution Functions (PDFs) which depend on the nebula Reynolds number and the spatial scales of interest (CHPD01). Because the nebula $Re$ is far higher than achievable with current 3D fluid models, we have developed and employed a ``cascade model" which, while not reproducing the physical structure of turbulence (vortex tubes and the like), has been shown to reproduce the PDFs of a number of attributes of turbulence (Menevaux and Sreenivasan 1991, Sreenivasan and Stolovitsky 1995). This model was described in detail by Hogan and Cuzzi (2007), and summarized by CHS08, so will be even more briefly sketched here. 

In turbulence, a number of properties (energy, velocity, vorticity, and particle abundance) can be thought of as being partitioned unequally and losslessly into sub-elements of eddies as they bifurcate. The partitioning fractions at each bifurcation are taken as $m$ and $1-m$, where the {\it ``multipliers"} $m$ are drawn from a PDF $p(m)$ which is generally independent of eddy scale throughout the turbulent inertial range (Meneveau and Sreenivasan 1991, Juneja et al 1994, Sreenivasan and Stolovitsky 1995; see however Bec et al 2007 where some evidence is presented for scale-dependence in the context of preferential concentration). We determine the PDFs of these multipliers $p(m)$ from our highest $Re$ 3D models, which still cover only a limited range of eddy scales or bifurcation levels (Hogan and Cuzzi 2007). Each bifurcation is thought of as a level in a cascade; in the cascade model, we extend the multiplier process to even deeper levels (which one can think of as the smaller eddy scales achieved at higher $Re$). Unless $m=0.5$, repeated application of asymmetrical partition fractions $(m, 1-m)$ constantly creates more extreme values (higher and lower) of all parameters as the number of levels increases; this is referred to as {\it intermittency} - the local value becomes not more well-defined, but more highly variable at smaller scales (see the readable introduction by Meneveau and Sreenivasan 1991). For a cube, three orthogonal 1D bifurcations, or {\it levels}, are needed to generate 8 subvolumes of linear size $l_{j+1} = l_j/2$, and thus $Re^{3/4} = L/\eta = 2^{N/3} =10^{ {\rm log2} \cdot N/3} \sim 10^{N/10}$, where $N$ is the total number of levels in the cascade. The general cascade relation giving the lengthscale associated with a given cascade level $N$, applied to a nebula situation with a large eddy scale $L$, is thus
\be
       l = 2^{-N/3} L  = 2^{-N/3}H \alpha^{1/2}.
\ee
Cascade models can achieve much higher $Re$ than Direct Numerical Simulations (DNS); to match our full 3D DNS simulations at $Re=2000$, the cascade model needs only about 15 levels, taking about 10 cpu-hours (for 1024 realizations) compared to over 90000 cpu hours to converge a single full 3D simulation. 

Our particle-gas cascade model (Hogan and Cuzzi 2007) simultaneously treats $\Phi$ and local enstrophy $S = \omega^2(l)$, where $\omega(l)$ is a vorticity on lengthscale $l$, using two distinct sets of multipliers, and allows for the observed spatial anticorrelation of $\Phi$ and $S$ on a statistical basis. The results of these cascades are binned into a second kind of (2D) PDF $P(\Phi,S)$. Examples are shown in {\bf figure 2}. The meaning of $P(\Phi,S)$ is volume fraction (volume per unit nebula volume) having a particular combination of particle mass loading factor $\Phi$ and relative enstrophy $S=\omega^2(l)/\left<\omega^2(l)\right>$, where $\left<\omega^2(l)\right>$ is the average enstrophy at scale size $l$. $P(\Phi,S)$ is given per unit log$_{10}(\Phi)$, per unit log$_{10}(S)$ and differs slightly in meaning from expressions in Hogan and Cuzzi (2007; see Appendix). $P(\Phi,S)$ is a function of level $N$ in the cascade, because going to deeper levels (smaller scales $l$) always enhances the variance of its properties (Meneveau and Sreenivasan 1991; {\bf figure 2}). The two-dimensional nature of  $P(\Phi,S)$ is also essential; we will show that the threshold conditions allowing planetesimal formation depend on both $\Phi$ and $S$, as well as level $N$ (section 3.2).  
\begin{figure*}[t]                                 
\centering                                                                   
\includegraphics[angle=0,width=3.4in,height=2.7in]{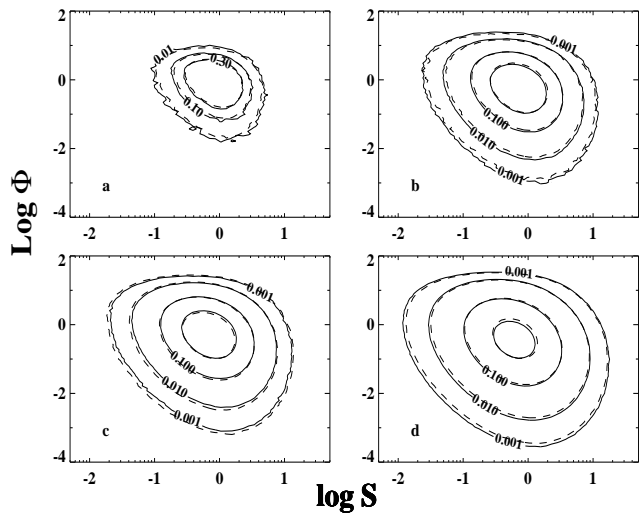}
\vspace{-0.1in}
\caption{\footnotesize{The PDFs $P(\Phi,S)$ for four different values of $Re$, computed from 3D direct numerical simulations (solid contours) are compared with cascade model predictions for the appropriate number of levels (dashed lines; $N$=9, 12, 15, and 18 respectively, corresponding to the various values of $Re$). Note that, as $Re$ and $N$ increase (from panel (a) to (d)) the variance of both $\Phi$ and $S$ increases (more extreme values of both are seen). Mass loading has begun to affect the PDFs at the higher $\Phi$ values and deeper cascades plotted. Figure from Hogan and Cuzzi (2007). }}
\end{figure*}

As the particle mass density increases relative to the gas mass density, it affects the physics of turbulent concentration.  Hogan and Cuzzi (2007) showed how particle mass loading affects the cascade; multipliers for mass loading $\Phi$ and enstrophy $S$ are shown to depend on the local mass loading itself. As mass loading increases towards $\Phi \sim 100$, multiplier PDFs $p(m)$ narrow towards a delta-function at $m=0.5$, implying an equal probability of partitioning and no further trend to intermittency (see the discussion in CHS08 or Hogan and Cuzzi 2007). In this situation, there can be no further increase of $\Phi$ as the cascade level increases, and $\Phi$ saturates near 100. The reason for this is not completely understood, but it is probably due to a combination of inertial effects (conservation of kinetic energy) and dissipation of turbulent kinetic energy (which may be smaller than in terrestrial experiments because of the tighter coupling of the particles to the gas in our regime). The mass-loaded cascade models of Hogan and Cuzzi (2007) showed good agreement with actual 3D two-phase, mass-loaded, DNS results ({\bf figure 2}). 

An important property of the cascade models is that the cascade level $N$ can be related to the corresponding nebula lengthscale $l$, for some given nebula $\alpha$ (equation 2 above), and we can thus calculate the mass of the planetesimal which forms from a given volume of material at each level. In section 3 we will show how the cascade model, combined with certain thresholds for the stability of dense clumps, leads to the primary mass distribution function, or IMF, of planetesimals.  The results noted in section 3.1 (see also CHS08) point to spatial scales of interest which are $10^3-10^4$ times larger than the Kolmogorov scale $\eta$ emphasized by CHPD01; that is, because of the role of mass loading, we no longer believe the {\it theoretically} high concentrations of CHPD01 are achievable at small scales, and thus, for reasons discussed in section 3.2.3, lengthscales as small as $\eta$ also become less relevant. There are fewer cascade samples in shallower cascades - larger $l$ means lower $N$ and the number of samples is about $2^N$ for a cascade with N levels or bifurcations. Thus, we needed to run many cases to obtain the proper statistics to explore the low-$P$ ranges of ($\Phi,S$) that exceeded our newly defined thresholds (section 3.2) and were capable of becoming planetesimals.  We ran cascade models for a period of several months on NASA's HEC Altix and Origins computers at Ames, ultimately running $10^3$, $10^6$, and $10^7$ cases at 24, 20, and 15 levels to build up statistics such as shown in {\bf figure 3} (next section). 
\section{Determination of Initial Mass Functions} 
We first review how the self-gravity of a clump enters, which is more subtle than usually believed. We then derive three different thresholds on different combinations of clump density and size, and local vorticity, that determine which dense clumps can become primary sandpile planetesimals in a turbulent nebula environment. 

\subsection{The role of self gravity}
Gravitational Instability (GI) or inexorable collapse on a dynamical timescale
\be
t_G = (4 G \Phi \rho_g)^{-1/2}
\ee
is a well-used tool in the cosmogonist's toolbox, but we have found that the traditional concept of GI is not appropriate for small particles which have stopping times $t_s$ much less than dynamical times $t_G$. CHS08 recently rediscovered numerically a result originally obtained analytically by Sekiya (1983), and since apparently forgotten: that gas pressure {\it stabilizes} dense clumps of particles against traditional gravitational instability on dynamical timescales (GI). In the regime where particle-gas coupling is strong, particles which begin to collapse under their self-gravity drag and compress the entrained gas, producing a radial gas density and pressure gradient, which in turn prevents the gas and tightly coupled particles from undergoing GI until the particle mass loading is {\it $10^3$ times larger} than the traditional GI criterion. Sekiya (1983) called the mode of particles and entrained gas that arises under these conditions, ordinarily assumed for traditional GI, a  3D ``incompressible mode" of instability. Within such blobs, Sekiya suggested and CHS08 showed that particles of radius $r$ can only sediment slowly inwards at their terminal velocities, on the timescale 
\begin{equation}
t_{sed} = 1/4G \Phi \rho_g t_s = c/4 G \Phi r \rho_s,
\end{equation}
on the order of $10^2-10^3$ orbit periods for typical chondrules and $\Phi = 100-10$.

CHS08 explored the ability of such dense clumps to resist disruptive forces for a time on the order of $t_{sed}$. As a clump settles vertically towards the midplane under the vertical component of solar gravity, or orbits at a velocity near Keplerian ($V_K)$, it incurs a ram pressure from the nebula gas. In the latter case the headwind arises because the gas, being pressure-supported, orbits more slowly than Keplerian at a speed of $(1-\beta)V_K$ where $\beta \sim 10^{-3}$ (see next section; Weidenschilling 1977, Nakagawa et al 1986; CHS08). Conservatively assuming each clump moves as a Keplerian object, the mean settling speed $V_z$ for a clump formed at altitude $z$ above the midplane is $V_z \sim (z/a)V_K \sim (z/H)\beta^{1/2}V_K$ since $\beta \sim (H/a)^2$. Then requiring the vertical headwind be smaller than the azimuthal headwind implies $z/H < \beta^{1/2}$.  Within a vertical distance $H\beta^{1/2}$ of the midplane, the vertical settling velocities are small compared to the orbital velocity difference between the pressure-supported gas, and the azimuthal ram pressure on a strengthless Keplerian clump dominates all other disruptive forces (CHS08). In section 3.3.1, we will restrict the volume in which plantesimals can form to this near-midplane region. 

CHS08 developed a toy model based on an analogy with the Weber number $We$ in the familiar raindrop problem, where $We$ is the ratio of surface tension to ram pressure forces acting on a fluid droplet  moving at velocity $\Delta V$ relative to a less dense fluid. They defined a ``gravitational Weber number" $We_G$ which balances the ram pressure force per unit area with the {\it self gravitational force per unit area} of a strengthless clump of initial diameter $l$ and particle density $\rho_p$.  The premise was that certain combinations of diameter $l$ and particle mass density $\rho_p = \Phi \rho_g$ would stabilize a clump against being disrupted by a headwind of magnitude $\Delta V =\beta V_K$.  CHS08 ran a range of numerical models of clumps experiencing a steady nebula headwind from the more slowly orbiting gas, to validate the toy model, and determined that stability was indeed achieved for $We_G$ greater than some critical value  $We^*_G$ of order unity. Viscous losses of material around the periphery of their numerical clumps limited their numerical runs, but such large viscous erosion is an artifact of the numerics and would not be present in the actual nebula case. They noted that the combination of parameters required for stability of a dense clump implied a substantial size for the ensuing sandpile (10-100km radius), and pointed out the similarity of this size to the ``fossil asteroid belt" modal size of Bottke et al (2005). Below we show simplistically, but quantitatively, how a combination of dense clump stability thresholds may determine the IMF of primary planetesimals.
\subsection{Thresholds for primary accretion in $(S,\Phi)$ space} 
Our prediction of primary object IMFs is based on mapping three different kinds of threshold onto the cascade probability contours ({\bf figures 2 and 3}). These are not thresholds at which any sort of traditional ``fast" instability occurs (section 3.1) - rather they are thresholds which allow dense particle clumps to avoid ram pressure disruption by the nebula gas for the long time ($t_{sed} \sim 10^2-10^3$ orbits) required for the particles in them to sediment into their mutual center, creating a sandpile planetesimal. The key step in deriving planetesimal IMFs is connecting the thresholds derived below (functions of lengthscale $l$) to the cascade model PDFs (functions of level $N$), which we do using the cascade relation given in equation 2. In the following sections we incorporate simple scaling of our criteria with distance $a$ from the sun, based on powerlaw approximations for the nebula gas surface mass density $\sigma(a) = \sigma_o(a/a_o)^{-p}$ and, to a less important extent, mean temperature $T(a) = T_o(a/a_o)^{-q}$. The combination of radial density and temperature gradients leads to a generally outward radial pressure gradient $dP/da$ which, normalized by the gravitational coriolis force, is specified by the nondimensional parameter $\beta = (dP/da)/(2 \rho_g a \Omega^2)$  (previous section, Weidenschilling 1977, Nakagawa et al 1986; possible complications are discussed in section 3.4). Then, following Cuzzi et al (1993, equations 54-59):
\begin{eqnarray}
\sigma(a) = 2 H(a) \rho_g(a)   \\ \nonumber  
\rho_g(a) = \rho_g(a_o)(a/a_o)^{-(2p-q+3)/2} \sim \rho_g(a_o)(a/a_o)^{-(p+3/2)} \\ \nonumber 
H(a) = H(a_o)(a/a_o)^{(3-q)/2} \sim H(a_o)(a/a_o)^{3/2} \\ \nonumber   
\beta(a) = \beta(a_o)(a/a_o)^{1-q}   \\ \nonumber  
\end{eqnarray}
where the $q$-dependence is weak for $q \sim 1/2$ and ignored for simplicity except in $\beta$; it can be easily allowed in more detailed studies. We also adopt $\Omega(a) = \Omega(a_o)(a/a_o)^{-3/2}$, and let $\Omega_o, \beta_o, H_o, \rho_{go} = \Omega(a_o), \beta(a_o),...$, {\it etc.} take their nominal values at $a_o$=2.5AU. 
\subsubsection{Threshold $\Phi_1$: Rotation and gravitational binding}
The first question most people have is, are the clumps rotating too quickly to be bound? This threshold is determined by comparing the local gravitational timescale $t_G$ (equation 3) and the local eddy timescale $1/\omega(l)$, where the local eddy frequency $\omega(l)$ is treated as a vorticity. This threshold dominates when the local vorticity exceeds the global value (see section 3.2.2). Requiring $t_G < 1/\omega(l)$ (Toomre 1964, Goldreich and Ward 1973) is conservative here because eddies don't truly ``rotate" (many times) with timescale $1/\omega(l)$; rather, $1/\omega(l)$ is their {\it existence} lifetime before bifurcating. Even though the dense zones of small particles of interest here cannot collapse on the timescale $t_G$ (CHS08; section 3.1), they can become bound entities based on a criterion close to this (Sekiya 1983). Then $t_G = (4 G \Phi \rho_g)^{-1/2} < 1/\omega(l)$ or $\Phi > \omega^2(l)/4 G  \rho_g$ determines our first threshold $\Phi_1$. To express $\Phi_1$ in terms of $S = \omega^2(l)/\left<\omega^2(l)\right>$ we use the inertial range mean enstrophy on scale $l$, $\left<\omega^2(l)\right> = \Omega_L^2 (L/l)^{4/3}$ (section 2.2) where $\Omega_L$ is the large eddy frequency, generally taken to be the orbit frequency $\Omega$. We then use $\Omega^2 = G M_{\odot}/a^3$ where $M_{\odot}$ is the Sun's mass and $a$ is the distance from the Sun, and also the definition of the Roche density $\rho_R \equiv 3 M_{\odot} / 4 \pi a^3$ (Safronov 1991) to get 
$ \Phi_1(S) = (\rho_R/\rho_g)(L/l)^{4/3}S$.  This relation is extended to arbitrary semimajor axes $a$ using the $a$-dependence of $\rho_R(a)/\rho_g(a)$. Then 
$
\rho_R/\rho_g = K_0 (a /a_o)^{p-3/2}$, where $K_0 \equiv (3 M_{\odot}/ 4 \pi \rho_{go} a_o^3)$.
We then use the cascade relation $l = 2^{-N/3}L$ to express $(L/l)^{4/3} = 2^{4N/9}$. In the nebula, $L = H \alpha^{1/2}$ is the large eddy scale. Combining these relations leads to 
\be 
\Phi_1(S,a) = 2^{4N/9} K_0 S \left({a \over a_o}\right)^{p-3/2}.
\ee 
Note above that $\Phi_1$ has no explicit $\alpha$-dependence, but each $N$ implies a lengthscale $l$ which does depend on $\alpha$. Each threshold $\Phi_1$ appears as a diagonal line in {\bf figure 3}, colored according to its value of $N$. 

The more refined stability analysis of Sekiya (1983) is easily generalized to this situation. Sekiya assumes the relevant rotation frequency is the orbital frequency $\Omega$ and finds that 3D incompressible modes become marginally bound at $4 \pi G \rho_p /\Omega^2 \sim 10$ (his section 4 and figure 3). We generalize this to $4 \pi G \rho_p /\omega^2(l) = 4 \pi G \Phi \rho_g /\omega^2(l) >10$, and follow the same logic as above, substituting $\omega(l)^2 = \left< \omega^2(l) \right> S =\Omega^2 (L/l)^{4/3} S = 2^{4N/9} \Omega^2  S $, and relating $\Omega^2$ to $\rho_R/\rho_g$ as above, to obtain a threshold value $\Phi_{Sek}= \frac{10}{3} \Phi_1$. Because even this more sophisticated analysis is idealized itself to some degree, we explore the implications of this factor of $10/3$ as one example of the uncertainty of the predictions in the figures and tables of section 3.4 below. 

\subsubsection{Threshold $S_{min}$: global rotation} Notice in {\bf figure 3} that the cascade PDFs extend to very low  values of relative enstrophy $S \sim$ few$\times 10^{-4}$. However, on the long timescales $t_{sed}$ a clump cannot be guaranteed of remaining in fluid zones with such low vorticity, and will experience the global rotation as a minimum. We thus impose a minimum local vorticity given by the global shear rate $\Omega$. We express this in terms of $S$ as 
\begin{equation}
S = \omega^2(l)/\left< \omega^2(l) \right> > S_{min} = \Omega^2/\left< \omega^2(l)\right>  = \Omega^2/(2^{2N/9}\Omega_L)^2 = 2^{-4N/9}.
\end{equation}
Thresholds of $S_{min}$ appear as vertical lines in {\bf figure 3}, colored according to $N$. Regions lying to their left can be disregarded as candidates, having unrealistically low $S$ to characterize the long timescales involved in sedimentation to sandpiles. 

\subsubsection{Threshold $\Phi_2$: Ram pressure and the Gravitational Weber Number}
As described in section 3.1 and CHS08, self-gravity of a dense clump can play the role of surface tension and stabilize a clump against the disruptive ram pressure of the nebula headwind if $\Phi_2 l > \beta a \Omega / (2 G \rho_g We_G^*)^{1/2}$ (CHS08 equation 4). CHS08 suggest that the most favorable region for clump survival is within some small vertical distance $\beta^{1/2}H$ of the midplane, where vertical settling of dense clumps under solar gravity is negligible and only the azimuthal headwind remains (section 3.1). The limited, coarsely gridded numerical simulations in CHS08 were unable to establish a precise value for $We_G^*$, but it appears to be of order unity, which we adopt here; for raindrops falling in Earth's atmosphere, $We_G^* =8$, so refining this constant in the nebula application is worthy of more attention.  
As above we substitute $\Omega^2 = G M_{\odot}/a^3$, and, closely approximating the definition of $\rho_R$ as $M_{\odot}/(4a^3)$, obtain 
$ \Phi_2(a) l = \beta a  (2 \rho_R / \rho_g We_G^*)^{1/2}$. 
Then using $\rho_R/\rho_g$ from section 3.2.1, $l = H \alpha^{1/2}2^{-N/3}$ from the cascade relation, and scaling $a/H$ and $\beta$ with $a$ as above, we obtain in a straightforward way
\be
 \Phi_2(a) = 2^{N/3}\left({\beta_o  a_o \over H_o } \right) 
                      \left({ 2 K_0 \over \alpha We_G^* } \right)^{1/2} 
                      \left({ a\over a_o} \right)^{(p-3/2)/2},
\ee
where $K_0$ is defined in section 3.2.1 and again we retain the $N$-dependence. Note that $\Phi_2$ is not a function of $S$, and thus appears as a horizontal line in {\bf figure 3} for each value of $N$, but is an explicit function of $\alpha$ because of the $l$ factor in the threshold equation for $\Phi_2l$ (CHS08 equation 4, and above). CHS08 discuss why other possible gas effects, such as turbulent pressure fluctuations, are negligible compared to simple ram pressure.

\subsection{Derivation of the Initial Mass Functions}  The two most important things about an IMF are (a) the {\it shape} of the mass distribution $P(M)$, in particular its modal value if any, and (b) its absolute value, giving the {\it rate} at which primary planetesimals of those masses are created. Our cascade model is the key to both. The cascade model PDFs $P(\Phi,S)$ (section 2.3) refer to  scale $l = 2^{-N/3}L$, where for the nebula $L \approx H \alpha^{1/2}$ is the largest eddy diameter. Thus the cascade model provides us with the volume density (the occurrence probability or volume per unit volume) of zones having a particular combination of density and vorticity on a specific nebula lengthscale $l$, and allows us to calculate both a mass associated with each bin (given by $M = \Phi \rho_g l^3$), and the abundance of these bins at any time (given by $P(\Phi,S)$). As we increase the cascade level $N$, we sample statistics at smaller $l$, where the PDF is more intermittent and the probability contours expand, limited by the constraint of saturation near $\Phi \sim 100$ (section 2.3; Hogan and Cuzzi 2007; figure 3). The values of $P(\Phi,S)$ depend explicitly on $N$ and the initial value of the total solid/gas ratio $A$, with canonical value $A_o$ which we take to be $10^{-2}$ everywhere for reference, comprising particles with sizes suitable for turbulent concentration ($t_s \sim t_{\eta}$; section 2.2). 

The thresholds $\Phi_1$, $\Phi_2$, and $S_{min}$ (equations 6-8) also increase with $N$ (a result of their $l$-dependence). For any $N$, the {\it most common} planetesimal mass is that corresponding to the clump (of size $l$) with the highest value of $P(\Phi,S) = P_N$ lying along the threshold lines $\Phi_1$, $\Phi_2$, for $S > S_{min}$. As $N$ changes, the contours and thresholds evolve at {\it different rates}; thus $P_N$  varies with $N$ and there is typically some maximum $P_N = P^*$ at some value of $N = N^*$. This defines the peak of the distribution at $N^*$, $\Phi^*$, $P^*$. Because there is also a mass $M$ associated with any $\Phi, l(N),$ and $\rho_g$, $P_N(M)$ provides the complete IMF and has a modal mass $M(\Phi^*)$. 

This situation is best perceived in a sequence of snapshots at different $N$, which are difficult to present in the format of a printed page (see online supporting material or http://spacescience.arc.nasa.gov/media/staff/jeff-cuzzi/IMF.ppt). We attempt to present it in {\bf figure 3} using two different colors for the contours and thresholds associated with two different values of $N$. The planetesimal diameters plotted in the right panel are derived from $M=\Phi^* \rho_g l^3$, assuming a planetesimal density of 2.0 g/cm$^{3}$.    

\begin{figure*}[t]                                 
\centering                                                                   
\includegraphics[angle=0,width=3.2in,height=2.9in]{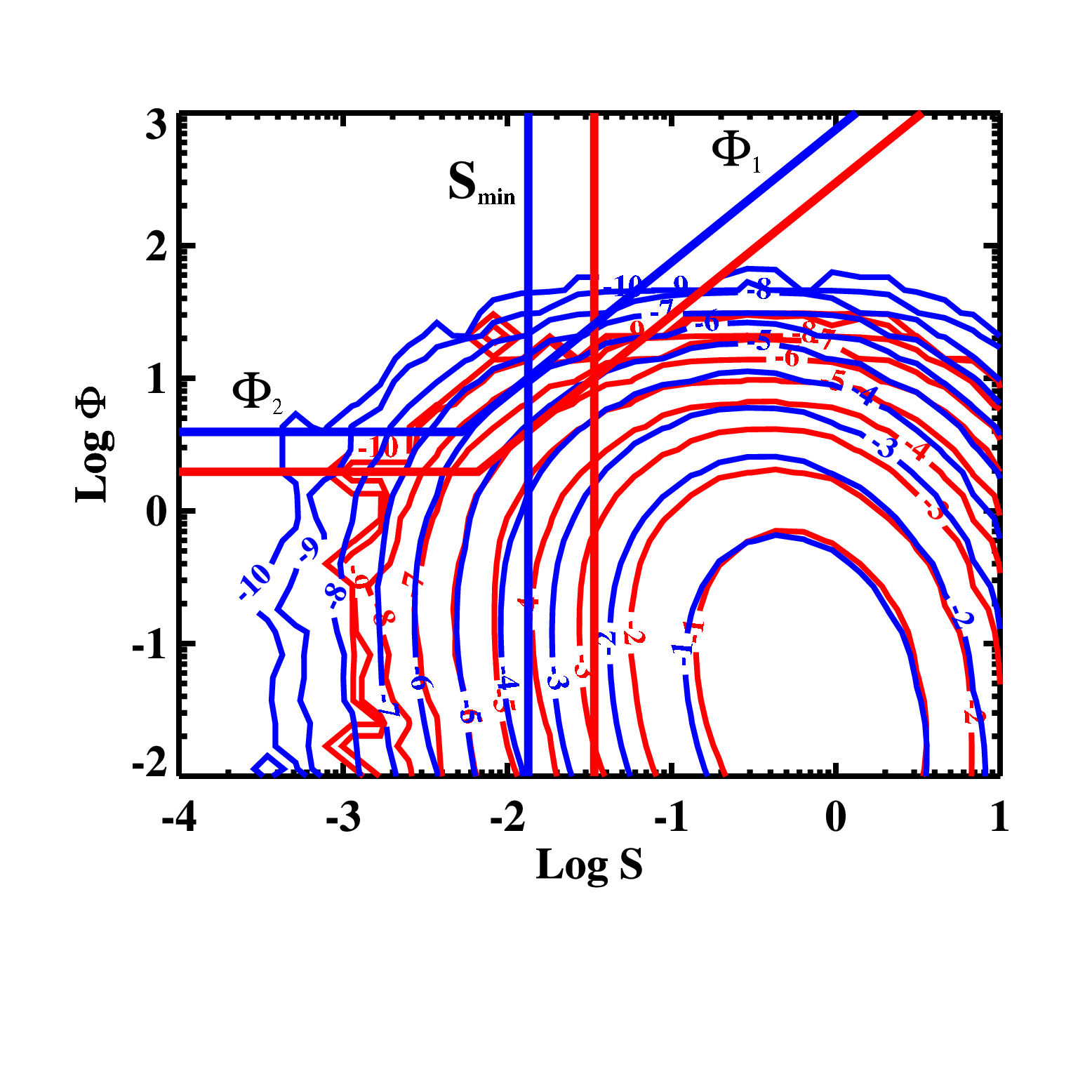}
\includegraphics[angle=0,width=3.2in,height=2.9in]{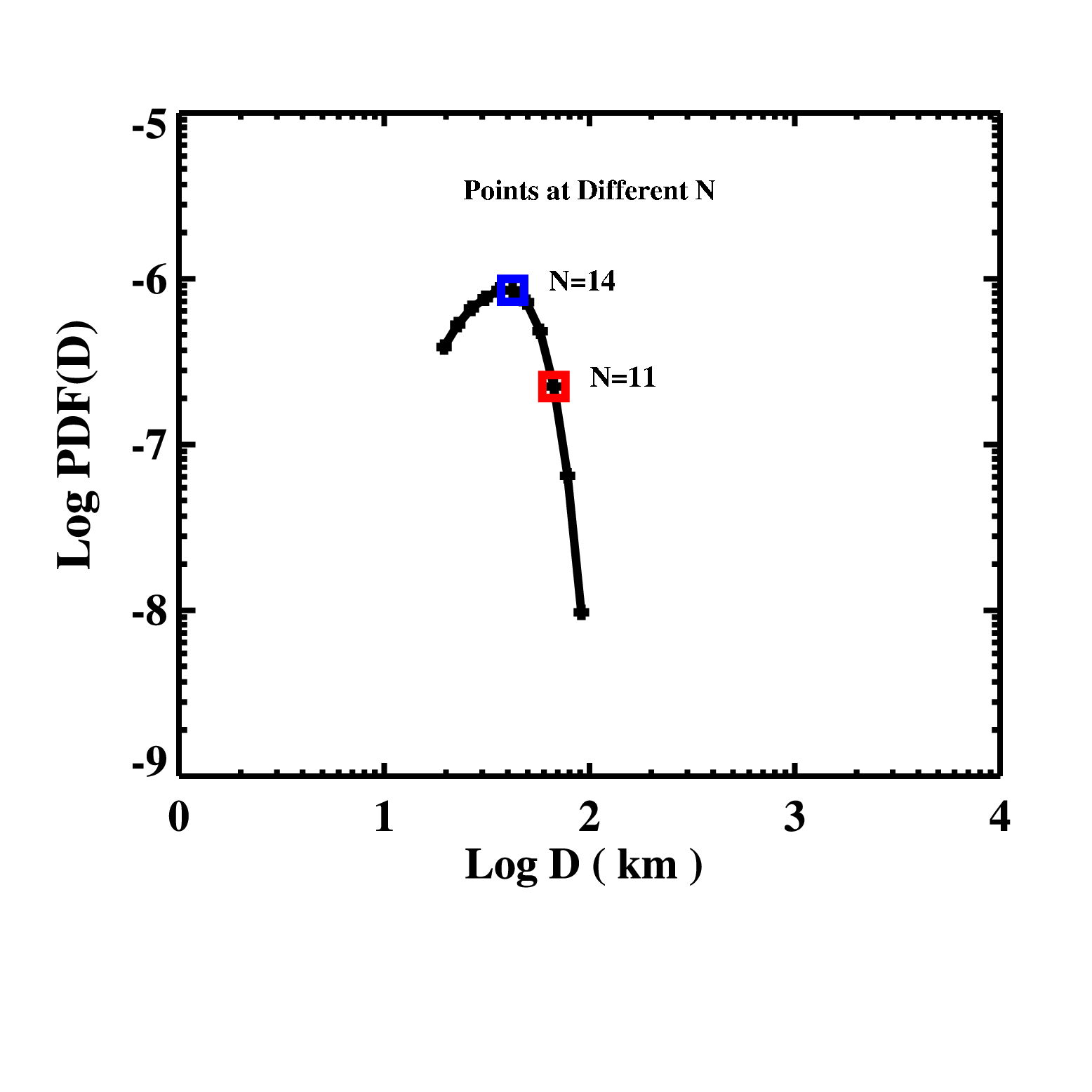}
\vspace{-0.0in}
\caption{\footnotesize{A closer look at how the IMF is determined from the PDFs $P(\Phi,S)$ and the thresholds for primary accretion (sections 3.2 and 3.3; see also figure 2 and section 2.3). {\bf Left}: Contours show the cascade model results for (the log of) fractional volume $P(\Phi,S)$, running from $10^{-1}$ to $10^{-10}$, at two different cascade levels $N$; the blue contours are for the larger $N$. This cascade refers to a case with $A=10A_o$. Also shown for the same two levels are the thresholds $\Phi_1 = \Phi_{sek}$ (diagonal), $\Phi_2$ (horizontal), and $S_{min}$ (vertical; see section 3.2). Note that, as the contours expand with increasing $N$, the thresholds recede up and to the left. {\bf Right}: the curve in the right panel (section 3.3) plots the maximum value of $P(\Phi,S)$ in the region of ($\Phi,S$) that exceeds {\it all three} thresholds, as a function of calculated diameter $D$, with one point for each value of $N$; red and blue points refer to the same cases in the left panel. The curve traced out as $N$ varies gives the primary accretion IMF, which has a peak at some $N=N^*$, defining $P^*$ and the associated $\Phi^*$. The curve in the right panel can be associated with the case $\rho_{go}=3\times 10^{-9}, \alpha=10^{-4}, \beta_o=10^{-4}$ (figure 4b); as noted in section 3.4, these values differ from canonical values.}}
\end{figure*}

To summarize, the placement of the {\it thresholds} $\Phi_1$, $\Phi_2$, and $S_{min}$ depends on $N$ and the physical parameters of the nebula model assumed: the nebula $\alpha$, the local gas density $\rho_g$ or surface density $\sigma$, and the headwind parameter $\beta$.  The placement of the {\it contours} $P(\Phi,S)$ depends on $N$ and the local solid/gas ratio $A$, which can be enhanced over cosmic abundance $A_o$ (here assumed to be 0.01). Primary IMFs vary accordingly. Extension of the theory to 30 AU is straightforward (section 3.2 and 3.3.1), depending on the radial dependence of $\sigma(a)$ and $\beta(a)$, as determined by powerlaw relationships plus whatever (ill-constrained) radial variation there might be of turbulent intensity $\alpha$.  In section 3.4 we show preliminary IMFs at 2.5 and 30 AU, for a number of nebula parameters. Before describing these, we outline our approach to constraining the vertical component of the IMFs - the actual creation rates as functions of size. 

\subsubsection{Planetesimal creation rate} 
A successful model must reproduce estimates of the mass originally created in primordial planetesimals in some region, over the time available; this is the average primary accretion rate $\dot{M}_{pa}$. We will compare $\dot{M}_{pa}$ from our models with expectations for the solar system. Primary accretion occurs when clumps having local density of solids $\Phi \rho_g$, occupying some small volume fraction of the nebula $P(\Phi,S)$, become stable against disruption and form a cohesive planetesimal in their sedimentation timescale $t_{sed} = 1/4G\Phi \rho_g t_s$ (section 3.1; CHS08). We can assume primary accretion is dominated by a region near the peak of each modeled distribution (a more refined approach is described in the Appendix). Then we require the set of parameters $(P^*,\Phi^*,N^*)$ at each peak or modal value to satisfy a stipulated primary accretion rate $\dot{M}_{pa}$ and solve for the value of $P^*$ which we refer to as the ``goal" value $P_{goal}$. We can then normalize the various distribution peak values $P^*$ by the corresponding $P_{goal}$ for the same parameter set, to assess how well the parameter set achieves the stipulated $\dot{M}_{pa}$. 
The available nebula volume between semimajor axes $a_1$ and $a_2$ is $ \pi (a_2^2-a_1^2)\cdot 2H \beta^{1/2}$, where only some vertical fraction $\beta^{1/2}$ may be suitable for this process (section 3.1). Thus  
\be
\dot{M}_{pa} = (\Phi^* \rho_g) P^*( 2 \pi (a_2^2-a_1^2) H \beta^{1/2})/t_{pa}.
\ee
In equation (9), the numerator represents the total amount of mass at any instant lying in zones which exceed the various thresholds ($\Phi_1, \Phi_2, S_{min}$) and can become planetesimals. The demominator $t_{pa}$ is the timescale on which primary accretion converts this mass into planetesimals. We then solve equation (9) for the values of $P^* \equiv P_{goal}$ which are needed to produce the estimated primary accretion rate as
\be
P_{goal}= { \dot{M}_{pa} t_{pa}\over ( 2 \Phi^* \rho_g \pi (a_2^2-a_1^2) H \beta^{1/2})} = { \dot{\sigma}_{pa}t_{sed} \over 2 \Phi^* \rho_g H \beta^{1/2}},
\ee
where we thereby define a primary accretion rate in terms of surface density: $ \dot{\sigma}_{pa}$. 

In equation (10), for specificity and to be conservative, we adopt numerical values of $t_{pa}=t_{sed}$ for the relevant mass production timescale; this is not a well-defined selection but assumes that all the physics of clump formation and dispersal, including various dynamical and fluid timescales, is captured by the ensuing average volume fractions $P(\Phi,S)$, such that the rate at which sandpile planetesimals appear is then simply the proto-sandpile mass so defined at any given time, divided by the time it takes them to become sandpiles. Because this choice is uncertain by a large factor, we will carry a final factor of $t_{pa}/t_{sed}$ which will illustrate the sensitivity of our results to the uncertainty in $t_{pa}$. While studying the results presented in the next section, the reader should keep in mind that the value of $P_{goal}$ would be significantly smaller, and thus the normalized IMFs in figures 4-6 would be considerably closer to unity, if the timescale in the denominator of equation (9) were, instead of $t_{sed}$, the {\it formation} time of a clump (plausibly on the order of $t_L \sim 1/\Omega_L$ or roughly the orbit time, which is a factor of $10^2-10^3$ shorter than $t_{sed})$. Indeed Chambers (2010) has assumed an even smaller timescale for $t_{pa}$, comparable to the (shorter) eddy timescale at lengthscale $l << L$. The question of the most appropriate approach to estimating $\dot{M}_{pa}$ is a fruitful subject for future consideration.  

Below, we will use crude estimates of $ \dot{\sigma}_{pa}$ at 2.5 and 30 AU to constrain our model predictions of $P(\Phi,S)$. First, we rewrite equation (10) in a more useable form, combining all occurrences of familiar nebula parameters. We assume the particle stopping time for preferential concentration $t_s$, which occurs in the definition of $t_{sed}$,
is equal to the Kolmogorov eddy timescale $t_{\eta}$ which depends on nebula properties (see section 2.3):
\be
t_s = t_{\eta}={ 1 \over \Omega Re^{1/2}}  = 
            {\nu_m^{1/2} \over \Omega (\alpha c H)^{1/2} } = 
            {K_1 \over \Omega (\alpha \rho_g H)^{1/2} },
\ee  
where $K_1 = 5.3\times 10^{-5}$ g$^{1/2}$ cm$^{-1}$, and we have expressed the gas kinematic viscosity as $\nu_m = m_{H_2} c / \sigma_{H_2} \rho_g$ where $m_{H_2}$ and $\sigma_{H_2}$ are the mass and cross section of a hydrogen molecule, respectively (Cuzzi et al 1993). Chondrule-like particles satisfy this relationship in the asteroid belt region, but in the outer nebula, much smaller or less dense particles will be optimally selected (CHPD01). Then substituting for $t_s$ we obtain
\be
P_{goal} = { \dot{\sigma}_{pa} \Omega (\alpha \rho_g H)^{1/2} \over 
                   8 G K_1 \Phi^{*2} \rho_g^2 H \beta^{1/2} }
			= { \dot{\sigma}_{pa} \Omega \alpha^{1/2} \over 
                   8 G K_1 \Phi^{*2} \rho_g^{3/2}H^{1/2}\beta^{1/2}}
            \left({ t_{pa}\over t_{sed}}\right) .
\ee
We now scale all radially variable quantities assuming nominal powerlaw nebula surface density and mean temperature discussed above, and obtain 
\be
P_{goal} = 
         \left({ \Omega_o \over 8 G K_1(\beta_o H_o \rho_{go}^3)^{1/2}} \right)
         \left({ \dot{\sigma}_{pa} \alpha^{1/2}(a) \over \Phi^{*2}} \right)
                   \left({ a\over a_o} \right)^{(3p +q-1)/2}
            \left({ t_{pa}\over t_{sed}}\right) .
\ee

Regarding $\dot{\sigma}_{pa}$, a consensus belief is that the 2-4AU region of the primordial asteroid belt (prior to dynamical clearing) contained planetesimals with a mass of about 2$M_{\oplus}$ (Petit et al 2001, Chambers 2004, and personal communication 2009), which isotopic age dating suggests formed over about 2Myr. The Kuiper belt is less well constrained but required perhaps 40 $M_{\oplus}$ in planetesimals between 16-30AU (Tsiganis et al 2005; see sections 2.1 and 3.4.2); for our scenario to be relevant this also must have happened before the gas vanished. Standard nebula lifetimes of about 3Myr (Haisch et al 2001) refer to the presence of warm dust, probably more relevant to the asteroid belt region than the Kuiper belt region. Currently, the lifetime of outer nebula dust can only be limited crudely to less than 10-30Myr (Carpenter et al 2005). For simplicity here, we simply assume the same accretionary lifetime for the outer nebula as for the inner nebula (2Myr); this is shorter than found by traditional incremental growth models (see section 2.1). Then $\dot{\sigma}_{pa}({\rm 2.5AU}) \sim 2.4 \times 10^{-14}$ g cm$^{-2}$ sec$^{-1}$, and $\dot{\sigma}_{pa}({\rm 30AU}) \sim  10^{-14}$ g cm$^{-2}$ sec$^{-1}$. After some algebra, equation (13) becomes 
\be
P_{goal}(2.5AU) \sim  10^{-5} 
             \left({\alpha \over 10^{-3}}\right)^{1/2} 
             \left( { 10 \over \Phi^*}\right)^2
             \left({ 10^{-9} \over \rho_{go}} \right)^{3/2} 
             \left( { 10^{-3} \over \beta_o}\right)^{1/2}
            \left({ t_{pa}\over t_{sed}}\right) .
\ee
At 30 AU the results depend on the radial scaling parameters $p$ and $q$ (now embedded in $K_2$ below); we assume $q=0.5$:
\begin{equation}
P_{goal}(30AU;p) = K_2(p) 
		 \left({\alpha \over 10^{-2}}\right)^{1/2} 
            \left( { 10 \over \Phi^*}\right)^2
            \left({ 10^{-9} \over \rho_{go}} \right)^{3/2} 
            \left( { 10^{-3} \over \beta_o}\right)^{1/2}
            \left({ t_{pa}\over t_{sed}}\right)  
\end{equation}
where $K_2(p)= 10^{-5}(a/a_o)^{(3p-0.5)/2}$; thus $K_2(0.5)=3.6 \times 10^{-5}$, $K_2(1.0)=2.4 \times 10^{-4}$, and $K_2(1.5)=1.5\times 10^{-3}$, and we have suggested (different) plausible values of $\alpha$ (sections 2.1 and 2.2) at 2.5 and 30 AU, and a typical overall value of $\Phi^*$, for scaling purposes. 

\subsection{Results}

{\bf Figure 4} shows preliminary IMFs we have derived at 2.5 AU, for a range of nebula parameters, based on the methods described in section 3.3. The modal diameters for primary planetesimals fall within the 20-200 km range of uncertainty spanned by models of subsequent stages of evolution leading to the observed asteroids (Morbidelli et al 2009a, Weidenschilling 2009; section 2.1 and figure 1). Each IMF is normalized by the ``goal" value $P_{goal}$ for the combination of parameters defining each curve, calculated using equation (14) or (15). If the peak of the normalized IMF approaches unity, it implies that the case is capable of producing enough mass in planetesimals, in the time available, to satisfy current expectations.  The actual values of $P^*$ and $P_{goal}$ are tabulated in Tables 2-4. Clearly, some cases are more successful than others in this regard, but it is intriguing that the model even comes close to satisfying both of these independent constraints at once. However, to do so, the results shown for 2.5AU prefer a local background solid mass enhancement over cosmic abundance $A/A_o =10$ (figures 4a, 4b), or $A/A_o =30$ (figure 4c), and a headwind parameter $\beta$ which is as much as 10 times lower than normally assumed ($\beta \sim 10^{-3}$; Nakagawa et al 1986; Cuzzi et al 1993). Enhancement of solids over canonical values, and suppression of the headwind speed below canonical values, are not only important, they are connected (see below). 

Comparison of {\bf figures 4a-c} is instructive regarding the effects of uncertainty in the models. All results are obtained applying all three thresholds: $\Phi_1$, $\Phi_2$, and $S_{min}$. However, figure 4a shows results derived assuming the simple derivation of $\Phi_1$ in section 3.2.1, with $A=10 A_o$. Figure 4b shows the implications of adopting, instead, the more refined threshold $\Phi_{sek} = (10/3)\Phi_1$ (see section 3.2.1) - the values of $P^*$ {\it and} $P_{goal}$ (and of the resulting normalized IMFs) decrease, while the modal sizes shrink slightly. In {\bf figure 4c}, we show that increasing $A/A_o$ by only a factor of three increases the IMF $P^*$ values dramatically, to the point that the normalized IMFs routinely exceed unity; thus for our assumption of $t_{pa}=t_{sed}$ (section 3.3.1), a degree of solids enhancement in the range $A/A_o \sim 10-30$ is apparently called for. On the other hand, the normalized IMFs would increase by several orders of magnitude if we were to adopt $t_{pa} \sim t_L$ instead of $t_{pa} \sim t_{sed}$ (see section 3.3.1 and Chambers 2010). In section 3.5 we discuss the ability of these cases to match other constraints, where $t_{pa}$ plays no apparent role. 
\begin{figure*}[htp!]                                 
\centering                                                                   
\includegraphics[angle=0,width=3.2in,height=3.2in]{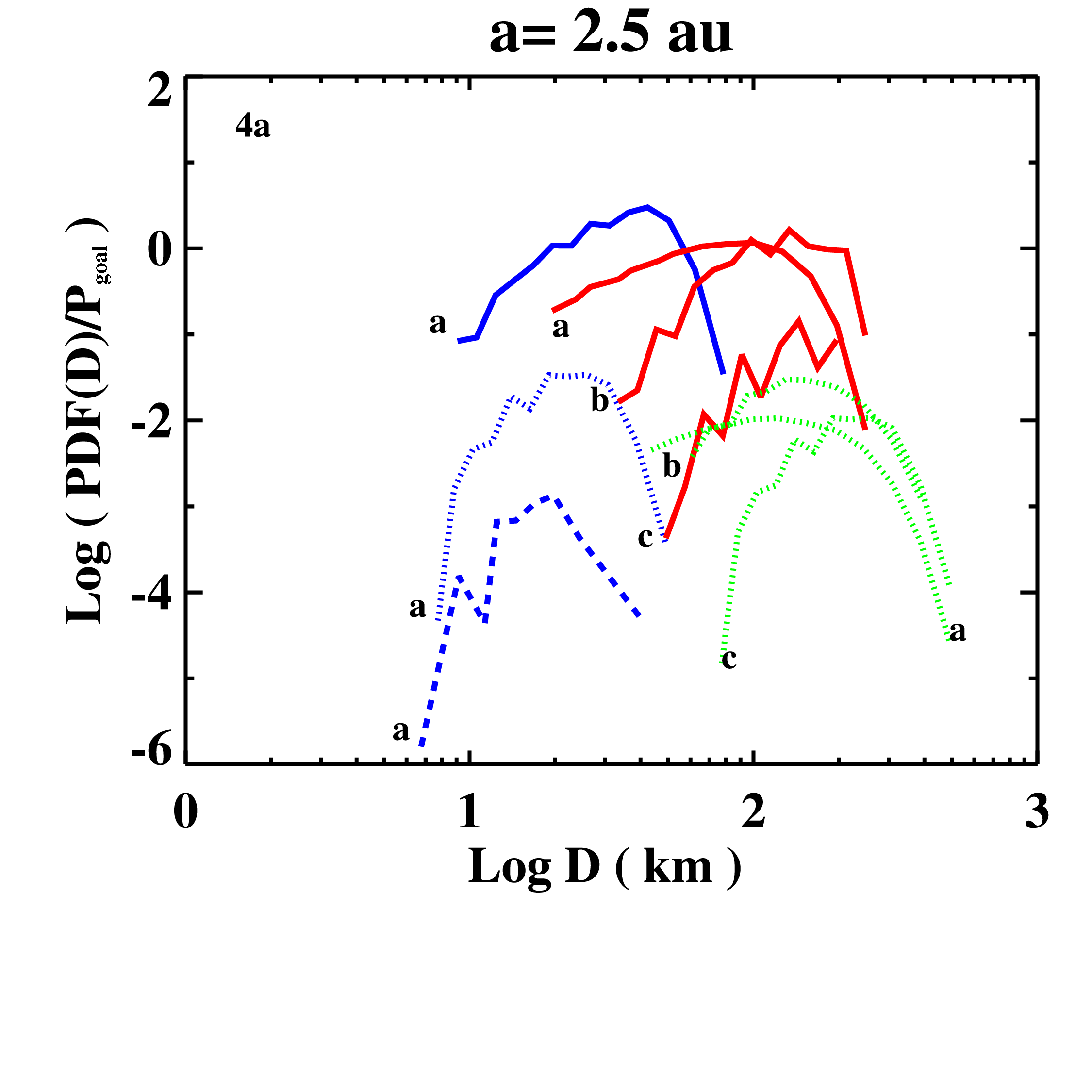}
\includegraphics[angle=0,width=3.2in,height=3.2in]{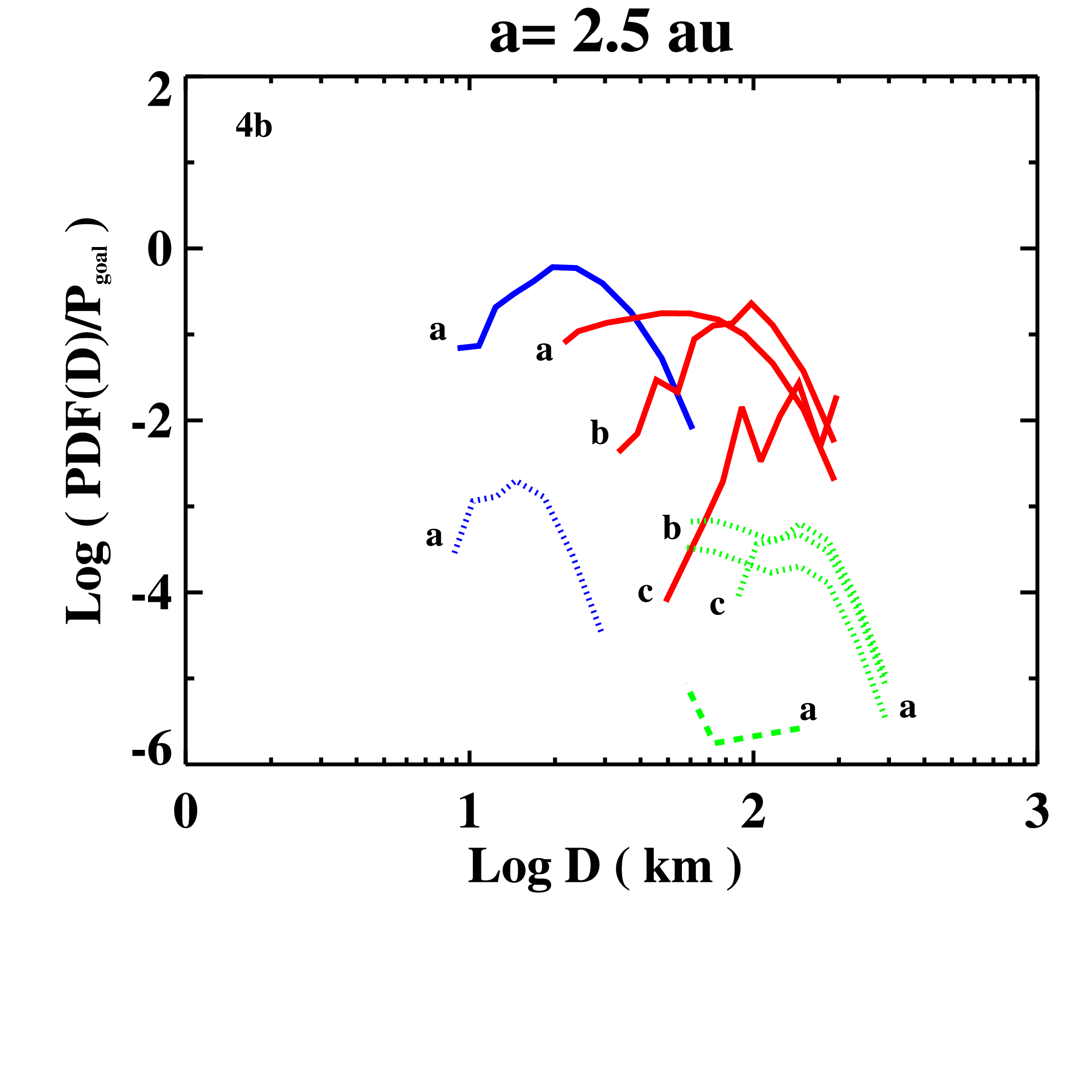}
\includegraphics[angle=0,width=3.2in,height=3.2in]{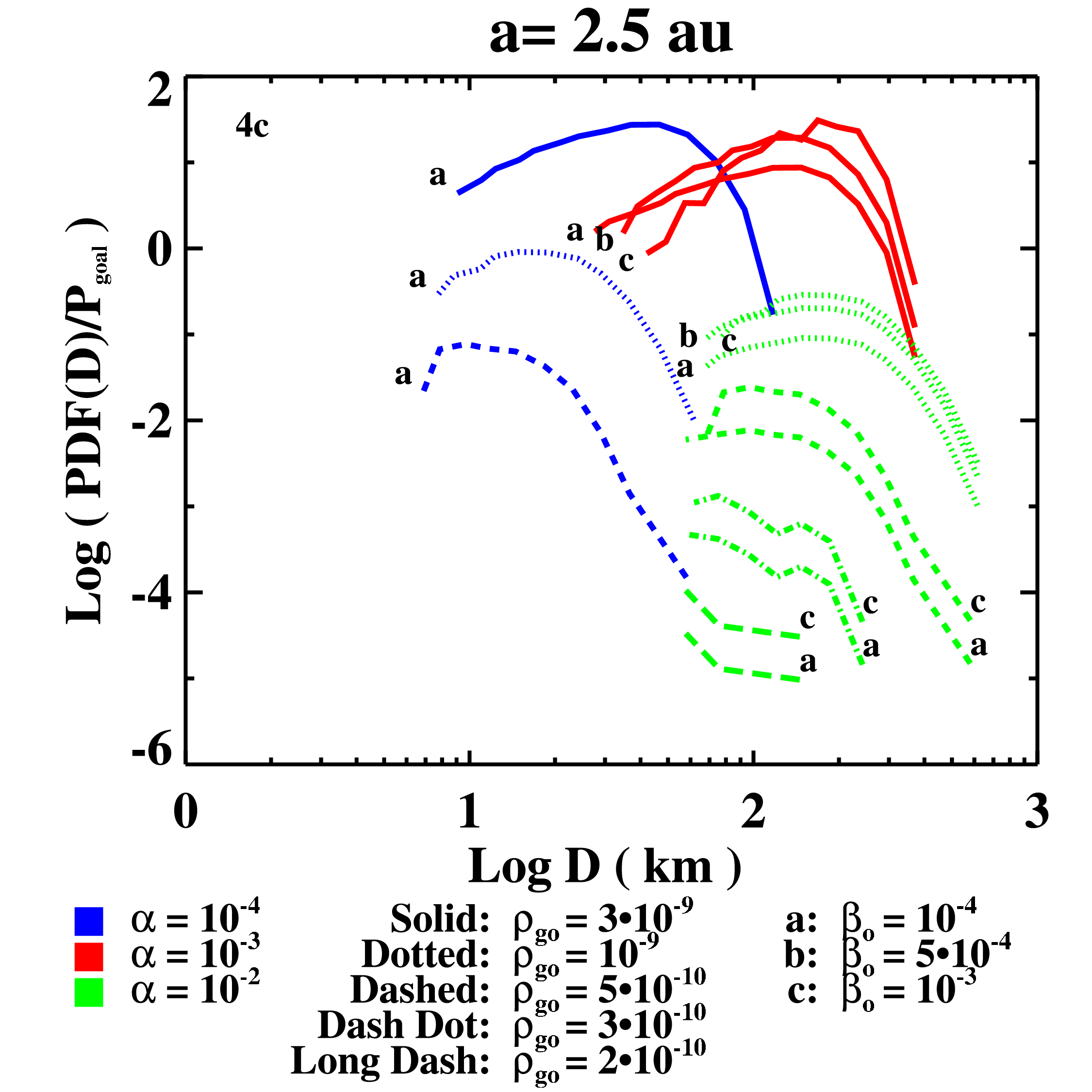}
\vspace{-0.0in}
\caption{\footnotesize{Initial Mass Functions (IMFs) at 2.5AU. The vertical axis plots the normalized function ${\rm log}(P(\Phi,S)/P_{goal})$ (section 3.3.1), which peaks at the mode value $P(\Phi,S)=P^*$ (section 2.3; right panel of figure 3). Peak values near unity indicate the scenario can produce the expected pre-depletion mass in planetesimals in the expected time. Figures 4a and 4b (top left and right) assume a background enhancement of solids over cosmic abundance of $A/A_o=10$, and figure 4c (bottom) assumes $A/A_o=30$. Blue: $\alpha=10^{-4}$, Red: $\alpha=10^{-3}$, and Green: $\alpha=10^{-2}$. Figures 4b and 4c (and 5 and 6) increase our simply derived value of $\Phi_1$ (equation 6) by a factor of 10/3 to align it with the result of Sekiya (1983). The normalizing gas densities $\rho_{go}$ refer to $a_o=$2.5AU. Curves are also labeled by the headwind parameter $\beta_o$($a_o$=2.5AU).  Roughly speaking, larger $\alpha$ produces larger planetesimals, and increasing $A/A_o$ or $\rho_{go}$ and/or decreasing $t_{pa}$ or $\beta_o$, leads to higher production rates. These cases are tabulated in Tables 2-5.}}
\end{figure*}
\begin{figure*}[htp!]                                 
\centering                                                                   
\includegraphics[angle=0,width=3.1in,height=3.1in]{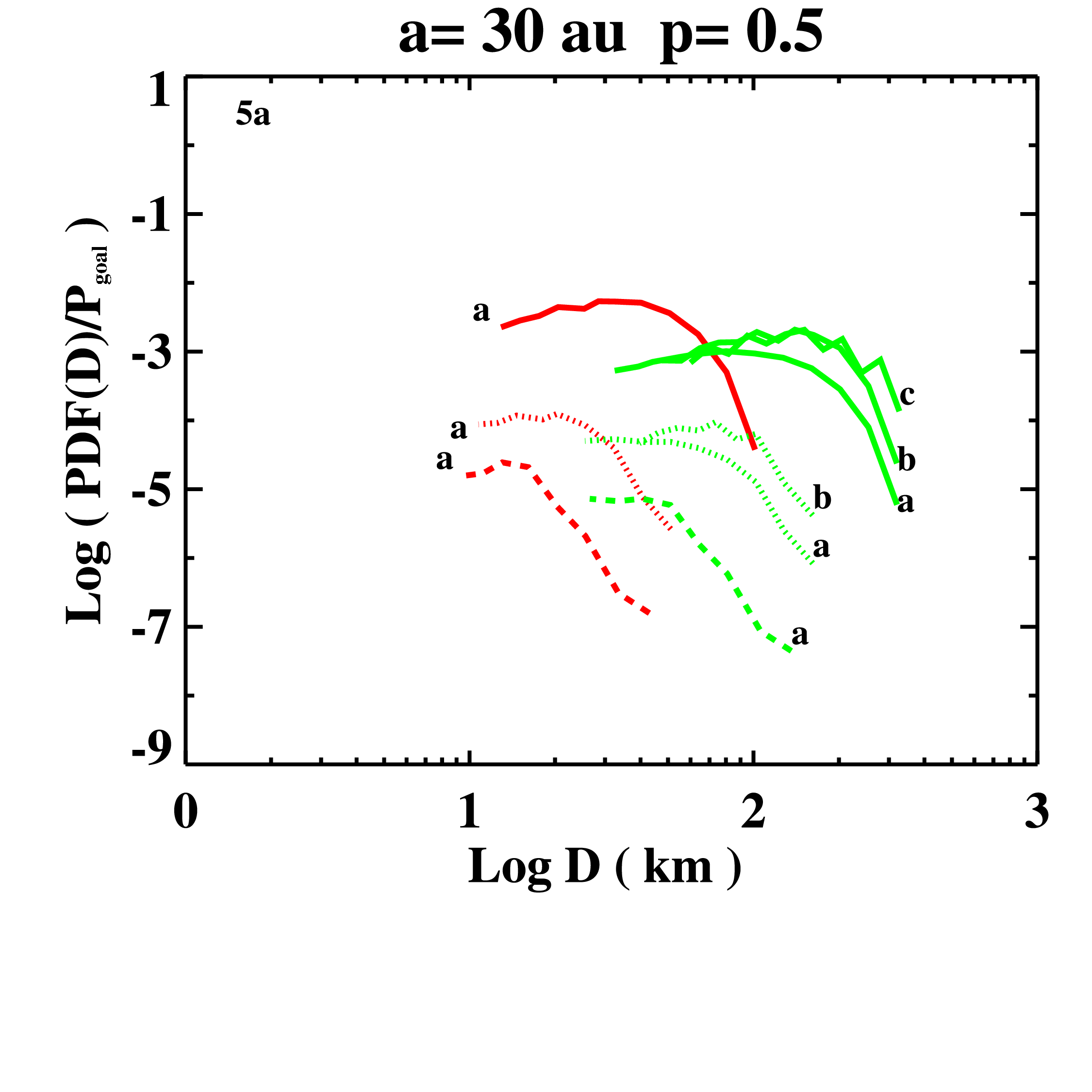}
\includegraphics[angle=0,width=3.1in,height=3.1in]{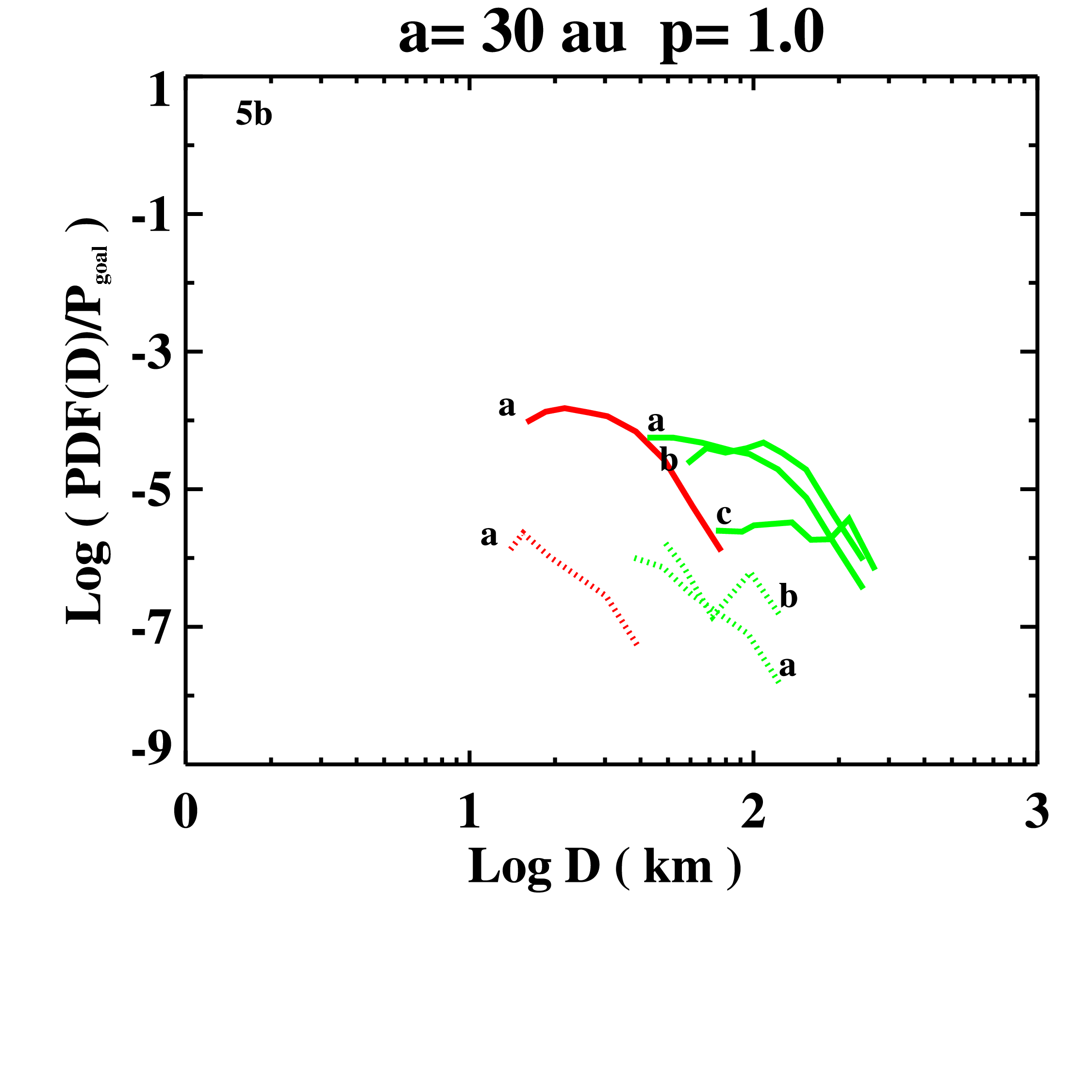}
\includegraphics[angle=0,width=3.1in,height=3.1in]{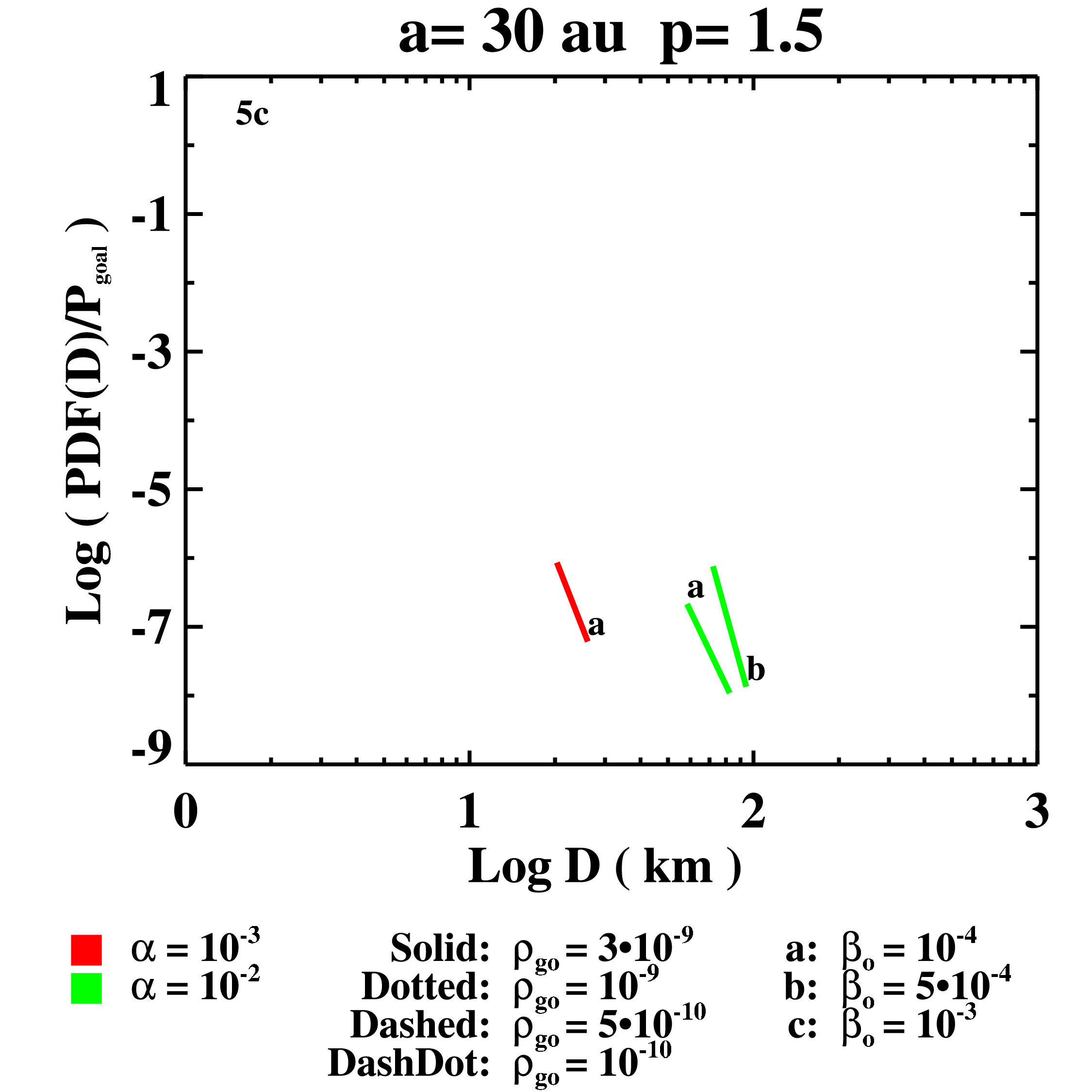}
\vspace{-0.0in}
\caption{\footnotesize{ Initial Mass Functions (IMFs) at 30AU, similar to figure 4; these  assume different nebula radial surface gas density powerlaws $\sigma(a_o)(a/a_o)^{-p}$ with $p=$ 0.5, 1.0, and 1.5. The vertical axis plots ${\rm log}(P(\Phi,S)/P_{goal})$ (section 3.3.1). The curves here assume normal cosmic abundance ($A=A_o$) and the Sekiya value of $\Phi_1$. The normalizing gas densities $\rho_{go}$ refer to 2.5AU, and curves are also labeled by the headwind parameter $\beta_o$(2.5AU). As in figure 4, larger $\alpha$ produces larger planetesimals, and increasing $\rho_{go}$ and/or decreasing $\beta_o$ leads to higher production rates. The scant results for $p=1.5$ suggest we would need to extend the cascade model to higher $N$ (smaller $l$) than we have so far, to capture the mode of the distribution, which would remain at low $P^*$. Allowing for a smaller $t_{pa}$ could raise these normalized IMFs by a factor of $10^2-10^3$ by decreasing $P_{goal}$ (section 3.3.1).}}
\end{figure*}
\begin{figure*}[htp!]                                 
\centering                                                                   
\includegraphics[angle=0,width=3.1in,height=3.1in]{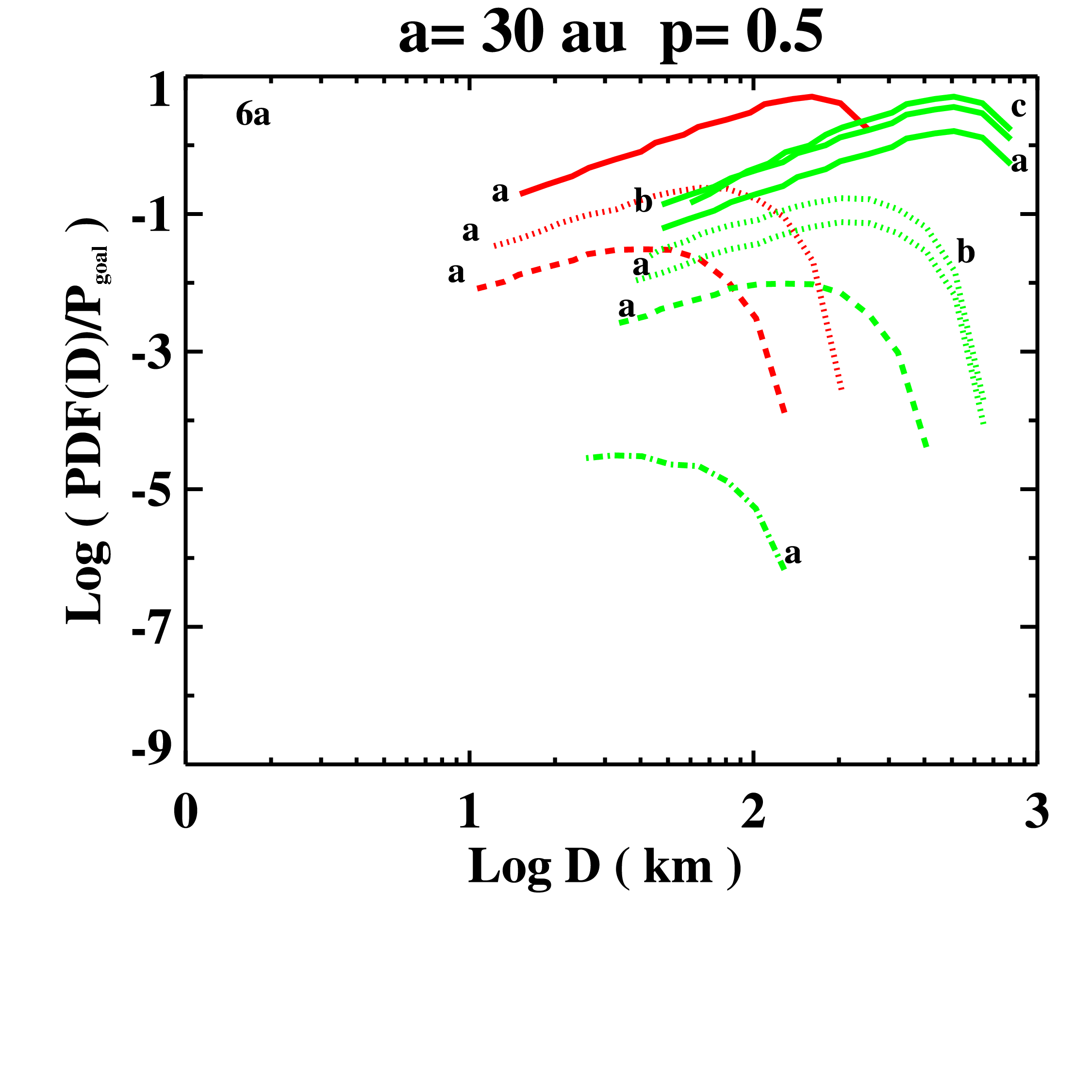}
\includegraphics[angle=0,width=3.1in,height=3.1in]{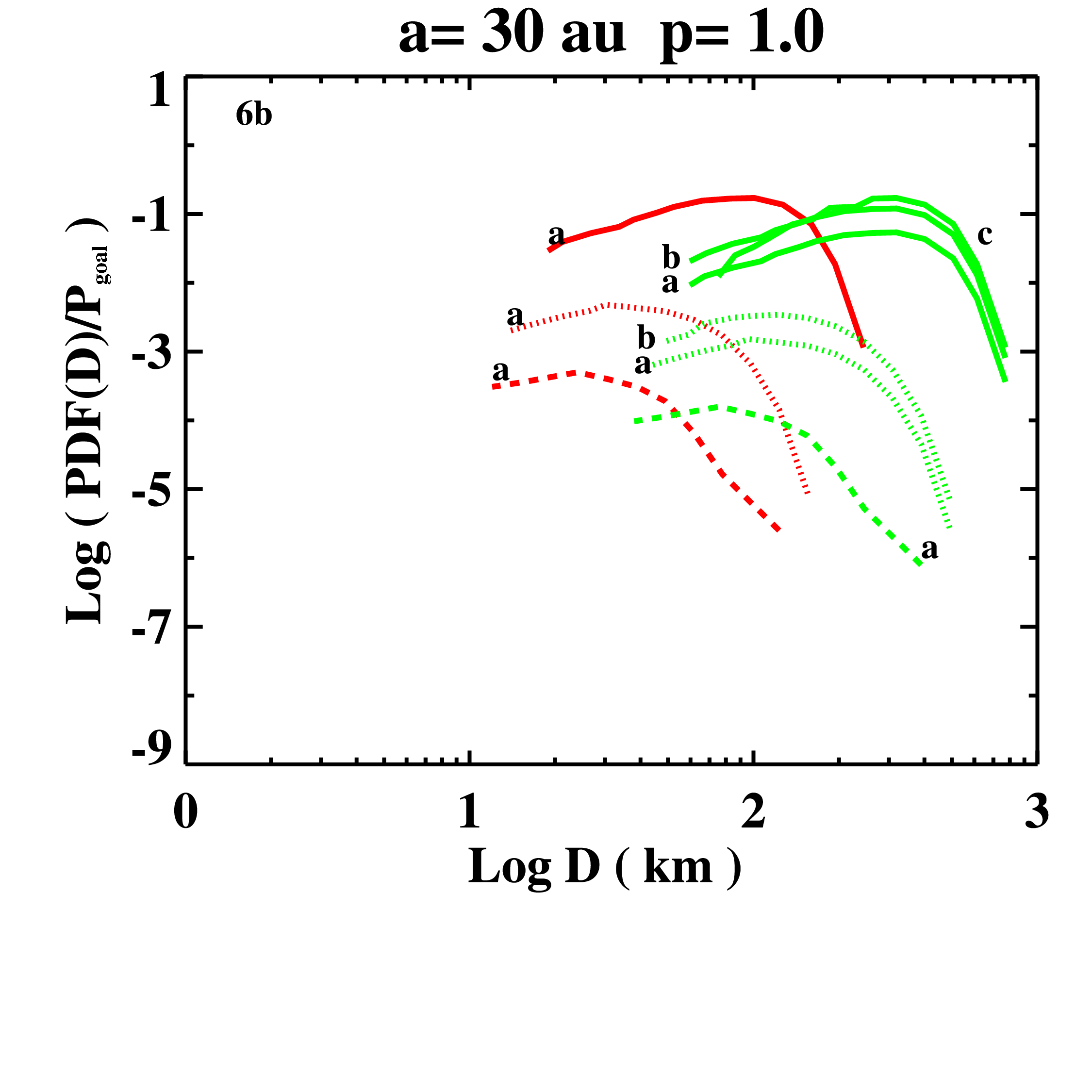}
\includegraphics[angle=0,width=3.1in,height=3.1in]{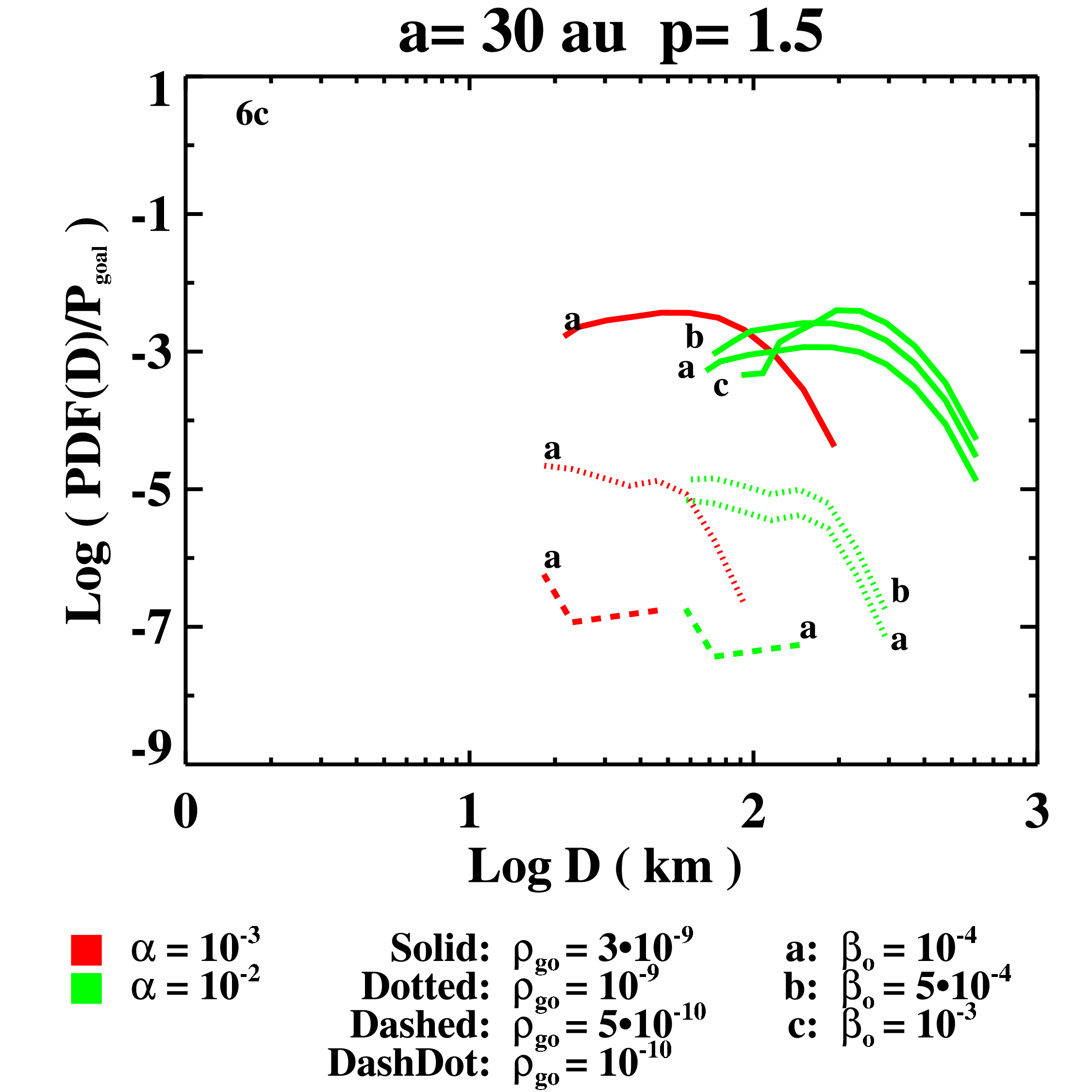}
\vspace{-0.0in}
\caption{\footnotesize{Same as figure 5; normalized IMFs at 30AU, assuming the Sekiya-adjusted value of $\Phi_1$ and different surface density profiles $p=$0.5, 1.0, and 1.5. These results differ from figure 5 by assuming a local abundance of solids enhanced over cosmic by a factor of 10 ($A=10A_o$). Normalized IMFs approach (or even exceed) unity  for flatter radial distributions, and all normalized IMFs can be increased by decreasing $t_{pa}$ (section 3.3.1). }}
\end{figure*}

\subsubsection{Enhancement of local solids by radial and vertical decoupling} Suggestions that $A>A_o$, where $A_o$ is the cosmic abundance, are not new in the context of primary accretion. Growth to dm-or-m size (but perhaps no further) may be robust in turbulence (Dominik et al 2007, Ormel et al 2008), especially in the outer solar system where water ice might increase particle ``stickiness". This will cause large amounts of mass to migrate from the outer solar system to the inner solar system much faster than the gas evolves, elevating the relative abundance of solids significantly (Stepinski and Valageas 1996, 1997, Cuzzi and Zahnle 2004, Ciesla and Cuzzi 2006, Kornet et al 2001). A similar process was advocated  by Youdin and Chiang (2004) in a nonturbulent nebula without particle growth. More recent models  (Zsom et al 2010) which account for experimental results for silicates (G{\"u}ttler et al 2010) find growth being frustrated at even smaller sizes, where radial drift would be much smaller - keeping material around longer in the inner solar system where ices are mostly absent.  A complementary process is vertical settling of clumps formed at high altitude, which brings material to lower altitudes faster than would otherwise be the case (Wang and Maxey 1993, Aliseda et al 2002, Bosse et al 2006). We are actively studying both of these processes, and feel that a combination of them could lead to background abundance of $A/A_o \sim 10-30$ within $z < H\beta^{1/2}$, where we have suggested primary accretion operates. Moreover, settling of solids towards the midplane can affect the local gas orbital velocity and thus the headwind experienced by particles (Nakagawa et al 1986). For plausible enhancements ($A/A_o <30$) this is unlikely to lead to more than an order unity effect; solving for the headwind velocity using equations in Nakagawa et al (1986), for particles with short stopping times such as those of interest, gives a reduction in effective $\beta$ by a factor $1-\rho_d/\rho_g$, or 0.7 for $A/A_o = 30$. Haghigipour and Boss (2003) showed in principle, and Johansen et al (2007) found in realistic 2D and 3D turbulent simulations, that the headwind parameter $\beta$ can essentially vanish in local pressure gradient reversals, which may be long-lived but might only occupy a small fractional volume. Testing these speculations with actual models is an important goal for future work. 

The 30AU cases generally fall well below their respective goals if $A=A_o$ is assumed ({\bf figure 5}), especially for the steeper radial density distributions. However, some enhancement of solids at 30AU is not obviously out of the question, at least near the midplane: if dense clumps are forming and settling, there will be some enhancement due to this process alone, as discussed above. Moreover, for nebulae that begin with $>$100 AU radial extent, some enhancement may occur due to particle growth at radii larger than 30AU, and subsequent inward drift, depending on the growth rate in these rarified regions. Stepinski and Valageas (1997), Weidenschilling (2004), and Kornet et al (2001) find enhancements of solids by factors of order unity by 0.1-1Myr, depending on $\alpha$. On the other hand, Garaud (2007), using different assumptions (very low particle sticking), finds a depletion of solids at these distances. Further studies of the growth of solids in the remote outer nebula are important to continue because of these discrepancies and their implications for primary accretion. Overall, we feel it is not implausible that some degree of enhancement of solids did indeed occur even at 30AU, although probably not as substantial as that which might have occurred at 3AU. Because our results are so strongly dependent on the abundance of solids (figure 4c), we ran the same set of cases at 30AU assuming $A=10A_o$ ({\bf figure 6}). Indeed the normalized IMFs now come closer to, or even exceed, unity. It does appear that moderately steep radial nebula profiles, such as  that of the traditional minimum mass nebula ($p=1.5$) are seriously challenged to match the mass production rate goals as adopted here. Again, we note that smaller values of $t_{pa}$ also have the potential to increase the normalized IMFs closer to unity by several orders of magnitude (section 3.3.1; Chambers 2010). It is interesting to note that, in {\bf figure 5c}, the normalized IMFs actually {\it increase} with increasing $\beta$, so for the larger values of $\alpha \sim 10^{-2}$ that may well apply to the outer solar system, nominal values of $\beta$ might be acceptable (somewhat in contrast to the cases at 2.5 AU). 

\subsubsection{Primary accretion efficiency and its implications} It is interesting to note that, at both 2.5 and 30 AU, only a small amount of the mass initially present actually gets accreted into planetesimals in this scenario; that is, our accretionary process is highly inefficient - quite distinct in this regard from the traditional 100\% efficient minimum mass nebula assumption. Consider a typical $p=1$ case with $\rho_{go}=10^{-9}$ g cm$^{-3}$ at 2.5 AU. Say our models achieving $P^* \sim P_{goal}$ had $A \sim 10 A_o$ on average. If this is a product of both radial and vertical settling, the solids {\it surface mass density} might be increased by a factor of 3 over our nominal nebula models (by, {\it eg.,} inward radial drift of particles from further out). This would result in available mass of about 120$M_{\oplus}$ in solids in the 2-4AU region and about 900$M_{\oplus}$ in the 16-30 AU region. Then, $P^* \sim P_{goal}$, or accretion of 2$M_{\oplus}$ in the 2-4 AU region, means that only $\sim$1.5\% of the solids there were captured into planetesimals, and accretion of 40$M_{\oplus}$ in the 16-30 AU region means that only $\sim$4\% of the solids there were captured. The balance (that is, almost everything) presumably escapes into the sun (or in the case of the outer nebula, escapes the solar system or helps feed the inner nebula). 

There are some especially interesting implications for the Kuiper Belt. These efficiencies, while small, are still larger than those of traditional models of incremental accretion of the current Kuiper Belt ({\it eg.} Stern and Colwell 1997, Kenyon 2002). In these models, tens of $M_{\oplus}$ of solids initially present in the region produced only the observed 0.01-.1$M_{\oplus}$ in KBOs - having an efficiency an order of magnitude smaller than that derived above because of the broad size distribution in which they are required to grow. Recent models of subsequent dynamical depletion (during evolution of Neptune) of the planetesimals that {\it did} form do not change this efficiency, but just require a larger starting mass.  On the other hand, perhaps the biggest remaining problem in the ``dynamical emplacement" theory of the KBOs (Levison et al 2008) is the cold classical population, which has lower eccentricity, and perhaps higher binary fraction, than might be expected from dynamical emplacement from within 30AU. That is to say, growing at least the cold classical KBO population {\it in situ}, while emplacing the rest, might have some appeal; we suggest below that this is not out of the question given an edge in the solids density near 30 AU. These issues, and others related to a large initial mass in the Kuiper belt region, are described in section 2.1.

As mentioned above, our primary accretion rates (and masses) are strongly dependent on $A/A_o$. Comparing figures 5 and 6 shows that a drop in $A/A_o$ by a factor of 10 leads to a drop in $P^*/P_{goal}$ by a factor of about 1000. Suppose there were a factor of 10 drop, or ``edge", in the abundance of solids at around 30 AU, and all other parameters were slowly varying ({\it eg.}, Weidenschilling 1997, Stepinski and Valageas 1996, 1997). We would then expect to form, over the 30-44AU region, 1000 times less mass in primary bodies than our goal value of 40$M_{\oplus}$ - about 0.04$M_{\oplus}$. Indeed this is about four times the mass of the cold classical population (Bernstein et al 2004). Even if the dynamical evolution of Neptune removes about 90\% of originally formed planetesimals (A. Morbidelli, personal communication 2010), the scenario described in this paper is only a factor of 2-3 away from forming the required $\sim 0.1 M_{\oplus}$ cold classical (primordial) population {\it in situ}. Moreover, this formation would not leave behind a massive population of other objects that would need to be eroded away and/or cause Neptune to migrate further than observed, because the bulk of the planetesimal mass formed by this scenario ``forms big" at the mode in the mass distribution. Given the simplicity of the scenario presented here, and its sensitivity to uncertain parameters, it seems this possibility may be worth further thought. 

In future, more detailed studies, allowance should be made for the fact that a primary accretion mechanism designed to produce (part or all of) the current crop of KBOs in 16-30 AU might also need to produce the roughly equal mass in planetesimals that grew into the cores of Uranus and Neptune, which would increase the mass accretion rate by a factor of 2 or so; this does not strike us as prohibitive given the observational and model uncertainties involved. On the other hand, if Uranus and Neptune formed inside of 17AU ({\it eg.} as in Tsiganis et al 2005) this requirement is relaxed. Finally, there is nothing to prevent the scenario described here from creating all the KBOs {\it in situ} between 30-44AU, if the solids mass were appropriately enhanced and nebula parameters slowly varying (at least for the $p=1$ models), but the cleanup and Neptune migration problems would remain. 

\subsection{Other comparisons between model predictions and observations}
We have noted several times that the normalized IMFs (figures 4-6) can be made to approach or exceed unity (that is, produce the needed mass in planetesimals in the allowed time) if our assumed value of $t_{pa}=t_{sed}$, which determines $P_{goal}$ is several orders of magnitude too large (see for instance Chambers 2010). In this section we discuss other constraints the model can be compared with, which are not dependent on this uncertain parameter.
\subsubsection{Clump encounter times and the age spread in chondrites}
As discussed in section 2.1, recent work suggests a wide spread in the formation times of chondrules in a given chondrite - a significant fraction of a Myr. Some concerns remain regarding the interpretation of these data as age differences, which will surely be addressed as more data emerge. Here we take the age differences at face value, and explore their implications for our models. Below we estimate how long a newly formed chondrule must wander through the nebula before, along with many other chondrules being independently formed and wandering about, entering one of the rare, dense clumps that is destined to become a sandpile planetesimal under the scenario presented here. We would expect this timescale to approximate the half-width in the formation age range observed in a particular chondrite, by the statistical nature of the process. That is, a few new chondrules will accrete shortly after their formation, most will accrete after a time $t_{enc}$, and some unlucky ones might need to wait another $t_{enc}$ or so to find their parent proto-sandpile clump. We assume a fixed particle density spatial distribution of proto-sandpile clumps of size $l$ and mass density $\Phi$ even if the individual clumps are everchanging. A wandering chondrule, being nearly tied to the turbulent gas, sweeps through the frame defined by the orbiting gas (and dense clumps) at a speed $V_p$ roughly equal to the turbulent velocity $V_L =c \alpha^{1/2}$ (Cuzzi and Hogan 2003). However, because the motion of preferentially concentrated particles is not random in space and does not sample all fluid volumes with equal probability, a simple random-walk, particle-in-a-box encounter calculation is inappropriate. 

Instead, we use a ``duty-cycle" approach similar to that described by CHPD01 (their section 6.2). In the Appendix we describe some of the details involved in translating the nomenclature of CHPD01 to that used here. CHPD01 integrate a two-dimensional function such as our $P(\Phi,S)$ over $S$ and distinguish $F_V(\Phi)$, the fraction of {\it volume} lying in zones of mass loading $\Phi$, from $F_p(\Phi)$, the fraction of {\it particles} lying in such zones. These are not equal because particles preferentially are found in dense zones, not randomly in space. CHPD01 demonstrated that the cumulative fraction of {\it particles} $F_p(>\Phi)$ lying in regions of density larger than $\Phi$ is the same as the fraction of {\it time} spent by a {\it given particle} in regions with density larger than $\Phi$, $F_t(>\Phi)$. This would be true for the differential functions $F_p(\Phi)$ and $F_t(\Phi)$ as well. We generalize here to the two-dimensional function $F_p(\Phi,S)$ because we are more carefully treating the role of enstrophy, but the same identification will hold between $F_p(\Phi,S)$ and $F_t(\Phi,S)$, and for their cumulatives. We also adopt a different treatment of the cumulative of $F_p$, as described in the Appendix: specifically, we calculate the fraction of particles $F_p(>T)$ lying in zones having properties anywhere within the stable region defined by the thresholds $S_{min}, \Phi_2,$ and $\Phi_1(S)$ (see figure 3; also see Appendix for derivation of $F_p(>T)$). As in CHPD01 we set $F_p(>T)$ equal to the fraction of time $F_t(>T)$ spent by any given particle in zones capable of becoming sandpile planetesimals. We note that $F_p(>T)$ is calculated at the level $N$ defining the mode, or maximum, in the IMF for each parameter case (figures 3-6), and is thus associated with a lengthscale $l$. 

For a wandering particle CHPD01 define $t_{in}=l/V_p$ as the time it spends traversing a clump of size $l$, and $t_{enc}$ as the time between encounters with such a clump. Setting $t_{in}=l/V_p$ neglects the ``peloton effect" in which some (but not most) particles are seen to follow a given clump for an extended period of time. Then the duty cycle, or fraction of time spent by a particle in regions of size $l$ capable of becoming sandpile planetesimals can be approximated (assuming $t_{in} \ll t_{enc}$) by 
\begin{equation}
F_t(>T) = { t_{in} \over t_{enc}} = { l \over V_p t_{enc}} = { l \over V_L t_{enc}}.
\end{equation}
We recall that only the {\it subset} of such regions which lie within a fraction $\beta^{1/2}$ of the nebula's vertical extent are candidates to become sandpile planetesimals (sections 3.1, 3.3.1). The time fraction spent by particles in this subset of regions is therefore $F'_t(>T) = \beta^{1/2}F_t(>T)$. We then obtain the encounter time of a particle {\it with proto-sandpiles} by setting $t_{in}/t_{enc} = F'_t(>T) = \beta^{1/2}F_t(>T) = \beta^{1/2}F_p(>T)$, and solving for $t_{enc}$:
\begin{equation}
t_{enc}= {l \over \beta^{1/2}F_p(>T) V_p}
       = {2^{-N/3}H \alpha^{1/2}\over \beta^{1/2}F_p(>T) c \alpha^{1/2}}
       \sim {2^{-N/3}\over \beta^{1/2}F_p(>T) \Omega}.
\end{equation}
In equation (17) above we have used $l = 2^{-N/3}L = 2^{-N/3}H \alpha^{1/2}$ and $V_L=c\alpha^{1/2}$, and $F_p(>T)$ is evaluated at the value of $N$ giving the peak of the IMF. In the Appendix we note that $F_p(>T)$ can be simply related to the modal peak value $P^*$. Of course, these arguments are simplified and need to be explored in more detail statistically and numerically. Nevertheless, the crude estimates shown for $t_{enc}$ in {\bf Tables 2-5}, ranging to values of a fraction of a Myr, confirm that, in this scenario, accretion is a drawn-out process with timescales compatible with those observed. Recall that the observed age dispersion half-widths in several different chondrite classes are a few$\times 10^5$ years (section 2.1).  Cuzzi et al (2010) show that the distribution of chondrule ages in two primitive chondrites is compatible with a Poisson arrival time distribution characterized by $t_enc$ = 0.2-0.4Myr.
\begin{table}
\begin{center}
\begin{tabular}{|c|c|c|c|c|c|c|c|c|c|c|}
\hline
$\alpha$&$\rho_{go}$&$\beta_o$&$N^*$&$\Phi^*$&$P^*$&$P_{goal}$&$F_p(>T)$&$t_{enc}$&${\Delta}a/a$&$\dot{M}$\\
\hline
1e-4&5e-10&1e-4&12&19.3&1.0e-8&7.6e-6&6.8e-9&5.8e+2&26.60&5.2e-9\\
1e-4&1e-9&1e-4&11&10.8&2.9e-7&8.5e-6&3.6e-7&1.4e+1&4.10&1.0e-8\\
1e-4&3e-9&1e-4&9&3.9&3.7e-5&1.2e-5&2.1e-5&3.7e-1&0.67&3.1e-8\\
1e-3&3e-9&1e-4&10&3.3&6.4e-5&5.5e-5&6.3e-5&1.0e-1&1.10&3.1e-7\\
1e-3&3e-9&5e-4&10&7.8&7.2e-6&4.4e-6&8.1e-6&3.5e-1&2.05&3.1e-7\\
1e-3&3e-9&1e-3&11&19.7&7.0e-8&4.9e-7&8.9e-8&1.8e+1&14.66&3.1e-7\\
1e-2&1e-9&1e-4&14&9.1&1.3e-6&1.2e-4&3.7e-6&6.7e-1&9.02&1.0e-6\\
1e-2&1e-9&5e-4&14&10.8&1.1e-6&3.8e-5&3.4e-6&3.2e-1&6.28&1.0e-6\\
1e-2&1e-9&1e-3&11&10.8&2.9e-7&2.7e-5&3.6e-7&4.4e+0&23.04&1.0e-6\\
\hline
\end{tabular}
\end{center}
\caption{\footnotesize{Summary of all predictions for the models of figure 4a (relevant to 2.5 AU, and assuming the $\Phi_1$ of equation 6), as designated by $\alpha_o, \rho_{go},$ and $\beta_o$. $N^*$, $\Phi^*$, and $P^*$ characterize the mode, or peak, of the IMF for each case. The encounter time of a chondrule with its ultimate planetesimal-forming clump is also tabulated ($t_{enc}$, in Myr; equation 17), as well as the corresponding value of the normalized radial diffusion width $\Delta a/a$ in a time $t_{enc}$ (sect. 3.5.2), and the implied mass accretion rate $\dot{M}$ ($M_{\odot}$/yr) given the other parameters (sect 3.5.3).  We assumed $\Omega=5 \times 10^{-8}$, appropriate at 2.5 AU. The value of $P_{goal}$ assumes $t_{pa}=t_{sed}$.}}
\end{table}
\begin{table}
\begin{center}
\begin{tabular}{|c|c|c|c|c|c|c|c|c|c|c|}
\hline
$\alpha$&$\rho_{go}$&$\beta_o$&$N^*$&$\Phi^*$&$P^*$&$P_{goal}$&$F_p(>T)$&$t_{enc}$&${\Delta}a/a$&$\dot{M}$\\
\hline
1e-4&5e-10&1e-4&15&61.0&2.0e-11&7.5e-7&5.0e-12&4.0e+5&696.40&5.2e-9\\
1e-4&1e-9&1e-4&15&30.6&2.1e-9&1.1e-6&4.2e-9&4.8e+2&24.05&1.0e-8\\
1e-4&3e-9&1e-4&14&12.5&7.4e-7&1.2e-6&2.0e-6&1.3e+0&1.24&3.1e-8\\
1e-3&3e-9&1e-4&15&11.1&8.6e-7&4.9e-6&3.1e-6&6.4e-1&2.78&3.1e-7\\
1e-3&3e-9&5e-4&12&12.4&4.0e-7&1.7e-6&4.7e-7&3.8e+0&6.76&3.1e-7\\
1e-3&3e-9&1e-3&11&19.7&1.3e-8&4.9e-7&4.6e-9&3.5e+2&64.82&3.1e-7\\
1e-2&1e-9&1e-4&19&30.7&3.5e-9&1.1e-5&1.5e-8&5.3e+1&80.38&1.0e-6\\
1e-2&1e-9&5e-4&18&30.6&3.3e-9&4.7e-6&1.2e-8&3.6e+1&66.31&1.0e-6\\
1e-2&1e-9&1e-3&15&30.6&2.1e-9&3.4e-6&4.2e-9&1.5e+2&135.27&1.0e-6\\
1e-2&5e-10&1e-3&15&61.0&2.0e-11&2.4e-6&5.0e-12&1.3e+5&3916.13&5.2e-7\\
1e-2&5e-10&1e-4&19&61.2&6.5e-11&7.5e-6&2.2e-11&3.5e+4&2069.48&5.2e-7\\
\hline
\end{tabular}
\end{center}
\caption{\footnotesize{See table 2 caption; this table replaces $\Phi_1$ of equation (6) by $\Phi_{Sek}$ (section 3.2.1), and assumes $A/A_o = 10$ (see figure 4b). In this set of models the encounter times are mostly too long, and the radial diffusion probably too large, to match the meteoritic constraints. Also, as shown in figure 4b, the desired condition $P^*/P_{goal}$ is not generally satisfied. Smaller values of $t_{pa}/t_{sed}$ would increase $P^*/P_{goal} \geq 1$  accordingly (section 3.3.1).}}
\end{table}
\begin{table}
\begin{center}
\begin{tabular}{|c|c|c|c|c|c|c|c|c|c|c|}
\hline
$\alpha$&$\rho_{go}$&$\beta_o$&$N^*$&$\Phi^*$&$P^*$&$P_{goal}$&$F_p(>T)$&$t_{enc}$&${\Delta}a/a$&$\dot{M}$\\
\hline
1e-4&5e-10&1e-4&17&72.7&4.1e-8&5.3e-7&1.5e-7&8.6e+0&3.23&5.2e-9\\
1e-4&1e-9&1e-4&15&32.4&8.6e-7&9.4e-7&3.1e-6&6.4e-1&0.88&1.0e-8\\
1e-4&3e-9&1e-4&10&10.5&4.8e-5&1.7e-6&4.0e-5&1.6e-1&0.44&3.1e-8\\
1e-3&3e-9&1e-4&10&10.5&4.8e-5&5.5e-6&4.0e-5&1.6e-1&1.39&3.1e-7\\
1e-3&3e-9&5e-4&10&10.4&4.8e-5&2.5e-6&4.0e-5&7.1e-2&0.93&3.1e-7\\
1e-3&3e-9&1e-3&10&15.7&2.4e-5&7.8e-7&2.3e-5&8.6e-2&1.02&3.1e-7\\
1e-2&1e-9&1e-4&15&32.4&8.6e-7&9.4e-6&3.1e-6&6.4e-1&8.79&1.0e-6\\
1e-2&1e-9&5e-4&15&32.3&8.6e-7&4.2e-6&3.1e-6&2.8e-1&5.88&1.0e-6\\
1e-2&1e-9&1e-3&15&32.4&8.6e-7&3.0e-6&3.1e-6&2.0e-1&4.94&1.0e-6\\
1e-2&5e-10&1e-3&17&72.7&4.1e-8&1.7e-6&1.5e-7&2.7e+0&18.15&5.2e-7\\
1e-2&3e-10&1e-3&18&111.6&2.0e-9&1.5e-6&5.8e-9&5.4e+1&81.31&3.1e-7\\
1e-2&2e-10&1e-3&19&153.5&1.6e-10&1.5e-6&3.2e-10&7.7e+2&306.49&2.1e-7\\
1e-2&5e-10&1e-4&17&72.7&4.1e-8&5.3e-6&1.5e-7&8.6e+0&32.27&5.2e-7\\
1e-2&3e-10&1e-4&19&111.7&2.3e-9&4.8e-6&7.5e-9&1.0e+2&112.86&3.1e-7\\
1e-2&2e-10&1e-4&19&153.5&1.6e-10&4.7e-6&3.2e-10&2.4e+3&543.95&2.1e-7\\
\hline
\end{tabular}
\end{center}
\caption{\footnotesize{See table 2 caption; this table replaces $\Phi_1$ of equation (6) by $\Phi_{Sek}$ (section 3.2.1) and assumes $A/A_o = 30$  (see figure 4c). Here, the meteoritic constraints of age variance ($t_{enc} <$ 1 Myr)and radial gradients ($\Delta a/a \leq 1$) may be satisfied by several parameter sets (rows 2-6); note how dramatically the results vary from those in Table 3 for only a factor of three change in solids abundance. Figure 4c also shows how the $P^*/P_{goal} \geq 1$ condition is now robustly satisfied.}}
\end{table}
\subsubsection{Radial diffusion and ``zoning" in the asteroid belt}
{\bf Tables 1-4} also show the extent of radial diffusion $\Delta a$ over the timescale $t_{enc}$, as normalized by semimajor axis $a$=2.5AU; a constant ratio $H/a=0.05$ is assumed here. The outcome of such a random walk in cylindrical geometry is a slightly non-gaussian profile centered on the starting position, having a halfwidth $\Delta a \sim 1.7({\cal D}t_{enc})^{1/2}$ where ${\cal D}\sim \alpha c H$ is the diffusion coefficient (Cuzzi et al 2003).  {\bf Figure 7} shows a slightly more detailed treatment, in which the constant viscosity Green's function for cylindrical geometry (Cuzzi et al 2003) is used to illustrate the diffusive spread of initial delta-functions of tracer ``chondrules" released at 2AU and 4 AU in a nebula with $\alpha=10^{-4}$, after periods of $10^5 - 10^6$ years. As indicated by the overlap of the curves in figure 7, some level of discrimination can be preserved over timescales of a few$\times 10^5$ years, but mixing is nearly complete by 1Myr for $\alpha = 10^{-4}$. Larger $\alpha$, of course, leads to more complete radial mixing, as shown in {\bf tables 2-5}. This sensitivity of the amount of radial mixing to time, across the 0.1-1Myr range, makes emerging developments in chondrule age dating (both the observations themselves and the interpretation of the results)highly relevant (section 2.1).
\begin{figure*}[t]                                 
\centering                                                                   
\includegraphics[angle=0,width=3.1in,height=3.1in]{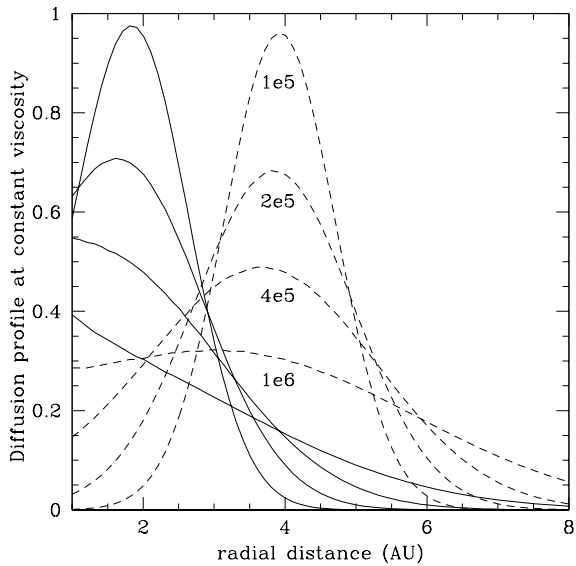}
\vspace{-0.0in}
\caption{\footnotesize{Diffusion profiles for narrow annular sources at 2AU and 4AU, after $10^5$, $2\times 10^5$, $4\times 10^5$, and $10^6$ years. Note that after $10^5$ years, nearly complete separation is retained between sources at the inner and outer edges of the current asteroid belt, but mixing rapidly increases with time and by $10^6$ years the region is fairly well mixed. }}
\end{figure*}
\subsubsection{Nebula mass accretion rate} Tables 2-5 also show the nebula gas mass accretion rate that would be implied by the various adopted parameters: $\dot{M} = 3 \pi \sigma \nu_T = 3 \pi \sigma \alpha c H$ (Lin and Papaloizou 1985). We assume $c=10^5$ cm/s and $H(a_o) = a_o/20 = 2\times 10^{12}$ cm at $a_o$=2.5 AU. The larger values of $\alpha$, combined with the large values of $\rho_{go}$ suggested by models where $P^*/P_{goal} \geq 1 $, produce mass accretion rates substantially larger than regarded as typical for Myr-old protoplanetary disks; a more canonical value is a few $\times 10^{-8}M_{\odot}$/yr. However, there is an order of magnitude scatter in these mass accretion rates which is apparently real (Calvet et al 2000, Hartmann 2005). It is interesting that the parameter range giving the most reasonable $\dot{M}$ for Myr-old disks also satisfies the $t_{enc}$ and diffusion length criteria the best (rows 2-3 for $A/A_o = 10$ and rows 2-5 for $A/A_o = 30$). Several combinations of large $\alpha$ and low $\rho_{go}$ can approach or at least suggest nominal values of $\dot{M}$ in the $10^{-7}M_{\odot}$/yr range; however, the values of $t_{enc}$ are very long for these, because of the low probabilities of the appropriate clumps, given by $F_p(>T)$, and combined with the large $\alpha$ values, radial mixing over these timescales precludes any distinction in properties between contemporaneously formed planetesimals a few AU apart. This constraint only applies to asteroids; no constraint of this type is yet known for KBOs where $\alpha$ might well be large.
\subsection{Optimal parameter range}
Looking at all the predictions together ({\it eg.} Table 5), one tends to favor the lower values of $\alpha$ combined with relatively large gas densities because the combination leads to plausible accretion rates, Myr-or-less encounter times, and small radial diffusion lengthscales. Somewhat larger $\beta$ values, closer to canonical, might be allowable if we adopted a shorter timescale for $t_{pa}$ than $t_{sed}$, which brings the normalized IMFs closer to unity by decreasing $P_{goal}$. Extremely large radial diffusion ranges tend to characterize the larger values of $\alpha \sim 10^{-2}$, and unless the currently inferred range of chondrule formation ages is misleading due to parent body resetting (section 2.1), large values of $\alpha$ at 2.5AU are probably inconsistent with evidence that ordinary chondrites and CO chondrites - having very different chemical and isotopic properties - are about the same age (Kunihiro et al 2004; Kurahashi et al 2008). We note that three grouped, H-like chondrites have been found which are unusual for ordinary chondrites in having abundant CAIs and CO-like matrix (Kimura et al 2002). These chimeric objects may be a sample of a parent body that accumulated at an intermediate location between the H chondrite parent(s) and the CO chondrite parent(s), containing a blend of components which dominated in the two locations. 

The region of parameter space which is consistent with all the constraints we have mentioned seems to be fairly small ($\alpha \sim 10^{-4}$, $\rho_{go} \sim 1-3 \times 10^{-9}$, $\beta_o \sim 10^{-4}$, and $A/A_o \sim 10-30$); in this sense the agreement of the model with expectations is sensitive to small changes in model parameters. However, we are trying to match a number of independent observations at once, with a very simple model, so we are encouraged that there is {\it any} reasonable combination of parameters that comes close to matching them all. 

\begin{table}
\begin{center}
\begin{tabular}{|c|c|c|c|c|c|c|c|c|c|}
\hline
\multicolumn{4}{|c|}{$2.5 AU$} & 
\multicolumn{3}{c|}{$A = 10A_o$} & 
\multicolumn{3}{c|}{$A = 30A_o$}\\
\hline
$\alpha$&$\rho_{go}$&$\beta_o$&$\dot{M}$&$F_p(>T)$&$t_{enc}$&${\Delta}a/a$&$F_p(>T)$&$t_{enc}$&${\Delta}a/a$\\
&g cm$^{-3}$&&$M_{\odot}$/yr&&Myr&&&Myr&\\
\hline
1e-4&5e-10&1e-4&5.2e-9&5.0e-12&4.0e+5&696.40&1.5e-7&8.6e+0&3.23\\
1e-4&1e-9&1e-4&1.0e-8&4.2e-9&4.8e+2&24.05&3.1e-6&6.4e-1&0.88\\
1e-4&3e-9&1e-4&3.1e-8&2.0e-6&1.3e+0&1.24&4.0e-5&1.6e-1&0.44\\
1e-3&3e-9&1e-4&3.1e-7&3.1e-6&6.4e-1&2.78&4.0e-5&1.6e-1&1.39\\
1e-3&3e-9&5e-4&3.1e-7&4.7e-7&3.8e+0&6.76&4.0e-5&7.1e-2&0.93\\
1e-3&3e-9&1e-3&3.1e-7&4.6e-9&3.5e+2&64.82&2.3e-5&8.6e-2&1.02\\
1e-2&1e-9&1e-4&1.0e-6&1.5e-8&5.3e+1&80.38&3.1e-6&6.4e-1&8.79\\
1e-2&1e-9&5e-4&1.0e-6&1.2e-8&3.6e+1&66.31&3.1e-6&2.8e-1&5.88\\
1e-2&1e-9&1e-3&1.0e-6&4.2e-9&1.5e+2&135.27&3.1e-6&2.0e-1&4.94\\
1e-2&5e-10&1e-3&5.2e-7&5.0e-12&1.3e+5&3916.13&1.5e-7&2.7e+0&18.15\\
1e-2&3e-10&1e-3&3.1e-7&-&-&-&5.8e-9&5.4e+1&81.31\\
1e-2&2e-10&1e-3&2.1e-7&-&-&-&3.2e-10&7.7e+2&306.49\\
1e-2&5e-10&1e-4&5.2e-7&2.2e-11&3.5e+4&2069.48&1.5e-7&8.6e+0&32.27\\
1e-2&3e-10&1e-4&3.1e-7&-&-&-&7.5e-9&1.0e+2&112.86\\
1e-2&2e-10&1e-4&2.1e-7&-&-&-&3.2e-10&2.4e+3&543.95\\
\hline
\end{tabular}
\end{center}
\caption{\footnotesize{Summary of Model results at 2.5AU, combining $A=10A_o$ and $A=30A_o$, from tables 3 and 4.}}
\end{table}
\subsection{Caveats and future work} 
The models presented here represent the most obvious and straightforward implications for primary accretion of the physics of turbulent concentration of small particles, and survival of dense clumps of them, as outlined qualitatively in a series of past papers (CHPD01, CW06, CHS08). In order to achieve even this first sanity check, a number of simplifying assumptions were made, and a broad range of nebula parameters was sampled. The caveats we feel are most important to mention (in rough priority order) are: 

1) Perhaps the most significant ``known unknown" is the timescale $t_{pa}$ used to constrain the primary mass accretion rate and assess which parameter sets can create the needed pre-depletion mass of planetesimals (section 3.3.1). In this paper we assumed $t_{pa}=t_{sed} \sim $ 100-1000 orbit periods, while Chambers (2010) has assumed $t_{pa}=t_e(l) \sim $ 0.01-0.1 orbit periods. An interesting compromise might be $t_{pa} \sim t_L$, or about an orbit period; this might be the timescale on which the density field is independently refreshed. A considerable amount of study, using 3D numerical models and following clumps for long amounts of time as they develop and dissipate, is needed to understand this timescale. 

2) The survival conditions for dense clumps based on our various thresholds (section 3.2) lie well down a steeply falling slope of the PDFs predicted by our cascade models; these cascade models must be checked and tested regarding some of the assumptions built into them. For instance, it has been assumed that the multipliers in these models, determined near the dissipation scale in numerical models, actually apply over a much wider range of scales up the inertial range. While our own tests to date have supported this assumption, some other results suggest a scale-dependence to the process. The form and scale-dependence (if any) of the multipliers is fundamental (the shapes of the PDFs on their steep edges are sensitive to details) and  must be checked more closely. The details of the cascade PDFs will have implications for other concerns below. 

3) Some of the physical assumptions made by the model (the density of a clump is so high that it moves at near-Keplerian velocity) are not fully compatible with the typical values of $\Phi \sim 10$ that emerge from requiring the IMFs to match the modal asteroid masses and mass production rate ``goals" (Tables 1-4, figure 4). For $\Phi \sim 10$, $t_{sed}$ is closer to 1000 orbits than 100. Fortunately, it seems that a slower contraction poses no obvious problems as long as the clump is stable against ram pressure disruption; once it is stable by the criteria of section 3.1, it only becomes more stable as it shrinks (CHS08). In fact, lower-$\Phi$ clumps might incur lower $\beta$ if their orbital speeds {\it are} subkeplerian. However, turbulent eddy variations on this long a timescale might assume a larger role in clump disruptions. Of course, higher-$\Phi$ clumps {\it are} also forming, but at lower volume fractions ({\it eg.}, figure 3). Future studies should explore this aspect of the apparent preferred parameter range more carefully with numerical models. 

4) The survival threshold $\Phi_1$ (or $\Phi_{Sek}$), involving centrifugal balance, applies the inverse of an eddy lifetime as if it were a true rotational frequency. This might be overly conservative, and if so, the diagonal lines associated with $\Phi_1$ (figure 3) shift to the right, increasing the values of $P(\Phi,S)$ in all the primary IMFs. 

5) The survival threshold $\Phi_2$ involves a poorly-determined parameter (the so-called gravitational Weber number $We_G^*$) that quantitatively affects $\Phi_2$ and thus the magnitude of $P(\Phi,S)$ in all the primary IMFs; numerical simulations of clump disruption must be carried out at higher resolution to better constrain $We_G^*$. 

6) Implicit in the primary IMF rates is that all solids at the modeled place and time have sizes and densities which are suitable for turbulent concentration; since this is unlikely to be the case, some other inefficiency factor must be allowed for, which will increase the $P(\Phi,S)$ needed to match mass production estimates (which are, of course, quite uncertain themselves). As seen in figure 4, order-of-unity changes in $A/A_o$ can provide IMFs where $P^*/P_{goal}$ increases significantly, allowing room for such inefficiency. Moreover, the value of $t_{pa}/t_{sed}$ also affects $P^*/P_{goal}$ (section 3.3.1); our choice of $t_{pa} = t_{sed}$ might be relaxed and increase $P^*/P_{goal}$ by 1-3 orders of magnitude, ample to compensate for inefficiency due to an initially broad particle size distribution. The recent work of Zsom et al (2010) incorporates extensive new experimental results and finds that, under a range of nebula conditions, grain aggregates in the asteroid belt region reach a ``bouncing barrier" at masses not too different from those of chondrule precursors. Such a moderately narrow size distribution of chondrule {\it precursors} would influence the narrow observed size distribution seen in chondrites, with or without turbulent concentration, and limit the degree of inefficiency inherent in a potentially broad pre-TC size distribution. Meteorite data show a range in mean chondrule size across chondrite types, even while the shape of the distribution remains apparently invariant (section 2.1). More data of this type is needed for more chondrule types, and for more types of objects {\it within} chondrites: metal particles, CAIs, and so on, to see how influential the limitation of {\it chondrule} sizes by limiting the size distribution of their {\it precursors} might be.  

7) If $\beta_o$ is actually much lower than the canonical $10^{-3}$ in regions where planetesimal precursor clumps form (as it appears), the role of turbulent nebula pressure fluctuations as an independent disruption mechanism should be reassessed using detailed numerical models (section 3.2). Moreover, if this were a global value instead of a being locally-determined {\it by} mass loading, as we have suggested, the supply of material from the outer nebula to the inner nebula (and thus $A/A_o$) will be affected (Cuzzi and Zahnle 2004, Ciesla and Cuzzi 2006). On the other hand, relaxation of our assumption that $t_{pa} = t_{sed}$ (section 3.3.1) might allow larger values of $\beta$ to satisfy the mass creation rate constraints. We note that recent studies of nebulae with "dead" or at least "dull" zones embedded within MRI-active layers show higher gas densities than canonical for the terrestrial planet region, and also suggest that radial pressure gradients might be more complex than expected from simple powerlaw radial dependence (Zhu et al 2010).

8) We have taken at face value several recent reports of nearly Myr-variance in the ages of chondrules found in the same chondrites. If uncontaminated by parent body resetting, these variances are critical constraints on primary accretion; a subset of our models indeed comes close to explaining this large variance while explaining other chondrite properties, but the parameters are not necessarily in the canonical range. It is important to extend relative age measurements on individual objects in the same meteorite to a greater variety of samples, and to address concerns that the inferred age differences merely represent parent body resetting events (section 2.1)
.
 
\section{Conclusions}

Traditional models of incremental accretion lead to powerlaw size distributions with equal mass per decade for particles ranging between centimeters and (at least) tens of km in size, where runaway gravitational growth sets in; such distributions are increasingly thought not to be compatible with the current asteroid size distribution (Bottke et al 2005, Morbidelli et al 2009a, Weidenschilling 2009). We have used a very simplified physical model to pursue the most obvious implications of the fate of dense clumps of mm-size particles, which are aerodynamically selected for preferred concentration in nebula turbulence. We follow the simplest physics determining which dense clumps avoid disruption and evolve, on periods of 100-1000 orbits, into objects with some physical cohesion. We find that some small fraction of these dense clumps (those forming near the nebula midplane) can proceed to become ``sandpile" planetesimals having diameters in the 20-200 km range observed for today's asteroids, in an abundance which is broadly consistent with poorly known estimates of the mass of the pre-depletion primordial belt. That is, it appears possible for most of the mass in primitive bodies to have simply skipped over the problematic m-km size range. 

The model is a simple one; except for the physics and statistics of the turbulent cascade model, it contains little more than physical scaling arguments. The parameter range studied is broad but not exhaustive, and the preferred range is not optimally centered on ``canonical" nebula conditions. Specifically, lower-than-expected headwind magnitudes, and higher-than-canonical gas densities and solid/gas ratios, are needed to achieve quantitative matches to (uncertain) estimates of required mass production rates (section 3.4). We have provided some thoughts on why these conditions might not be unrealistic in view of the uncertainties, but more work along these lines is surely needed. For instance, the critical timescale $t_{pa}$ remains poorly understood even in principle (sections 3.3.1 and 3.7; Chambers 2010). 

In spite of the fact that the optimum parameter set is perhaps not the canonical one, we are encouraged that such a simple model can achieve so much with any plausible range of parameters. The model scenario potentially explains many things simultaneously: the mean size, the size distribution, and the age dispersion of chondrules in chondrites, {\it and} the modal mass and mass production rate of primary asteroids. While doing this, the scenario also provides a plausible radial mixing length for planetesimal constituents between their creation and their accretion, allowing for some radial zoning of primitive asteroids, while retaining a canonical nebula mass accretion rate for Myr-old protoplanetry nebulae (for which considerable scatter does exist, however). This is not to say that we believe a final solution is at hand; there are many serious uncertainties and unresolved issues (section 3.7). Moreover, this scenario may not work alone; planetesimal formation is likely to have been very complicated. Aggregation effects acting on small particle scales before chondrules were even melted (most recently G{\"u}ttler et al 2010 and Zsom et al 2010) and particle-gas dynamics acting on larger particles than we consider (Johansen et al 2007) both operate in the same (turbulent) environment we have studied. 

The scenario presented here identifies a path which - while still fraught with hazards - leads directly from freely-floating, mm-size nebula particulates, to sizeable (tens to hundreds of km diameter) sandpile planetesimals formed almost entirely of size-sorted particles, which is in reasonable accord with the meteorite record. The typical encounter time of any particular chondrule with the clump in which it becomes a planetesimal can be a significant fraction of a Myr - comparable to formation age variance observed isotopically in several different chondrite groups.  The constituents can be narrowly sorted physically because primary accretion reflects local properties, but can be diverse chemically and isotopically, with well-defined group properties only in the ensemble, because production/alteration regions can be separated in space by several $H$, and in time by nearly 1 Myr. The primary sandpiles, made of constituents with a wide range of formation ages and chemical and isotopic properties, are in the size range of today's ``fossil" asteroids, and we imagine they go on to experience an extended sequence of compaction, heating, and sintering, perhaps even before the dynamical depletion stage when more violent, erosive and destructive impacts lead to the objects we see today. The process is capable of starting very early in nebula history and proceeding for a long time, as nebula solids evolve. Primary objects of tens-hundreds of km diameter forming early will almost certainly melt extensively, while those forming more than 1.5-2.5 Myr after CAIs will be able to remain unmelted and ``primitive". 

In this scenario, one would expect {\it some} degree of {\it both} radial mixing {\it and} temporal evolution in the properties of chondrules ending up in a particular chondrite. Along these lines it is useful to recall that, while the properties of a chondrite {\it group} are well-defined, there is not only the formation age diversity mentioned above, but also a substantial variance in petrological, chemical, redox, and isotopic properties {\it amongst} the constituent chondrules in any single chondrite (Scott and Krot 2005; Brearley and Jones 1998). The well-defined properties of a chondrite group may only manifest the ensemble homogeneity of their common parent body - which itself might represent a grab-sample of constituents which came together at a given time and place. Other planetesimals forming nearby and contemporaneously might have very similar properties, but planetesimals forming at a different time, or in a different location, would draw from a slightly evolved mixture of essentially the same building blocks, perhaps modified by ongoing alteration, remelting, changing oxidation environment, and/or mixture with particles from adjacent regions, resulting in different ensemble properties.  Cuzzi et al (2005) discuss other meteoritics implications in more detail, including the concept of ``complementarity". Some models of the subsequent stage of accretion, characterized by gravitational scatterings and collisional mergers of primary objects, suggests even more radial diffusion (Bottke et al 2006, Levison et al 2009).

As a cautionary remark, we note that the meteorite data, interpreted in the context of our scenario, also suggest that combinations of nebula $\rho_g$ and $\alpha$ must have varied in time and/or space over the region and duration of primary accretion. This is implied because different chondrite groups with similar accretion ages (ordinary and CO chondrites) have noticeably different modal chondrule {\it sizes} - varying by a factor of several (King and King 1978, Rubin 1989, Scott and Krot 2005; Brearley and Jones 1998; Kurahashi et al 2008). In the context of turbulent concentration, because all chondrules are after all made mainly of silicates and have comparable densities, variations in $\alpha$ and $\rho_g$ are the most obvious way to do this (CHPD01, figure 1). 

Extension of the scenario to the Kuiper Belt region has also been explored; IMFs show a similar preference for 10-100km diameter objects. Mass creation rates assuming canonical cosmic abundance tend to fall short of estimates for KBOs (which are rather poorly known, however), but enhanced solid abundances over cosmic dramatically improve the agreement with mass production estimates. Moderately flat nebula gas density distributions are substantially more favorable to the extension of this scenario from the asteroid belt to 30AU.  Under such conditions, and given the strong dependence of primary accretion on local solids abundance, the diversity of KBOs - from thoroughly melted, water-ice-mantled objects such as Haumea to those retaining abundant ``supervolatiles" - might be explained by the same drawn-out accretionary process as we envision for the asteroids. The specific scenario explored here assumed KBO initial formation between 16-30 AU and subsequent dynamical emplacement to 30-44AU; this is not a requirement of our model, however.  Crude scaling estimates suggest that the mass of the ``cold classical" KBOs could be formed {\it in situ}, by the physics discussed here, from a local mass density that would not lead to excessive migration of Neptune or a problematic subsequent cleanup. More refined future models of planetesimal formation in the 16-30AM region should, for self-consistency, strive to produce roughly twice the mass we assumed for this first assessment, to allow for the cores of the ice giants themselves. 

\vspace{0.5in}

{\bf Acknowledgements:} We thank Conel Alexander, Mike Brown, John Chambers, Eugene Chiang, Fred Ciesla, Kees Dullemond, Paul Estrada, Will Grundy, Carsten G{\"u}ttler, Scott Kenyon, Noriko Kita, Erika Kurahashi, Bill McKinnon, Hal Levison, Alessandro Morbidelli, Chris Ormel, Alan Rubin, Stu Weidenschilling, and Andras Zsom for helpful conversations and comments on earlier manuscripts, and for providing results in advance of publication. We thank John Chambers and another (anonymous) reviewer for helpful comments that improved the presentation. This work was supported by grants from NASA's Planetary Geology and Geophysics and Origins of Solar Systems Programs. Our group has profited greatly from generous allocations of cpu time on the NASA High-End Computing (HEC) machines at Ames. In addition to raw cycles, expert consultants have provided invaluable help in visualization, parallelization, and optimization. 

\section*{Appendix}
Here we provide some notational clarifications to better connect the distribution functions $P(\Phi,S)$ and $F_p(\Phi,S)$ of this paper to similar functions in CHPD01 and Hogan and Cuzzi(2007). We first note that we have defined $P(\Phi,S)$ in this paper as a probability per unit 
log$_{10}(\Phi)$ and log$_{10}(S)$ (as in Chambers 2010; a hand check for identical parameters shows that our PDF contours are in very good agreement with those of Chambers). Hogan and Cuzzi (2007) are not specific about the use of ${\rm log}_{10}$ {\it vs.} ${\rm log}_{e} = {\rm ln}$, but here we are more explicit. We define $P'(\Phi,S)$, which Hogan and Cuzzi (2007) call $P(\Phi,S)$, as a true differential probability density per unit $\Phi$ and per unit $S$. Since, for instance, $S = e^{{\rm ln} S}$, ${\rm log}S = {\rm log}e \cdot {\rm ln}S$ and $d{\rm log}S = {\rm log}e \cdot dS/S$. We require the functions $P(\Phi,S)$ and $P'(\Phi,S)$ to be separately normalized: $\int_0^{\infty}\int_0^{\infty}P'(\Phi,S)d\Phi dS = 1 = \int_{0}^{\infty}\int_{0}^{\infty}P(\Phi,S)
d{\rm log}\Phi d{\rm log}S$. Since $P'(\Phi,S)d\Phi dS = P'(\Phi,S) (\Phi \cdot d{\rm log}\Phi/{\rm log}e) (S \cdot d{\rm log}S/{\rm log}e)$, it follows that 
$$
P'(\Phi,S) \Phi S = {\rm log}^2e P(\Phi,S).
$$
The quantity $P'(\Phi,S) \Phi S $ was proposed by Hogan and Cuzzi (2007) as a convenient estimate of effective volume fraction ``at" $(\Phi,S)$, as it covers the range $(\Phi \pm \Delta\Phi/2, S \pm \Delta S/2)$ with $\Delta \Phi = \Phi$ and $\Delta S = S$, but here we use instead the quantity $P(\Phi,S)$ in this role. The difference is a factor of ${\rm log}^2e$, and can be explained by $P(\Phi,S)$ having a larger effective binning size $\Delta \Phi = \Phi/{\rm log}e, \Delta S= S/{\rm log}e$. In fact, we find that $P(\Phi,S)$ actually better approximates the more formally exact integrals (below) than does $P'(\Phi,S) \Phi S$. 

To be even more specific, we define $P(\Phi,S)$ by taking the number of cascade outcomes $H$ which lie in bins at $(\Phi,S)$ within some bin size $ d{\rm log}\Phi \cdot d{\rm log}S = \delta^2$, and normalizing by the total number $n$ of cascade outcomes so $P(\Phi,S) = H(\Phi,S,\delta)/n \delta^2$. This makes $P(\Phi,S)$ also a probability (fractional volume) density, but per unit ${\rm log}_{10}(\Phi)$, per unit ${\rm log}_{10}(S)$. It is easy to show that using this definition and $P'(\Phi,S) \Phi S = {\rm log}^2e P(\Phi,S)$, $P(\Phi,S)$ and $P'(\Phi,S)$ are both automatically normalized: $\int_0^{\infty}\int_0^{\infty} P'(\Phi,S)d\Phi dS = \int_0^{\infty}\int_0^{\infty} P(\Phi,S)d{\rm log}\Phi \cdot d{\rm log}S = 1$. 

We also note that the particle concentration $C$ of CHPD01 is simply related to the mass loading factor $\Phi$ of this paper and CHS08. Specifically, $C=n_p/\left< n_p \right> = \rho_p/\left< \rho_p\right>$, where $n_p$ is the number of particles per unit volume, $\left< n_p \right>$ is its global average, and all the particles are of equal mass in our simple treatment. Thus $\Phi = \rho_p/\rho_g =A' C$, where $A' = 0.01 A/A_o$. As in CHPD01 we must distinguish between the fraction of {\it volume} found at some $(\Phi,S)$ and the fraction of {\it particles} found there, because the particles are not randomly distributed in $(\Phi,S)$ space. Because the volume and particle fractions of CHPD01 ($F_V(C),F_p(C)$) are normalized, they can also be written as $F_V(\Phi),F_p(\Phi)$. Then recall that $F_V(\Phi) = \int_0^{\infty}P'(\Phi,S)dS$ so $F_V(>\Phi)=\int_{\Phi}^{\infty}\int_0^{\infty}P'(\Phi',S)d\Phi' dS \rightarrow 1$ as $\Phi \rightarrow 0$, as in CHPD01. We define a particle fraction $F_p(\Phi,S) \propto \Phi P'(\Phi,S)$, which is the 2D extension of the function $F_p(C) \propto C F_V(C)$ of CHPD01, and like it, must be separately normalized. We define a normalization constant $c_p$ such that  $c_p \int_0^{\infty}\int_0^{\infty} \Phi P'(\Phi,S)d\Phi dS =1$; it can easily be shown that the same $c_p$ implies that 
$c_p \int_0^{\infty}\int_0^{\infty} \Phi P(\Phi,S)d{\rm log}\Phi \cdot d{\rm log}S =1$.  

The closest analog to $F_p(C)$ of CHPD01 is $F_p(\Phi) = c_p \Phi F_V(\Phi)= c_p\Phi \int_0^{\infty}P'(\Phi,S)dS$, and has the equivalent cumulative function $F_p(>\Phi)$. 
However, for the purpose of the current paper we are working in a more profoundly 2D regime, where both $\Phi$ and $S$ are important. Thus we will work with the cumulative of $F_p(\Phi,S) = c_p \Phi P'(\Phi,S)$ that represents the fraction of particles lying in clumps that exceed all of our thresholds $(\Phi_1, \Phi_2, S_{min})$ for sandpile formation or primary accretion. {\it Any} clump lying above the entire $S_{min}-\Phi_1$-$\Phi_2$ threshold line is capable of becoming a sandpile (section 3.3.1), and encounters of a wandering particle with all of them should be included in estimation of $t_{enc}$ (section 3.5.1). Thus we define $F_p(>T)$ - the particle fraction lying in all proto-sandpile-clumps at any given time - as an integral over the 2D segment of $(\Phi,S)$ space lying above the threshold value $\Phi_T(S) = max(\Phi_2, \Phi_1(S))$ and to the right of $S_{min}$ (see figure 3). That is, 
\begin{equation}
F_p(>T) = { \int_{S_{min}}^{\infty} \int_{\Phi_T(S)}^{\infty} \Phi P'(\Phi, S) d\Phi dS \over { \int_0}^{\infty} \int_0^{\infty} \Phi P'(\Phi, S) d\Phi dS} = c_p \int_{S_{min}}^{\infty} \int_{\Phi_T(S)}^{\infty} \Phi P'(\Phi, S) d\Phi dS, 
\end{equation}
where the constant $c_p = A_o/A$. This cumulative measure will be dominated by clumps falling closest to the threshold line, and in particular at the peak of the IMF (figure 3) which we characterize by $\Phi = \Phi^*$ and $P(\Phi^*,S^*)=P^*$. For example (see section 3.3.1 and Chambers 2010) the total primary accretion rate of planetesimals $\dot{M}_{pa}$ can be written as 
$$
\dot{M}_{pa} = { 2 \pi (a_2^2-a_1^2) H \beta^{1/2} \over t_{pa}} 
\int_{S_{min}}^{\infty} \int_{\Phi_T(S)}^{\infty} (\Phi \rho_g) P'(\Phi, S) d\Phi dS,
$$
where the double integral is over the entire range of $\Phi,S$ where sandpiles can form, the integrand is the product of particle mass per unit volume {\it in} a clump $\Phi \rho_g$ times the volume {\it fraction in clumps}, and the numerator is the volume in which this transpires using only a narrow region of thickness $\beta^{1/2}$ near the midplane. The denominator $t_{pa}$ is some formation timescale, which we take as $t_{sed}$ but could be shorter (section 3.3.1; Chambers 2010). We can rewrite this as
$$
\dot{M}_{pa} = { 2 \pi (a_2^2-a_1^2) H \beta^{1/2} \rho_g \over t_{pa}} 
\int_{S_{min}}^{\infty} \int_{\Phi_T(S)}^{\infty} \Phi P'(\Phi, S) d\Phi dS.
$$
From expressions given above, it is clear that $\int_{S_{min}}^{\infty} \int_{\Phi_T(S)}^{\infty} \Phi P'(\Phi, S) d\Phi dS = F_p(>T)/c_p = F_p(>T)A/A_o$. Meanwhile we could also approximate $\dot{M}_{pa}$ as
$$
\dot{M}_{pa} = { 2 \pi (a_2^2-a_1^2) H \beta^{1/2} \over t_{pa}} (\Phi^* \rho_g) P^*(\Phi^*,S^*)
$$
where the peak of the IMF is at $(N^*, \Phi^*, S^*)$ and has associated $P(\Phi,S) = P^*$. Physically this amounts to saying that the bulk of the primary accretion {\it is} that which occurs near the peak of the IMF and neglecting the contributions from further down the IMF, but treating the volume fraction as a binned value over a range $\Delta \Phi^* = \Phi^*/{\rm log}e, \Delta S^*= S^*/{\rm log}e$ as described above, based on our definition of $P(\Phi,S)$. Comparison of the above equations shows that taking $\Phi^* P^*(\Phi^*,S^*) = F_p(>T)A/A_o$ makes the expressions equal. We make use of this in section 3.3.1, using numerical validation (in tables 1-4) that $\Phi^* P^*(\Phi^*,S^*) = F_p(>T)A/A_o$ is indeed valid to tens of percent. As described in section 3.5.1 and CHPD01, we can also associate $F_p(>T)$ with the fractional time $F_t(>T)$ spent by a given particle in protosandpile clumps. 
\newpage

\section{References}

\begin{list}{}{\leftmargin \parindent \itemindent -\parindent \itemsep 0in}

\item Alexander, C. A. (2005) From Supernovae to Planets: the view from Meteorites and IDPs; in ``Chondrites and the Protoplanetary Disk", ASP Conference Series, Vol. 341; Edited by Alexander N. Krot, Edward R. D. Scott, and Bo Reipurth. San Francisco: Astronomical Society of the Pacific, 2005., p. 972-1002

\item Aliseda, A., A. Cartellier, F. Hainaux, and J. C. Lasheras (2002) Effect of preferential concentration on the settling velocity of heavy particles in homogeneous isotropic turbulence; J. Flu. Mech. 468, 77-105

\item Bec, J., L. Biferale, M. Cencini, A. Lanotte, S. Musacchio, and F. Toschi (2007) Heavy particle concentration in turbulence at dissipative and inertial scales; Phys. Rev. Lett. 98, 084502 

\item Binzel, R. P., D. Lupishko, M. DiMartino, R. J. Whitely, and G. J. Hahn
(2002) Physical properties of near-Earth objects; in Asteroids III; W. F.
Bottke, jr., A. Cellino, P. Paolicchi, and R. P. Binzel, eds; Univ. of Arizona
Press

\item Bockel{\`e}e-Morvan, D., D. Gautier, F. Hersant, J.-M. Hur{\' e},
and F. Robert (2002) Turbulent radial mixing in the solar nebula as the source
of crystalline silicates in comets; Astron Astrophys. 384, 1107-1118

\item Bosse, T., L. Kleiser, and E. Meiburg (2006) Small particles in homogeneous turbulence: settling velocity enhancement by two-way coupling; Phys. Fluids 18, 027102

\item Bottke, W. F. jr., D. D. Durda, D. Nesvorny, R. Jedicke, A. Morbidelli,
D., Vokrouhlicky, and H. Levison (2005); The fossilized size distribution of
the main asteroid belt; Icarus, 175, 111-140

\item Bottke, W. F., Nesvorny, D., Grimm, R. E., Morbidelli, A., O'Brien, D. P. (2006) Iron meteorites as remnants of planetesimals formed in the terrestrial planet region; Nature 439, 821-824

\item Brauer, F.; Dullemond, C. P.; Henning, Th. (2008) Coagulation, fragmentation and radial motion of solid particles in protoplanetary disks; Astronomy and Astrophysics, 480, 859-877

\item Brearley, A. J. (1993) Matrix and fine-grained rims in the unequilibrated CO3 chondrite, ALHA77307 - Origins and evidence for diverse, primitive nebular dust components; Geochim. Cosmochim. Acta 57, 1521-1550

\item Brearley, A.J., Jones, R.H. (1998). Chondritic meteorites. In: Papike, J.J.
(Ed.), Planetary Materials. In: Rev. Mineral., vol. 36. Mineralogical Society
of America, Washington, DC. Chapter 3 

\item Calvet, N. , Hartmann, L.; Strom, S. E. (2000)  Evolution of Disk Accretion; in ``Protostars and Planets IV" University of Arizona Press; eds Mannings, V., Boss, A.P., Russell, S. S.), p. 377

\item Carballido, A., J. N. Cuzzi, and R. C. Hogan (2009) Relative velocities and radial diffusion of solids in a turbulent protoplanetary disc; submitted to M.N.R.A.S. 
 
\item Carpenter, J. M., S. Wolf, K. Schreyer, R. Launhardt, and Th. Henning (2005) Evolution of cold circumstellar dust around solar type stars; Astrophys. J. 129, 1049-1062

\item Chambers, J. E. (2004) Planetary Accretion in the inner solar system; E.
P. S. L. 223, 241-252

\item Chambers, J. E. (2010) Planetesimal Formation by Turbulent Concentration; submitted to Icarus

\item Chiang, E.; Lithwick, Y.; Murray-Clay, R.; Buie, M.; Grundy, W.; Holman, M.(2007) A Brief History of Transneptunian Space; in Protostars and Planets V, B. Reipurth, D. Jewitt, and K. Keil (eds.), University of Arizona Press, 895-911

\item Chiang, E. and A. Youdin (2009) Ann. Revs. Astron. Astrophys, submitted

\item Ciesla, F. (2009) Two-dimensional transport of solids in viscous protoplanetary disks; Icarus, 200, 655-671.

\item Ciesla, F. J. and J. N. Cuzzi (2006) The evolution of the water
distribution in a viscous protoplanetary disk; Icarus, 181, 178-204

\item Clark, B. E., B. Hapke, C. Pieters, and D. Britt (2002) Asteroid Space
Weathering and regolith evolution; Asteroids III; W. F. Bottke, jr., A.
Cellino, P. Paolicchi, and R. P. Binzel, eds; Univ. of Arizona Press

\item Cuzzi, J. N. (2004) Blowing in the wind: III. Accretion of dust rims by chondrule-sized particles in a turbulent protoplanetary nebula; Icarus, 168,  484-497. 

\item Cuzzi, J. N.; Ciesla, F. J.; Petaev, M. I.; Krot, A. N.; Scott, E. R. D.; Weidenschilling, S. J. (2005) Nebula Evolution of Thermally Processed Solids: Reconciling Models and Meteorites; in ``Chondrites and the Protoplanetary Disk", ASP Conference Series, Vol. 341; Edited by Alexander N. Krot, Edward R. D. Scott, and Bo Reipurth. San Francisco: Astronomical Society of the Pacific, 2005., p.732-773

\item Cuzzi, J. N., A. R. Dobrovolskis, and J. M. Champney (1993) Particle-gas
dynamics near the midplane of a protoplanetary nebula; Icarus, 106, 102-134

\item Cuzzi, J. N. and R. C. Hogan (2003) Blowing in the wind: I. Velocities of
Chondrule-sized Particles in a Turbulent Protoplanetary Nebula; Icarus, Icarus,
164, 127-138.

\item Cuzzi, J. N., Hogan, R. C., and Bottke, W. F., 2010. Towards Initial Mass Functions for Asteroids and Kuiper Belt; 41st L. P. S. C., The Woodlands, Texas. Contribution No. 1533, p.1861

\item Cuzzi, J. N., R. C. Hogan, J. M. Paque, and A. R. Dobrovolskis (2001; CHPD01)
Size-selective concentration of chondrules and other small particles in
protoplanetary nebula turbulence;  Astrophys. J., 546, 496-508

\item Cuzzi, J, ; Hogan, R, C.; Shariff, K. (2008; CHS08) Toward Planetesimals: Dense Chondrule Clumps in the Protoplanetary Nebula, ApJ, 687, 1432-1447

\item Cuzzi, J. N. and S. J. Weidenschilling (2006) Particle-Gas Dynamics and
Primary Accretion; a chapter in ``Meteorites and the Early Solar System, II"; D.
Lauretta and H. McSween, eds.;

\item Cuzzi, J. N. and K. J. Zahnle (2004) Material Enhancement in Protoplanetary Nebulae by Particle Drift through Evaporation Fronts; The Astrophysical Journal, 614, 490-496.

\item Dominik, C. P., J. Blum, J. N. Cuzzi, and G. Wurm (2007) Growth of dust
as initial step towards planet formation; in ``Protostars and Planets V",
University of Arizona Press, B. Reipurth, S. Krot, and E. Scott, eds.;  
\newline http://spacescience.arc.nasa.gov/users/cuzzi/Dominiketal\_PPV.pdf

\item Dubrulle, B., G. E. Morfill, and M. Sterzik (1995) The dust sub-disk in
the protoplanetary nebula; Icarus 114, 237-246

\item Dullemond, C. P. and C. Dominik (2004) The effect of dust settling on the
appearance of protoplanetary disks; Astron. Astrophys. 421, 1075-1086

\item Dullemond, C. P. and C. Dominik (2005) Dust coagulation in protoplanetary
disks: A rapid depletion of small grains; Astron. Astrophys.434, 971-986

\item Elkins-Tanton, L. T. and B. P. Weiss (2009) Chondrites as samples of differentiated planetesimals; 40th Lunar and Planetary Science Conference, The Woodlands, Texas

\item Falkovich, G. and K. R. Sreenivasan (2006) Lessons from hydrodynamic
turbulence; Physics Today 59, 43-49.

\item Farinella, P.; Davis, D. R. (1992) Collision rates and impact velocities in the Main Asteroid Belt; Icarus 97, 111-123.

\item Fernandez, J.A., and Ip, W.H. (1984) Some dynamical aspects of the accretion
of Uranus and Neptune: the exchange of orbital angular momentum
with planetesimals. Icarus 58, 109Ð120.

\item Fleming, T. and J. M. Stone (2003) Local magnetohydrodynamic models of
layered accretion disks; Astrophys. J. 585, 908-920

\item Ford, E. B. and E. I. Chiang (2007)  The Formation of ice giants in a packed oligarchy: instability and aftermath; Astrophys. J. 661, 602-615

\item Fraser, W., and M. E. Brown (2009) Quaoar: a rock in the Kuiper belt; AAS/DPS \#41, 65.03. 

\item Gammie, C. F. (1996) Layered Accretion in T Tauri Disks;Astrophysical Journal 457, 355-362

\item Garaud, P. (2007) Growth and Migration of Solids in Evolving Protostellar Disks. I. Methods and Analytical Tests; Astrophys. J., 671, 2091-2114. 

\item Goldreich, P. and W. R. Ward (1973) The formation of planetesimals;
Astrophys. J. 183, 1051-1061

\item Gomes, R. S. (2003) The origin of the Kuiper Belt high-inclination population; Icarus,  161, 404-418.

\item Gomes, R. S.; Morbidelli, A.; and Levison, H. F. (2004) Planetary migration in a planetesimal disk: why did Neptune stop at 30 AU? Icarus, 170, 492-507

\item  G{\"u}ttler, C., J. Blum, A. Zsom, C. W. Ormel, and C. P. Dullemond (2009) The outcome of protoplanetary dust growth: pebbles, boulders, or planetesimals? I. Mapping the zoo of laboratory collision experiments; Astron. Astrophys. in press

\item Grimm, R. E., W. F. Bottke, D. Durda, E. R. D. Scott, E. Asphaug, and D.
Richardson (2005) Joint thermal and collisional modeling of the H-chondrite
parent body; 36th LPSC, League City, Texas, Abstract 1798.

\item Haghighipour, N. and Boss, A. P. (2003) On gas drag-induced rapid 
migration of solids in a nonuniform solar nebula.  ApJ 598, 1301-1311

\item Haisch, Karl E., Jr.; Lada, Elizabeth A.; Lada, Charles J. (2001) Disk Frequencies and Lifetimes in Young Clusters; ApJ, 553, L153-L156 

\item Hartmann, L. (2005)  Astrophysical Observations of Disk Evolution around Solar Mass Stars; in ``Chondrites and the Protoplanetary Disk" ASP Conference Series, Vol. 341, 2005 A. N. Krot, E. R. D. Scott, and B. Reipurth, eds. 131-144

\item Hevey, P. H. and I.S. Sanders (2006) A model for planetesimal meltdown by 26Al and its implications for meteorite parent bodies; Meteoritics and Planetary Science, 41, 95-106

\item Hogan, R. C. and J. N. Cuzzi (2007) A cascade model for particle
concentration and enstrophy in fully developed turbulence with mass loading
feedback; Phys. Rev. E. 75, 056305 

\item Ida, S.; Guillot, T.; Morbidelli, A. (2008) Accretion and Destruction of Planetesimals in Turbulent Disks; ApJ 686,  1292-1301

\item Jedicke, R., J. Larsen, and T. Spahr (2002) Observational selection effects in asteroid surveys. In {\it Asteroids III} (Eds. W.F. Bottke, A. Cellino, P. Paolicchi, and R.P. Binzel), Univ. of Arizona Press, Tucson, 71-87.

\item Johansen, A.; Oishi, J. S.; MacLow, M.-M.; Klahr, H.; Henning, T.; Youdin, A. (2007) Rapid planetesimal formation in turbulent circumstellar disks; Nature, 448, 1022-1025 

\item Johnson, B. M. and Gammie, C. F. (2005) Vortices in thin, compressible,
unmagnetized disks; Astrophys. J. 635, 149-156

\item Juneja, A. , Lathrop, D. P., Sreenivasan, K. R. and Stolovitsky, G.
(1994) Synthetic turbulence; Physical Review E., 49, 5179 - 5194

\item Kato, S. and A. Yoshizawa (1997) A steady hydrodynamical turbulence in
differentially rotating disks; Publ. Ast. Soc. Jap. 49, 213-220

\item Kenyon (2002) Planet Formation in the outer solar system; P. A. S. P. 114, 265-283

\item Kenyon, S. J.; Bromley, B. C.; O'Brien, D. P.; Davis, D. R. (2008) Formation and Collisional Evolution of Kuiper Belt Objects; in The Solar System Beyond Neptune, M. A. Barucci, H. Boehnhardt, D. P. Cruikshank, and A. Morbidelli (eds.), University of Arizona Press, Tucson, p.293-313

\item Kenyon, S. and Luu, J. X. (1998) Accretion in the early Kuiper belt. I. Coagulation and velocity evolution;  Astron. J., 115, 2136-2160

\item Kimura, M., Hiyagon, H., Palme, H., Spettel, B., Wolf, D., Clayton, R. N., Mayeda, T. K., Sato, T., Suzuki, A., and Kojima, H. (2002), Yamato 792947, 793408, and 82038: The most primitive H chondrites, with abundant refractory inclusions; Meteorit. Planet. Sci., 37, 1417-1434

\item King, T. V. V. and E. A. King (1978) Grain size and petrography of C2 and C3 carbonaceous chondrites; Meteoritics 13, 47-72

\item Kita, N. T.; Nagahara, H.; Togashi, S.; Morishita, Y. (2000)  A short duration of chondrule formation in the solar nebula: evidence from $^{26}$Al in Semarkona ferromagnesian chondrules; Geochimica et Cosmochimica Acta, 64, 3913-3922.

\item Kita, N. T., G. R. Huss, S. Tachibana, Y. Amelin, L. E. Nyquist, and I.
D. Hutcheon (2005) Constraints on the origin of chondrules and CAIs from
short-lived and long-lived radionuclides; Conference on Chondrites and the
Protoplanetary Disk, Kauai, Hawaii, November 2004; ASP Conference Series vol
341, 558-587

\item Kleine, T.; Mezger, K.; Palme, H.; Scherer, E.; MŸnker, C. (2005) Early core formation in asteroids and late accretion of chondrite parent bodies: Evidence from 182Hf-182W in CAIs, metal-rich chondrites, and iron meteorites; Geochimica et Cosmochimica Acta, 69, 5805-5818. 	
\item Kornet, K.; Wolf, S.; Rozyczka, M. (2001) Diversity of planetary systems from evolution of solids in protoplanetary disks; Astronomy and Astrophysics, 378, 180-191

\item Krot, A.N.; Amelin, Y.; Cassen, P.; Meibom, A. (2005) Young chondrules in CB chondrites from a giant impact in the early Solar System; Nature, 436, 989-992

\item Kunihiro, T., Rubin, A. E.; McKeegan, K. D.; Wasson, J. T. (2004) Initial $^{26}$Al/$^{27}$Al in carbonaceous-chondrite chondrules: too little $^{26}$Al to melt asteroids; Geochimica et Cosmochimica Acta, 68, 2947-2957. 

\item Kurahashi, E.; Kita, N. T.; Nagahara, H.; Morishita, Y. (2008) $^{26}$Al-$^{26}$Mg systematics of chondrules in a primitive CO chondrite; Geochimica et Cosmochimica Acta, 72, 3865-3882.

\item LaTourrette, T., and G. J. Wasserburg (1998) Mg diffusion in anorthite: implications for the formation of early solar system planetesimals; Earth Planet. Sci. Lett., 158, 91-108

\item Levison, H.F.; Bottke, W. F.; Gounelle, M.; Morbidelli, A.; Nesvorny, D.; Tsiganis, K, (2009) Contamination of the asteroid belt by primordial trans-Neptunian objects; Nature, 460, 364-366

\item Levison, H. F., A. Morbidelli, C. VanLaerhoven, R. Gomes, and K. Tsiganis (2008) Origin of the structure of the Kuiper Belt during a dynamical instability in the orbits of Uranus and Neptune; Icarus 196, 258-273

\item Lin, D. N. C.; Papaloizou, J. (1985)  On the dynamical origin of the solar system; ``Protostars and Planets II" University of Arizona Press, Tucson, AZ, 981-1072. 

\item Markiewicz, W. J.; Mizuno, H.; V\"{o}lk, H. J.	(1991) Turbulence induced relative velocity between two grains; Astronomy and Astrophysics 242, 286-289
 
\item Malhotra, R. (1995) The origin of Pluto's orbit: implications for the solar
system beyond Neptune. Astron. J. 110, 420Ð429.

\item Markowski, A., Quitt{\'e}, G.; Kleine, T., Halliday, A. N., Bizzarro, M., Irving, A.  J. (2007)  Hafnium-tungsten chronometry of angrites and the earliest evolution of planetary objects; E. P. S. L., 262, 214-229.

\item Marti, K. and T. Graf (1992) Cosmic ray exposure history of ordinary chondrites; Ann. Revs. Earth Plan. Sci. 20, 221-243

\item McKinnon, W. B.; Prialnik, D.; Stern, S. A.; Coradini, A. (2008) Structure and Evolution of Kuiper Belt Objects and Dwarf Planets; in The Solar System Beyond Neptune, M. A. Barucci, H. Boehnhardt, D. P. Cruikshank, and A. Morbidelli (eds.), University of Arizona Press, 213-241

\item McSween, H. Y., Ghosh, A., Grimm, R. E., Wilson, L., and Young, E. D.
(2002) Thermal evolution models of asteroids. Asteroids III; W. F. Bottke, jr.,
A. Cellino, P. Paolicchi, and R. P. Binzel, eds; Univ. of Arizona Press

\item Meneveau, C. and Sreenivasan, K. R. (1991) The Multifractal Nature of
Turbulent Energy Dissipation; J. Fluid Mech. 224, 429-484

\item Metzler, K., A. Bischoff, and D. St{\"o}ffler (1992) Accretionary dust
mantles in CM chondrites: evidence for solar nebula processes; Geochim.
Cosmochim. Acta 56, 2873-2897

\item Morbidelli, A.; Levison, H. F.; Gomes, R. (2008) The Dynamical Structure of the Kuiper Belt and Its Primordial Origin; in The Solar System Beyond Neptune, M. A. Barucci, H. Boehnhardt, D. P. Cruikshank, and A. Morbidelli (eds.), University of Arizona Press, p.275-292

\item Morbidelli, A. W. Bottke, D. Nesvorny, and H. Levison (2009a) Asteroids were born big; Icarus, 204, 558-573.

\item Morbidelli, A.; Levison, H. F.; Bottke, W. F.; Dones, L.; Nesvorny, D. (2009b) Considerations on the magnitude distributions of the Kuiper belt and of the Jupiter Trojans;  Icarus, 202, 310-315.
	
\item Mostefaoui, S.,  Kita, N. T.; Togashi, S.; Tachibana, S.; Nagahara, H.; Morishita, Y. (2002) The relative formation ages of ferromagnesian chondrules inferred from their initial aluminum-26/aluminum-27 ratios; Meteoritics and  Planetary Science, 37, 421-438 

\item Nakagawa, Y., Sekiya, M., and Hayashi, C. (1986) Settling and growth of
dust particles in a laminar phase of a low-mass solar nebula; Icarus, 67,
375-390.

\item Ormel, C. and J. N. Cuzzi (2007) Closed-form expressions for particle relative
velocities induced by turbulence; Astron. Astrophys. 466, 413-420

\item Ormel, C. W.; Cuzzi, J. N.; Tielens, A. G. G. M. (2008) Co-Accretion of Chondrules and Dust in the Solar Nebula; Astrophysical Journal, 679, 1588-1610.

\item Petit, J.-M., A. Morbidelli, and J. E. Chambers (2001) The primordial
excitation and clearing of the asteroid belt; Icarus 153, 338-347

\item Prinn, R. G. (1990) On neglect of angular momentum terms in solar nebula
accretion disk models; Astrophys. J. 348, 725-729

\item Rubin, A. E. (1989) Size-frequency distributions of chondrules in CO3 chondrites; Meteoritics 24,  179-189.

\item Russell, S. S., L. A. Hartmann, J. N. Cuzzi, A. N. Krot, and S. J.
Weidenschilling (2006) Timescales of the protoplanetary disk; a chapter in
``Meteorites and the Early Solar System, II"; D. Lauretta and H. McSween, eds.

\item Ryu, R. and J. Goodman (1992) Convective instability in differentially rotating disks; Astrophysical Journal, 388, 438-450.

\item Safronov, V. (1991) Kuiper Prize Lecture - Some problems in the formation
of the planets; Icarus, 94, 260-271.

\item Scott, E. R. D.; Krot, A. N. (2005) Chondritic Meteorites and the High-Temperature Nebular Origins of Their Components; in ``Chondrites and the Protoplanetary Disk", ASP Conference Series, Vol. 341, Edited by Alexander N. Krot, Edward R. D. Scott, and Bo Reipurth. San Francisco: Astronomical Society of the Pacific, 2005., 15-53

\item Sekiya, M. (1983) Gravitational instabilities in a dust-gas layer and
formation of planetesimals in the solar nebula; Prog. Theor. Physics, 69,
1116-1130

\item Sreenivasan, K. R. and G. Stolovitsky (1995) Turbulent Cascades; J. Stat.
Physics 78, 311-333

\item Stephens, D. C. and K. S. Noll (2006) Detection of six trans-Neptunian binaries with NICMOS: a high fraction of binaries in the cold classical disk; Astron. J. 131, 1142-1148

\item Stepinski, T. F. and P. Valageas (1996) Global evolution of solid matter
in turbulent protoplanetary disks. I. Aerodynamics of solid particles;
Astron. Astrophys. 309, 301-312

\item Stepinski, T. F. and P. Valageas (1997) Global evolution of solid matter
in turbulent protoplanetary disks. II. Development of icy planetesimals.
Astron. Astrophys. 319, 1007-1019

\item Stern, S. A. and J. E. Colwell (1997) Accretion in the Edgeworth-Kuiper Belt: Forming 100-1000 km radius bodies at 30 AU and beyond. Astron. J. 114, 841 - 884 

\item Stewart, S. T. and Leinhardt, Z. M. (2009) Velocity-Dependent Catastrophic Disruption Criteria for Planetesimals; ApJ, 691, L133-L137 

\item Stone, J. M.; C. F. Gammie, S. A. Balbus, and J. F. Hawley (2000)
Transport Processes in Protostellar Disks; in Protostars and Planets IV;
p589-599; V. Mannings, A. P. Boss, and S. S. Russell, eds. Univ. of Arizona
Press

\item Sugiura, N.; Krot, A. N. (2007)  $^{26}$Al-$^{26}$Mg systematics of Ca-Al-rich inclusions, amoeboid olivine aggregates, and chondrules from the ungrouped carbonaceous chondrite Acfer 094; Meteoritics and Planetary Science, 42, 1183-1195

\item Sunshine, J. M.; Bus, S. J.; McCoy, T. J.; Burbine, T. H.; Corrigan, C. M.; Binzel, R. P. (2004) High-calcium pyroxene as an indicator of igneous differentiation in asteroids and meteorites; Meteoritics and Planetary Science, 39, 1343-1357

\item Teitler, S. A.; Paque, J. M.; Cuzzi, J. N.; Hogan, R. C. (2009) Statistical Tests of Turbulent Concentration of Chondrules; 40th LPSC, March 23-27, 2009 in The Woodlands, Texas, id.2388; Meteoritics and Planetary Science, submitted. 

\item Toomre, A. (1964) On the gravitational stability of a disk of stars; 
Astrophys. J., 139, 1217-1238 

\item Trieloff, M., Jessberger, E. K., Herrwerth, I., Hopp, J., Fi{\'e}ni, C., Gh{\'e}llis, M., Bourot-Denis, M., and Pellas, P. (2003) Surface and thermal history of the H-chondrite parent asteroid revealed by thermochronometry; Nature, 422, 502-506

\item Tsiganis, K., Gomes, R., Moriddelli, A., Levison, H. (2005) Origin of the orbital architecture of the giant planets of the Solar System; Nature, 435, 459-463

\item  Turner, N. J., Sano, T., Dziourkevitch, N. (2007) Turbulent Mixing and the Dead Zone in Protostellar Disks; ApJ, 659, 729-737. 

\item  Turner, N. J. and Sano, T. (2008) Dead Zone Accretion Flows in Protostellar Disks; ApJ 679,  L131-L134. 

\item Villeneuve, J. M. Chaussidon, and G. Libourel (2009)  Homogeneous Distribution of $^{26}$Al in the Solar System from the Mg Isotopic Composition of Chondrules; Science 325, 985-988

\item V\"{o}lk, H. J., F. C. Jones, G. E. Morfill, and S. R{\"o}ser (1980)
Collisions between grains in a turbulent gas; Astron. Astrophys. 85, 316-325

\item Wang, L-P and M. R. Maxey (1993) Settling velocity and concentration distribution of heavy particles in homogeneous isotropic turbulence; Journal of Fluid Mechanics (1993), 256, 27-68

\item Wasson, J. T. and G. W. Kallemeyn (1990) Allan Hills 85085 - A subchondritic meteorite of mixed nebular and regolithic heritage; E.P.S.L. 101, 148-161.

\item Weidenschilling, S. (1997) The Origin of Comets in the Solar Nebula: A
Unified Model; Icarus, 127, 290-306

\item Weidenschilling, S. J. (2000) Formation of Planetesimals and Accretion of the Terrestrial Planets; Sp. Sci. Rev. 92, 295-310

\item Weidenschilling, S. J. (2004) From icy grains to comets; in ``Comets II", M. C. Festou, H. U. Keller, and H. A. Weaver (eds.), University of Arizona Press, Tucson, 97-104

\item Weidenschilling, S. J. (2009) How Big Were the First Planetesimals? Does Size Matter? 40th LPSC, March 23-27, 2009 in The Woodlands, Texas, id.1760

\item Woolum, D.  and P.M. Cassen (1999) Astronomical constraints on nebular
temperatures: Implications for planetesimal formation; Meteoritics and
Planetary Science, 34, 897-907

\item Youdin, A. N.; Chiang, E. I. (2004) Particle Pileups and Planetesimal Formation; Ap. J. 601, 1109-1119.

\item Youdin, A. and J. Goodman (2005) Streaming Instabilities in Protoplanetary Disks; ApJ 620, 459-469.

\item Zhu, Zhaohuan; Hartmann, Lee; Gammie, Charles, 2010. Long-term Evolution of Protostellar and Protoplanetary Disks. II. Layered Accretion with Infall; arXiv:1003.1756; Astrophys. J., accepted

\item Zsom. A., C.W. Ormel, C. G{\"u}ttler, J. Blum, and C. P. Dullemond (2009) The outcome of protoplanetary dust growth: pebbles, boulders, or planetesimals? II. Introducing the bouncing barrier; Astron. Astrophys. in press

\end{list}

\end{document}